\documentclass[acmsmall]{acmart}
\renewcommand\footnotetextcopyrightpermission[1]{} 
\usepackage[english]{babel}
\usepackage{color}
\usepackage{xcolor}
\usepackage{inputenc}
\usepackage[ruled,vlined]{algorithm2e}
\SetKwInput{KwInput}{Input}
\SetKwInput{KwOutput}{Output}
\SetKwComment{tcp}{$\triangleright$\ }{}

\usepackage{subcaption}

\usepackage{fp}
\usepackage{xspace}
\usepackage{amsthm}
\usepackage{mathtools}
\usepackage{braket}
\usepackage[T1]{fontenc}
\usepackage{listings}

\usepackage{hyperref}
\usepackage[acronym, hyperfirst=false]{glossaries}

\makeglossaries

\newcommand{\mygls}[1]{%
  \ifglsused{#1}{%
    \glshyperlink[\glsentryshort{#1}]{#1}%
  }{%
    \glstarget{#1}{\gls{#1}}%
  }%
}

\newcommand{\myglspl}[1]{%
  \ifglsused{#1}{%
    \glshyperlink[\glsentryshortpl{#1}]{#1}%
  }{%
    \glstarget{#1}{\glspl{#1}}%
  }%
}
\definecolor{lfdblack}{HTML}{000000}
\definecolor{lfdyellow}{HTML}{E69F00}
\definecolor{lfddgrey}{HTML}{999999}
\definecolor{lfdgreen}{HTML}{009371}
\definecolor{lfdhgrey}{HTML}{beaed4}
\definecolor{lfdred}{HTML}{ed665a}
\definecolor{lfdblue}{HTML}{1f78b4}
\definecolor{bggray}{gray}{0.9}

\newboolean{anonymous}
\setboolean{anonymous}{false}

\addto\extrasenglish{%

}

\makeatletter
\def\calcLength(#1,#2)#3{%
  \pgfpointdiff{\pgfpointanchor{#1}{center}}%
  {\pgfpointanchor{#2}{center}}%
  \pgf@xa=\pgf@x%
  \pgf@ya=\pgf@y%
  \FPeval\@temp@a{\pgfmath@tonumber{\pgf@xa}}%
  \FPeval\@temp@b{\pgfmath@tonumber{\pgf@ya}}%
  \FPeval\@temp@sum{(\@temp@a*\@temp@a+\@temp@b*\@temp@b)}%
  \FProot{\FPMathLen}{\@temp@sum}{2}%
  \FPround\FPMathLen\FPMathLen5\relax
  \global\expandafter\edef\csname #3\endcsname{\FPMathLen}
}
\makeatother

\lstdefinestyle{query}{
  language=SQL,
  stepnumber=1,
  numbersep=10pt,
  tabsize=4,
  showspaces=false,
  showstringspaces=false,
  basicstyle=\linespread{1}\fontfamily{lmtt}\selectfont\small,
  keywordstyle=\color{blue},
  stringstyle=\color{purple},
  upquote=true,
  breaklines=true,
  commentstyle=\color{CadetBlue}
}

\definecolor{mygray}{rgb}{0.643,0.643,0.643}

\newtheoremstyle{property}%
{3pt}%
{3pt}%
{}%
{}%
{\bfseries}%
{:}%
{.5em}%
{}%

\theoremstyle{property}
\newtheorem{theorem}{Property}

\newcommand{\ie}{\emph{i.e.}\xspace}
\newcommand{\eg}{\emph{e.g.}\xspace}
\newcommand{\etal}{\emph{et al.}\xspace}

\newcommand{\suppweb}{\href{https://github.com/lfd/Fast_Quadratisation}{supplementary website}\xspace}
\newcommand{\repropkg}{\href{https://doi.org/10.5281/zenodo.14245587}{reproduction package}\xspace}

\setcopyright{acmlicensed}
\copyrightyear{2026}
\acmYear{2026}
\acmDOI{XXXXXXX.XXXXXXX}

\newacronym{qaoa}{QAOA}{Quantum Approximate Optimisation Algorithm}
\newacronym{lrqaoa}{LR-QAOA}{Linear Ramp QAOA}
\newacronym{qubo}{QUBO}{Quadratic Unconstrained Binary Optimisation}
\newacronym{pubo}{PUBO}{Polynomial Unconstrained Binary Optimisation}
\newacronym{pbf} {PBF} {Pseudo-Boolean Function}
\newacronym{cop} {COP} {Combinatorial Optimisation Problem}
\newacronym{nisq}{NISQ}{Noisy Intermediate-Scale Quantum}
\newacronym{lsr} {LSR} {Local Structure Reduction}
\newacronym{pc}{PC}{Pair Combination}
\newacronym{upc}{UPC}{Unique pair combination}
\newacronym{sat}{SAT}{Satisfiability}
\newacronym{cnf}{CNF}{Conjunctive Normal Form}
\newacronym{dPubo}{\textit{direct PUBO}}{}
\newacronym{optPubo}{\textit{optimised PUBO}}{}
\newacronym{dQubo}{\textit{direct QUBO}}{}
\newacronym{optQubo}{\textit{optimised QUBO}}{}

\begin{document}


\title{It's Quick to be Square:\\ Fast Quadratisation for Quantum Toolchains}

\author{Lukas Schmidbauer}
\email{lukas.schmidbauer@othr.de}
\orcid{0009-0001-7171-0865}
\affiliation{%
  \institution{Technical University of Applied Sciences}
  \city{Regensburg}
  \country{Germany}
}

\author{Elisabeth Lobe}
\email{elisabeth.lobe@dlr.de}
\orcid{0000-0002-3473-8906}
\affiliation{%
  \institution{German Aerospace Center
(DLR), Institute of Software
Technology, Department
High-Performance Computing}
  \city{Braunschweig}
  \country{Germany}
}

\author{Ina Schaefer}
\email{ina.schaefer@kit.edu}
\orcid{0000-0002-7153-761X}
\affiliation{%
  \institution{KIT, Institute of Information Security and Dependability (KASTEL)}
  \city{Karlsruhe}
  \country{Germany}
}

\author{Wolfgang Mauerer}
\email{wolfgang.mauerer@othr.de}
\orcid{0000-0002-9765-8313}
\affiliation{%
  \institution{Technical University of Applied Sciences}
  \city{Regensburg}
  \country{Germany}
}
\affiliation{%
    \institution{Siemens AG, Technology}
    \country{Germany}
}

\renewcommand{\shortauthors}{Schmidbauer et al.}

\begin{abstract}
Many of the envisioned use-cases for quantum computers involve
optimisation processes. While there are many algorithmic primitives to
perform the required calculations, all eventually lead to quantum gates
operating on quantum bits, with an order as determined by the structure
of the objective function and the properties of target hardware.
When the structure of the problem representation is not aligned with structure
and boundary conditions of the executing hardware, various overheads degrading the 
computation may arise, possibly negating any possible quantum advantage.

Therefore, automatic transformations of problem representations play
an important role in quantum computing when descriptions (semi-)targeted at humans must be cast into forms that can be
``executed'' on quantum computers. Mathematically equivalent formulations are known
to result in substantially different non-functional properties depending on hardware,
algorithm and detail properties of the problem. Given the current state
of noisy intermediate-scale quantum (NISQ) hardware, these effects are considerably
more pronounced than in classical computing. Likewise, efficiency of the
transformation itself is relevant because possible quantum advantage may easily be
eradicated by the overhead of transforming between representations. In this paper,
we consider a specific class of higher-level representations, that is, PUBOs, and devise novel automatic transformation mechanisms 
into widely used QUBOs that 
substantially improve efficiency and versatility over the state of the art.
In addition, we conduct a comprehensive investigation of industry-relevant problem formulations and their conversion into a quantum-specific representation, identifying significant obstacles in scaling behaviour and demonstrating how these can be circumvented.
\end{abstract}




\maketitle
\thispagestyle{empty}
\pagestyle{empty}

\section{Introduction}
\label{sec:intro}
\myglspl{cop} encode practically relevant problems, such as finding optimal time schedules or routes in planning and logistics.
Many practically relevant \myglspl{cop} cannot be solved classically in polynomial time and thus need to be approximated.

There are a multitude of possibilities for (a)~encoding a problem mathematically, (b)~transforming the encoding into an 
equivalent representation that can be processed by quantum
algorithms (quadratic polynomials are very frequently used
for this purpose), and (c) transforming the quantum representation and the description of algorithmic 
processing steps into hardware-specific instructions.
Many of the choices that must be taken during this chain of transformations influence properties like size of the problem representation, the size and structure of required interactions, and eventually also the obtained solution quality and performance. 

Quantum compilers (or, more precisely: transpilers) transform a quantum circuit into a hardware-executable representation, which requires, among others, to 
consider the native hardware gate set into which logical
operations must be transformed~\cite{Monz_2009, Shapira_2020, Tanburn_2015, Yue2023}, or which physical qubits
can be brought into direct interaction~\cite{Siraichi_2018, Yamanaka_2015, Hirata_2009, Cowtan_2019, Zhang_2021, Safi_2023}.
Transformations that address part~(b) in the above list may
have to deal with problem representations on higher
abstraction layers. For example, a problem may be formulated
in the form of a mathematical optimisation problem,  
like the representation of the join-ordering problem as a \emph{mixed-integer linear program (MILP)}, which can be transformed by discretization into a \textit{binary integer linear program (BILP)}, which can be transformed into a \mygls{qubo} problem~\cite{schoenberger:23:leap}. 
Starting from a \mygls{qubo} representation, one has again many choices regarding a concrete solver strategy (\eg, \mygls{qaoa}~\cite{Campbell_2021, Majumdar_2021, Harrigan_2021, thelen:24:noisy-qaoa}, Annealing~\cite{Lidar:2018, krueger_2020_SAT, sax_2020_approximate}, Grover search~\cite{Grover_1996}), which then need to be transformed to hardware-compatible representations.

Our paper is concerned with a particular transformation to \myglspl{qubo}, which are a relevant abstraction for quantum computers, since many available hardware vendors only support interactions between a maximum of two qubits~\cite{Hauke2020}, which (non-trivially) translate to at maximum quadratic interactions in problems represented as polynomials. 
The more general form of \mygls{qubo} is called \mygls{pubo}, which allows for higher-degree interactions.

On the one hand, it is possible to directly encode higher-degree terms in quantum circuits and then use later transpilation steps to decompose them to hardware compatible gates.
On the other hand, one can also reduce the degree of higher-degree interactions to quadratic ones and then encode the now quadratic terms in a quantum circuit.
We compared these methods for a specific industry-relevant Job-Shop Scheduling problem with regard to \mygls{qaoa} circuits in a previous work~\cite{schmidbauer_2024_reductions} by using the framework \texttt{\href{https://gitlab.com/quantum-computing-software/quark/-/tree/v1.1}{quark}}~\cite{Lobe_2023, Windgtter_2025} and found beneficial effects for the latter reduction variant on quantum circuit metrics (\ie, number of gates, circuit depth and gate distribution). 
In particular, the reviewed existing reduction method in \texttt{\href{https://gitlab.com/quantum-computing-software/quark/-/tree/v1.1}{quark}}~\cite{Lobe_2023, Windgtter_2025} is able to generate good structural properties.
Meanwhile, the classical effort to compute a reduction also needs to be considered to enjoy any advantage gained from using quantum computers. 
We also showed in \cite{schmidbauer_2024_reductions} that this classical effort is unfeasible with the current implementation for practically relevant problem sizes.
In this paper, by choosing a suited data structure that only changes locally during updates and considers subsequent steps, we are able to alleviate the influence of classical preparatory effort for general reductions from \mygls{pubo} to \mygls{qubo}.
We also want to expand upon the application of our approach beyond a particular Job Shop Scheduling formulation in this work by considering a versatile set of problems --- namely \mygls{sat}.

In classical computing, fault tolerance is a given property of hardware. 
For a quantum computer --- independent of its realisation \cite{Carbonelli2024} --- it is believed that fault tolerance is the missing integral part of enjoying advantages gained from the fundamentally different computational model~\cite{Greiwe_2023, Steane1998, Babbush_2021, Resch_21}.
Even if fault tolerant systems are available, it is still necessary to optimise properties of hardware-executable representations, such as above mentioned circuit metrics, since shorter execution times also come with a financial benefit.
Even more, early fault-tolerant systems are limited and it is unknown if they can be made immune to environmental noise in a macroscopic scale.
Hence, the importance for optimising transformations from high-level representations to low-level hardware-executable representations is pronounced for these upcoming systems.

The rest of this paper is structured as follows: 
\autoref{sec:mathBackground} formalises our problem, establishes
notational conventions and reviews existing reduction methods. 
\autoref{sec:GraphRepresentation} introduces the data structures
that our approach is based on and analyses them mathematically, which paves the way for complexity-theoretical performance gains.
\autoref{sec:mechanism} goes into more detail about algorithmic steps and therefore lays the ground for a correctness argument, 
as well as for a complexity-theoretical analysis in \autoref{sec:Analysis}. 
Furthermore, \autoref{sec:Analysis} shows empirically how a concrete implementation performs in comparison to a currently available implementation.
Additionally, we show in \autoref{sec:EffectsOnKSat} how our approach fits into transformation paths for quantum computers and explicitly consider the effects of transformations on $k$-\mygls{sat} instances that are similar to industry instances in depth.
We conclude our study in \autoref{sec:concl}.
The paper is augmented by a \suppweb and a comprehensive \repropkg~\cite{mauerer_22_QSaner} 
(link in PDF) that allows for extending our work.

\section{Mathematical Background}
\label{sec:mathBackground}
\subsection{Pseudo-Boolean Functions}
\label{ssec:MBackPBF}
A prominent technique to formulate \myglspl{cop} includes \myglspl{pbf}~\cite{pelofske2024short, Bian_2010, Zielinski:2023}.
On the one hand, they are suitable for encoding optimisation problems with $\mathcal{NP}$-complete decision variants, given their complexity-theoretic properties~\cite{cipra2000ising, Glover_2018}.
On the other hand, they provide a consolidated interface to encode many \myglspl{cop} in quantum frameworks~\cite{Goech_2020, Aramon_2019} --- making them a fitting choice for abstraction and ease of integration in a quantum toolchain.

A \mygls{pbf} is a function
\begin{equation}
    f: \{0,1\}^n \rightarrow \mathbb{R}.
\end{equation}
Every \mygls{pbf} assumes a multi-linear polynomial representation~\cite{Boros_2002}:
\begin{equation}
  \label{eq:multi_linear_polynomial}
  f(x_1, \ldots , x_n) = \sum_{S\subseteq \{1,\ldots ,n\}} \alpha_S \prod_{j\in S}x_j,
\end{equation}
where $\alpha_S \prod_{j\in S}x_j$ is called a monomial of $f$ and $\alpha_S \in \mathbb{R}$. 
We always refer to this representation in the following, since it is unique with respect to monomials with non-zero coefficients $\alpha_S$.
For example, 
\begin{equation}
\label{eq:examplePBF1}
    f(x_1, \ldots, x_6) = \pi x_1x_2x_3 - 13 x_2x_4x_5x_6 + 7 x_1x_3 
\end{equation}
is a \mygls{pbf} and $\pi x_1x_2x_3$, $- 13 x_2x_4x_5x_6$ and $7 x_1x_3$ are monomials of $f$. 
We say $m \in f$ for a monomial $m= \alpha_S \prod_{j \in S} x_j$ with index set $S$, if we have $\alpha_S \neq 0$ in the representation of $f$ according to  \autoref{eq:multi_linear_polynomial}. 
We also use this notation in particular for `unweighted' monomials with $m= \prod_{j \in S} x_j \in f$.
Analogously, we say $x_j \in m$, if $j \in S$ for the corresponding index set $S$ of $m$.
Moreover, we define the degree-$k$ density of $f$ by the ratio of actually present to possible monomials of degree~$k$ in~$f$, 
\(d_{k} = t_{k}/\binom{n}{k}\), 
where the degree of a monomial\footnote{A monomial is itself also a \mygls{pbf}.} $m$ is the number of variables it contains.
For example, for $m = - 13 x_2x_4x_5x_6$, we have $\operatorname{deg}(m) = 4$. 
A short notation for the degree of a monomial is $|m| \coloneq \operatorname{deg}(m)$.
Furthermore, the degree of a \mygls{pbf} $f$ is the maximum degree of its monomials.
For the example of Eq.~\ref{eq:examplePBF1}, we thus have
\begin{equation*}
    \operatorname{deg}(f) = \max\{\operatorname{deg}(x_1x_2x_3),\operatorname{deg}(x_2x_4x_5x_6), \operatorname{deg}(x_1x_3)\} = 4.
\end{equation*}

In the context of quantum computing, \mygls{qubo} problems are a highly-used abstraction from hardware-specific peculiarities~\cite{Glover_2018, Kochenberger_2014, Punnen_2022}. 
They are a standard interface to widely used frameworks, such as quantum annealers \cite{Goech_2020} and digital annealing \cite{Aramon_2019}.
\myglspl{qubo} are minimisation problems of quadratic \myglspl{pbf} $f$: 
\begin{equation*}
    \min_{\vec{x} \in \{0,1\}^n} f(\vec{x}).
\end{equation*}
Usually a possibly existing constant term in $f$ (i.e. when $\alpha_\emptyset \neq 0$) is omitted directly from the optimisation problem, since it only shifts the optimisation landscape.

Typically, when formulating optimisation problems for real world applications, higher-degree terms can provide better expressivity and are sometimes necessary to encode constraints \cite{bayerstadler_21}.
\text{(In-)equality} constraints can be encoded in \myglspl{qubo} by adding suiting penalty terms~\cite{Glover_2018}.
For example, it is not trivial to include the absolute value of terms in a \mygls{pbf}. 
However, one can square terms to achieve a similar effect that, when applied to a series of degree-$2$ monomials, results in degree-$3$ and degree-$4$ monomials.
The resulting higher-degree monomials can no longer be directly mapped to \mygls{qubo} problems and instead require an additional transformation step --- also called \emph{quadratisation}.

\subsection{Quadratisation}
\label{ssec:MBackQuadratisations}
Starting from a higher-degree \mygls{pbf} $f$, there are many methods, reviewed by Dattani~\cite{Dattani_2019}, to reduce the degree of $f$.
For example, it is possible to split the objective function~\cite{Okada_2015} or to pre determine variable assignments and then exclude monomials in special cases~\cite{Hiroshi_2014}.
A versatile \emph{quadratisation} method, reviewed by Boros \cite{Boros_2002}, can reduce the degree of an arbitrary \mygls{pbf} $f$ to degree-$2$, \ie, quadratise $f$, and is thus suited for an automatic transformation.
In essence, it works on the multi-linear representation of $f$ by iteratively choosing a variable pair $x_ix_j$ and replacing it by a new binary variable $y_h$. 
By introducing a constraint term~\cite{Boros_2002}
\begin{equation}
\label{eq:PenaltyTermIntro}
    \operatorname{p}(x_i,x_j,y_h) = 3y_h + x_ix_j - 2x_iy_h -2x_jy_h,
\end{equation} 
which fulfils
\begin{equation}
\label{eq:PenaltyRequirement}
  \begin{split}
    &x_ix_j = y_h \Rightarrow \operatorname{p} = 0 \\
    &x_ix_j \neq y_h \Rightarrow \operatorname{p} > 0,
  \end{split}
\end{equation}
it is possible to preserve the values of $f$ under the minimisation of the newly introduced variable. 
The constraint term $\operatorname{p}$ (also called the penalty) may need to be scaled by a constant $c \in \mathbb{R}^+$ when added to the objective function $f$ to achieve the value preservation\footnote{One could for example define $c$ as the sum of all positive monomial coefficients in the non-reduced function $f$, which may however not be optimal for the optimisation landscape in particular with regard to quantum annealers.}.
Newly introduced variables can be replaced as well, such that, at the end of the iteration, the new \mygls{pbf} is just quadratic but represents the original one. 
More technically, a \mygls{pbf} $f'(\vec{x},\vec{y})$ is a \emph{quadratisation} of $f(\vec{x})$, if $f'(\vec{x},\vec{y})$ is a quadratic \mygls{pbf} ($\operatorname{deg}(f') = 2$) in $\vec{x} = x_1,\ldots ,x_n$ and $\vec{y} = y_{1},\ldots ,y_{m}$\footnote{Note that it may be necessary to shift variable indices when the same variable name is used.}, and satisfies:
\begin{equation}
  \label{eq:QuadratizationMinimumPreserving}
  f(\vec{x}) = \min_{\vec{y}\in \{0,1\}^m} f'(\vec{x},\vec{y}) \; \forall \vec{x} \in \{0,1\}^n.
\end{equation}

\begin{algorithm}[H]
\caption{Basic steps for iterative \emph{quadratisation}.}
\label{alg:quark}
\SetAlgoLined
\LinesNumbered
\DontPrintSemicolon
    \KwInput{\mygls{pbf} $f$}
    \KwOutput{\mygls{pbf} $f'$ with $\deg(f') \leq 2$, penalty \mygls{pbf} $p$}
    $h \gets 1$\;
    $p \gets 0$\;
    \While{$\deg(f) > 2$}{
        $\{x_i,x_j\} \gets \textsc{get\_next\_var\_pair}(f)$\;
        $f \gets \textsc{replace\_var\_pair}(f, x_i, x_j, y_h)$\;
        $p \gets p + \operatorname{p}(x_i, x_j, y_h)$\;
        $h \gets h+1$\;
    }
    \Return{$f$, $p$}
\end{algorithm}

However, this choice can influence the structural properties of the resulting quadratic \mygls{pbf}~$f'$, which can thus translate to varying properties in quantum programs that solve the quadratic \mygls{pbf}~$f'$.
\autoref{alg:quark} shows the basic structure of an iterative \emph{quadratisation}, as for example done in \texttt{\href{https://gitlab.com/quantum-computing-software/quark/-/tree/v1.1}{quark}}~\cite{Lobe_2023, Windgtter_2025}, where we iteratively choose the next candidate variable pair with \textsc{get\_next\_var\_pair(.)}, replace it in all monomials of $f$ with \textsc{replace\_var\_pair(.)} and add the constraint term (\autoref{alg:quark}: l.~6). 
The function \textsc{get\_next\_var\_pair(.)} is the decisive part of varying structural properties in $f'$ and at the same time influences the run time decisively as evaluating the number of occurrences involves inefficient, repeated searching for a variable pair in all monomials of $f$ per iteration in the shown implementation.

In \cite{schmidbauer_2024_reductions} we have already discussed the different choices of the next variable pair in each iteration, that lead to vastly different degree-$2$ densities $d_2$ for $f'$:
\begin{enumerate}
    \item \textit{Dense}: Choosing the variable pair that appears most often among all monomials.
    \item \textit{Medium}: Choosing the variable pair that appears most often among all highest degree monomials.
    \item \textit{Sparse}: Choosing the first variable pair of a monomial with highest degree.
\end{enumerate}
The \textit{Dense} selection leads to $d_2$ tending towards $1$ and the \textit{Sparse} selection leads to $d_2$ tending towards $0$ for an increasing size of the tested polynomials of degree~$4$, which stem from a Job-Shop Scheduling problem. 
This means that the resulting quadratic polynomials are densely and sparsely packed with terms of degree-$2$ respectively.
Whether this convergence behaviour also holds for other polynomials is of interest for this work.
However, the time to compute the \emph{quadratisation} using the \textit{Dense} and \textit{Medium} selection type even for small problem instances was shown in~\cite{schmidbauer_2024_reductions} to be already in the order of days. 
Hence, we develop an efficient and more versatile algorithm for reductions and for this introduce an efficient graph structure in \autoref{ssec:Fundamentals} and prove important properties that lay the basis for a complexity theoretic performance gain in \autoref{ssec:PropertiesLemmata}.

\section{Graph Representation}
\label{sec:GraphRepresentation}
\subsection{Fundamentals}
\label{ssec:Fundamentals}

Polynomials can be represented in a variety of ways~\cite{Knuth:1997}.
In our implementation, to efficiently store the relevant information about the input \mygls{pbf} $f$, we create a multi-graph $G_f = (V_f,E_f)$ by iterating over all monomials $m \in f$ and adding an edge between nodes~$i$ and~$j$ if the variable pair $x_ix_j$ is part of $m$ (\ie, $x_i \in m \land x_j \in m$).
We enrich the graph with further information such that each edge refers to the monomial it stems from via an edge label to be able to differ between multi-edges. 
This is realised by firstly assigning an arbitrary but fixed and unique index to each monomial $m \in f$, as in \autoref{tab:QuarkPolyDict}, and secondly using these indices as edge labels.  
Internally, a dictionary represents this relation --- allowing for fast average case access.

\begin{table}[htb]
    \centering
    \caption{Internal representation of example \mygls{pbf} of \autoref{eq:examplePBF1} with the associated running index $z$, which uniquely identifies each monomial.}
    \label{tab:QuarkPolyDict}
    \begin{tabular}{cccc}\toprule
        Running index $z$   & Index set $S$     & $\prod_{j \in S} x_j$ &  $\alpha_S$ \\ \midrule
                        $1$ & $\{1,2,3\} $      & $ x_1x_2x_3$          & $\pi$ \\
                        $2$ & $\{2,4,5,6\}$     & $x_2x_4x_5x_6$        & $-13$ \\
                        $3$ & $\{1,3\}$         & $x_1x_3$              & $7$   \\
         \bottomrule
    \end{tabular}
\end{table}

\begin{figure}[htb]
    \centering
    \includegraphics[]{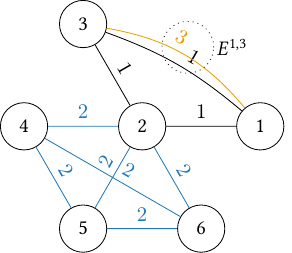}
    \caption{Multi-graph representation of $f(x_1, \ldots, x_6) = \pi x_1x_2x_3 -13 \textcolor{lfdblue}{x_2x_4x_5x_6} + 7 \textcolor{lfdyellow}{x_1x_3}$, where edge labels correspond to monomial indices according to \autoref{tab:QuarkPolyDict}.}
    \label{fig:Graph_example}
\end{figure}

More formally, we define the set of edges $E_f$ by including the edge label such that
\(E_f \subseteq V_f \times V_f \times \mathbb{N}\).
For the example of Eq.~\eqref{eq:examplePBF1}, recall that $f(x_1, \ldots, x_6) = \pi x_1x_2x_3 -13 \textcolor{lfdblue}{x_2x_4x_5x_6} + 7 \textcolor{lfdyellow}{x_1x_3}$ and let its monomial index mapping be defined as in \autoref{tab:QuarkPolyDict}.
Then, the set of edges $E_f$ is filled by iterating over all monomials present in~$f$, calculating their variable pair combinations and adding the index to each edge according to the respective monomial:
\begin{equation*}
    \begin{aligned}
        E_f = \big\{&(1,2,1), (1,3,1), (2,3,1),\\
        &\textcolor{lfdblue}{(2,4,2),(2,5,2),(2,6,2),(4,5,2),(4,6,2),(5,6,2)}, \\
        &\textcolor{lfdyellow}{(1,3,3)}\big\}.
    \end{aligned}
\end{equation*}
It suffices to consider undirected edges for \myglspl{pbf} due to the commutative property of multiplication. 
In the following, we suppose that for any edge $e = (i,j, z) \in E_f$, it holds that \(i < j\). We also exclude self-edges, since $x^2=x\; \forall x \in \{0,1\}$.
We let 
$E_f^{i,j} = \{(i,j,z) \in E_f\} \subseteq E_f$ denote the set of edges between nodes $i,j \in V_f$.
Furthermore, we let the multiplicity $\beta_f(i,j) = \big|E_f^{i,j}\big|$ denote the number of edges between nodes $i,j \in V_f$. 
An example for the multi-graph representation is depicted in \autoref{fig:Graph_example}.

Let $f: \{0,1\}^n \to \mathbb{R}$ be a \mygls{pbf} and let $G_f = (V_f,E_f)$ be the corresponding graph.
We consider the set of all multiplicities in $G_f$: $B_f = \{\beta_f(i,j) \,|\, i,j \in V_f\}$.
Firstly, $f$ might be constant ($\operatorname{deg}(f) = 0$) or only consist of single variable monomials ($\operatorname{deg}(f) = 1$).
In both cases, $G_f$ has no edges and thus $B_f = \{0\}$.
Secondly, if $\operatorname{deg}(f) \geq 2$, monomials in $f$ introduce edges to $G_f$ and therefore $a \in |B_f|, a \in \mathbb{N}$.
Since a variable pair $x_ix_j$ can at maximum occur in every (at least quadratic) monomial in~$f$, the corresponding multiplicity $\beta(i,j) \leq T_f$, where $T_f$ denotes the total number of monomials in $f$, and $B_f \subseteq \{1,\ldots,T_f\}$.
Furthermore, the number of different multiplicities $|B_f|$ is upper bounded by the number of monomials $m \in f$ with $|m| \geq 2$.

We can now define a function $R_f$ that retrieves the node-pairs with multiplicity $\beta \in B_f$:
\begin{equation*}
\begin{aligned}
    R_f:
        \beta &\mapsto \left\{\{i, j\}: i, j \in V_f, \beta_f(i,j) = \beta\right\}.
\end{aligned}
\end{equation*}
Take into consideration that $R_f$ maps to disjoint subsets of node-pairs, that is $R_f(\beta_1) \cap R_f(\beta_2) = \emptyset$ $\forall \beta_1 \neq \beta_2 \in B_f$.
By construction, the size of all sets that $R_f$ maps to is given by $\sum_{\beta \in B_f} |R_f(\beta)| = |V_f|^2$.
\autoref{tab:SortedDictofDicts1} continues the example from \autoref{fig:Graph_example} and shows the mapping $R$ for its graph.
Note that function $R$ ranks variable pairs based on their occurrence in other monomials to compute the next candidate variable pair.
It is motivated by the fact that reducing a variable pair which occurs in many monomials reduces the number of following iterations in the \emph{quadratisation} process on the one hand. 
On the other hand, we can deliberately select a variable pair, which occurs in less monomials --- increasing the number of variables and lowering the resulting \mygls{pbf}'s density.
Let $B_f = \{\beta_1, \ldots, \beta_{|B_f|}\}$, where $\beta_1 < \ldots < \beta_{|B_f|}$.
Hence, $\beta_n \in B_f$ lets us compute the $n$-th multiplicity via a percentile $q$:
\begin{equation}
\label{eq:BetaTildePercentile}
    \tilde{\beta}_q \coloneq \beta_{\lceil q \cdot |B| \rceil}.
\end{equation}
Then, function $R_f$ allows access to node-pairs with such multiplicity:
\begin{equation}
\label{eq:RfBetaTildeNodePairs}
    R_f(\tilde{\beta_q}).
\end{equation}
With the isomorphic nature of variable pair occurrence in monomials and multiplicity in the graph, we can therefore choose suiting variable pairs based on their occurrence via a percentile $q$ in each iteration. 

Furthermore, we propose an algorithm that neither needs to consider $R_f(0)$ (unconnected node-pairs) nor $R_f(1)$ (node-pairs connected by a single edge).
This is an important insight for functions $f$ that induce a sparse graph $G_f$, since the mapping $R_f$ changes during reduction steps --- thus saving computational effort, when omitting $R_f(0)$ and $R_f(1)$ \footnote{We also exclude $0$ and $1$ from $B_f$.}.

During a reduction, monomials change and thus edges in the graph are removed or added. 
This leads to changing multiplicities on node-pairs in the graph. 
Therefore, set $B_f$ changes and node-pairs need to be reallocated to the correct set in $R(\beta)$.
Consider that this additional local effort recompenses when searching for the next variable pair, based on its number of occurrences.

\begin{table}[htb]
    \centering
    \caption{Mapping $R_f$ of increasing multiplicity to sets of node-pairs for example of \autoref{fig:Graph_example}.}
    \label{tab:SortedDictofDicts1}
    \begin{tabular}{cp{0.6\linewidth}}\toprule
        Multiplicity $\beta \in B_f$ & Set of node-pairs with 
        multiplicity $\beta$ \ie, $R_f(\beta)$\\ \midrule
        0   & $\{\{1,4\},\{1,5\},\{1,6\},\{3,4\},\{3,5\},\{3,6\}\}$ \\
        1   & $\{\{1,2\},\{2,3\},\{2,4\},\{2,5\},\{2,6\},\{4,5\},\{4,6\},\{5,6\}\}$ \\
        2   & $\{\{1,3\}\}$ \\ \bottomrule
    \end{tabular}
\end{table}

\subsection{Properties}
\label{ssec:PropertiesLemmata}
In the following, we propose a reduction algorithm that iteratively selects a multi-edge (\ie, $e \in E_f^{i,j}$ with $\beta_f(i,j) > 1$) of a starting \mygls{pbf} $f_0 \vcentcolon = f$ with $\operatorname{deg}(f) > 2$, reduces the corresponding monomials and updates the graph structure until no multi-edges are left --- ending in a \mygls{pbf} $f_t$ after $t$ steps. 
However, $f_t$ might not be quadratic (\ie, $\operatorname{deg}(f_t) > 2$) as there might be monomials left that do not share variable pairs\footnote{For instance:  $x_1x_2x_3 \in f_t$ and $x_3x_4x_5x_6 \in f_t$.} and thus do not introduce multi-edges in $G_{f_t}$.
Any remaining necessary reduction steps do not introduce multi-edges to the graph corresponding to a subsequent \mygls{pbf} $f_{t+i}, i \in \mathbb{N}$~\cite{schmidbauer_2024_reductions}.
We go into more detail about the inner workings of this algorithm in \autoref{sec:mechanism}.

In the following, we list important properties that pave the way for algorithmic optimisation of \autoref{alg:quark} and are the cornerstone for our proposed algorithm in \autoref{sec:mechanism}, as well as for bounding runtime in \autoref{sec:Analysis}.
For all properties, we assume that $f: \{0,1\}^n \mapsto \mathbb{R}$ is a \mygls{pbf} and $G_f(E_f, V_f)$ its corresponding graph.
Furthermore, we let $x_ix_j$ be the variable pair that is going to be replaced in $f$.

\begin{theorem}
\label{thm:MultiEdgeGuaranteedHigherOrderMonomial}
    If $G_f$ has a node-pair $i, j \in V_f$ with multiplicity $\beta_f(i,j) > 1$, $f$ contains a monomial $m$ with $\operatorname{deg}(m) > 2$.
    For $z$ being the index of $m$, the edge $(i, j, z) \in E_f^{i,j}$, that is, it is one of the multiple edges between $i$ and $j$. 
\end{theorem}
\begin{proof}
    By construction only degree-$k$ monomials with $k \geq 2$ introduce edges to the graph. 
    With the representation of a PBF from \autoref{eq:multi_linear_polynomial}, monomials are unique\footnote{Monomials are totalled. For example, let $f(x_1,x_2,x_3) = x_2x_1$. Then, $f(x_1,x_2,x_3) + x_1x_2 = 2 x_1x_2$.}.
    That means, apart from $x_ix_j$, there is no other degree-$2$ monomial containing $x_i$ and $x_j$.
    Hence, any further edge between nodes~$i$ and~$j$ must stem from a monomial with degree larger than two, providing the multiplicity larger than one. Let this be $m$ with index $z$. By construction of $G_f$, $m$ produces edge $(i, j, z)$ in $G_f$. 
\end{proof}

\begin{theorem}
\label{thm:c4alreadybetweenij}
    All monomials that need to be updated during that reduction step, correspond to indices on edges between nodes $i$ and $j$. 
\end{theorem}
\begin{proof}
    Monomials that do not contain $x_i$ and $x_j$ are by definition invariant under the effect of reduction (see \autoref{ssec:MBackQuadratisations}).
    Hence, it suffices to show that all monomials containing $x_i$ and $x_j$ already occur on edges $e \in E_f^{i,j}$ (\ie, edges between nodes $i$ and $j$).
    Without loss of generality, let $m = x_1x_2x_3\ldots x_ix_j, \, i<j$ be an arbitrary monomial $m$, such that $x_i \in m \land x_j \in m$ and let $z$ be its corresponding index.
    It introduces an edge $(i,j,z)$. 
    Furthermore, it introduces edges $(a, i, z)$ and $(a, j, z)\; \forall a \in \{1,\ldots,i-1\}$ or in other words edges that are not between nodes $i$ and $j$, but are already associated to $m$ on $(i, j, z)$ via $z$.
\end{proof}

\begin{theorem}
\label{thm:PenaltyInvariant}
    The edges introduced by the penalty term $\operatorname{p}(x_i,x_j,y_h) = 3y_h + x_ix_j - 2x_iy_h -2x_jy_h$, that is, $(i,j, z_1), (i,h, z_2), (j,h, z_3)$ for some $z_1, z_2, z_3$ not equal to existing running indices (see \autoref{tab:QuarkPolyDict}) stay invariant under the effect of further reduction steps.
\end{theorem}
\begin{proof}
    Since $y_h$ is the newly added variable, $(i,h,z_2), (j,h,z_3)$ are unique (\ie, $\beta(i,h) = \beta(j,h) = 1$). 
    $y_h$ replaces the variable pair $x_ix_j$.
    Consequently, $x_ix_j$ only occurs in the penalty term, which is quadratic ($\operatorname{deg}(p) = 2$).
    Hence, $(i,j,z_1)$ represents a single edge in the graph and is thus not a valid choice for an algorithm that only selects multi-edges.
    Furthermore it is not a valid choice for any subsequent steps, since it is already quadratic.
\end{proof}
As as side note, suppose one chooses the single-edge $(i,j,z_1)$ and therefore the variable pair $x_ix_j$ after it has been reduced. 
Even then, the edges from the previous reduction remain invariant.
The same applies for any other degree-$2$ monomial from the penalty term (\ie, $x_iy_h$ and $x_jy_h$).

\begin{theorem}
\label{thm:ChangingEdgesMustConnectToiOrj}
    Edges $e \in E_{f_t}$, not connected to $i$ or $j$, are invariant under reduction of $x_ix_j$ in $E_{f_{t+1}}$.
\end{theorem}
\begin{proof}
    Let $m_{t} = x_1x_2\ldots x_kx_ix_j$ ($k<i<j$) be a monomial and let $m_{t+1} = x_1x_2\ldots x_ky_h$ be its reduced version.
    Let $P_S = \{\{i,j\} \, |\, i \in S \land j \in S \land i \neq j \land \alpha_S \neq 0\}$ be the two combination set of a monomial specified by the subset of indices $S$.
    Then, 
    \begin{equation}
        P_{\{1,2,\ldots ,k\}} = P_{m_{t}} \setminus \{\{1,i\}, \{2,i\}, \ldots , \{k,i\}, \{1,j\}, \{2, j\}, \ldots , \{k,j\}\} = P_{m_{t+1}} \setminus \{\{1,h\}, \{2,h\}, \ldots \}.
    \end{equation}
    Remark: $m_{t+1}$ introduces edges to $E_{f_{t+1}}$ between node-pairs $\{\{1,h\}, \{2,h\}, \ldots \}$.
\end{proof}

\begin{theorem}
\label{thm:NodeOnlyConnectedToEitherIorJEdgesInvariant}
    When a node $a \in V_f$ is connected to exactly on of the nodes $i$ or $j$, its edges are invariant under reduction.
\end{theorem}
\begin{proof}
    If an edge $(a,i,z) \in E_f$ is affected by reduction, $z$ must refer to a monomial $m$ containing both $x_i$ and $x_j$.
    Without loss of generality, let $m = \ldots x_ax_ix_j$. 
    Therefore, $(a,j,z) \in E_f$, which contradicts the assumption that node $a$ is connected to either $i$ or $j$.
    This argument is analogous for $(a,j,z) \in E_f$.
\end{proof}

\begin{theorem}
\label{thm:MaxNumberOfEdgesBetweenNodes}
    We consider nodes $i$, $j$ and their multiplicity $\beta_f(i,j)$.
    Then, 
    \begin{equation}
        \beta_f(i,j) \leq \sum_{k=0}^{\operatorname{deg}(f)-2} \binom{n-2}{k} \leq \sum_{k=0}^{n-2} \binom{n-2}{k}.
    \end{equation}
\end{theorem}
\begin{proof}
    $\beta_f(i,j)$ depends on the number of monomials $m \in f$ with $|m| \geq 2$ containing the variable pair $x_ix_j$. 
    There are $\sum_{k=0}^{\operatorname{deg}(f)-2} \binom{n-2}{k}$ many such degree-$k+2$ monomials in $f$ at maximum.
    Therefore, $\beta_f(i,j) \leq \sum_{k=0}^{\operatorname{deg}(f)-2} \binom{n-2}{k} \leq \sum_{k=0}^{n-2} \binom{n-2}{k}$.
\end{proof}

Let $f$ denote a higher-degree \mygls{pbf} and let $T_f$ denote the number of monomials in $f$.
At maximum there are 
\begin{equation}
\label{eq:MaximumNumberofIntroducedVariables}
    \sum_{\substack{m \in f,\\|m| \geq 2}}{(|m|-2)} = -2 T_f + \sum_{\substack{m \in f,\\|m| \geq 2}} {|m|}
\end{equation} newly introduced variables --- or similarly iterations in the reduction process.
Thus, reducing multiple monomials at once, that is, when the candidate variable pair occurs in multiple monomials, decreases the remaining steps by the number of influenced monomials.
For example, let $f(x_1,x_2,x_3,x_4) = x_1x_2 + x_1x_2x_3 + x_1x_2x_3x_4$. 
The maximum number of introduced variables is $0 + 1 + 2 = 3$.
When choosing, $x_3x_4$ as the first reduction pair, $x_2x_3$ as the second and $x_2y_0$ as the third, then the above term is sharp: $x_1x_2 + x_1x_2x_3 + x_1x_2x_3x_4 \rightarrow x_1x_2 + x_1x_2x_3 + x_1x_2y_1 \rightarrow x_1x_2 + x_1y_2 + x_1x_2y_1 \rightarrow x_1x_2 + x_1y_2 + x_1y_3$.
However, when choosing $x_1x_2$ as the first reduction pair and $x_3x_4$ as the second, we can quadratise $f$ in $2$ steps:
$x_1x_2 + x_1x_2x_3 + x_1x_2x_3x_4 \rightarrow y_0 + y_0x_3 + y_0x_3x_4 \rightarrow y_0 + y_0x_3 + y_0y_1$.\footnote{These examples do not show the already quadratic penalty term.}
The size of the highest degree monomial gives the minimum number of reduction steps or introduced variables via Boros' method~\cite{Boros_2002}, that is, 
\begin{equation}
\label{eq:MinimumNumberofIntroducedVariables}
    \max_{\substack{m \in f,\\|m| \geq 2}}(|m|-2) = \max_{\substack{m \in f,\\|m| \geq 2}}(|m|) -2.
\end{equation}

\begin{figure}[htb]
    \centering
    \includegraphics[]{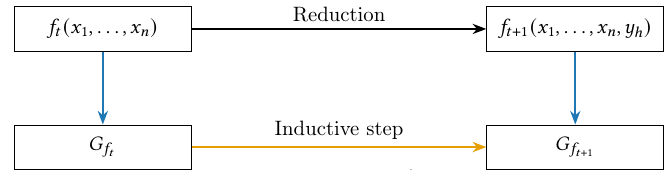}
    \caption{Known reduction (black), introduced construction (\textcolor{lfdblue}{blue}) and graph evolution (\textcolor{lfdyellow}{yellow}).}
    \label{fig:InductiveGraphEvolution}
\end{figure}

Although the graph structure is by construction intertwined with the multi-linear polynomial representation of $f$, we show the necessary steps to arrive at graph $G_{f_{t+1}}$ based on $G_{f_t}$.
This is to simplify the algorithm by concentrating on the graph representation.
Consider an arbitrary reduction step from $f_t$ to $f_{t+1}$. 
$f_t$ and $f_{t+1}$ both induce a graph, that is $G_{f_t}$ and $G_{f_{t+1}}$ by construction.
We now consider how $G_{f_t}$ can be transformed to arrive at $G_{f_{t+1}}$ without explicitly considering the effects of reduction on monomials in $f_t$.
\autoref{fig:InductiveGraphEvolution} shows the open relation.

Following \autoref{thm:ChangingEdgesMustConnectToiOrj}, we only need to consider edges connected to nodes $i$ or $j$.
Following the argument of \autoref{thm:c4alreadybetweenij}, the set of indices on edges between nodes $i$ and $j$ refers to all monomials that need to be updated:
\begin{equation*}
    Z \coloneq \big\{z\, |\, (i,j,z) \in E_{f_t}^{i,j}\big\}.
\end{equation*}
Since $z \in Z$ refers to a monomial that is affected by the reduction of $x_ix_j$ to $y_h$, it suffices to remap edges connected to node $i$ or $j$ that contain $z$ to the newly introduced node $h$.
Any neighbouring edge that contains $\bar{z} \notin Z$ is invariant under the effect of reduction, since its corresponding monomial does not contain the variable pair $x_ix_j$. 
Furthermore, it suffices to consider neighbouring nodes that are connected to both $i$ and $j$ (see \autoref{thm:NodeOnlyConnectedToEitherIorJEdgesInvariant}).
\begin{figure}[htb]
    \centering
    \includegraphics[]{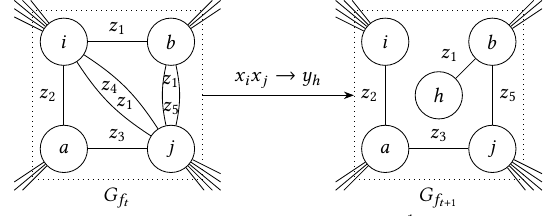}
    \caption{Shows the inductive graph evolution for changing edges from $G_{f_t}$ to $G_{f_{t+1}}$. The penalty term induced edges $\{(i,h,z_k), (j,h,z_{k+1}), (i,j,z_{k+2})\}$ for a fitting $k\in \mathbb{N}$ can be excluded from $G_{f_{t+1}}$ (see \autoref{thm:PenaltyInvariant}) and are not shown.}
    \label{fig:InductiveGraphEvolutionCases}
\end{figure}
\autoref{fig:InductiveGraphEvolutionCases} shows the above stated for a node $a$ that is connected to both nodes $i$ and $j$ but does not contain $z \in Z = \{z_1, z_4\}$ on its edges $(a,i,z_2)$ and $(a,j,z_3)$. 
Hence edges from node $a$ to $i$ and $j$ are invariant under the effect of reduction.
Conversely, node $b$ is connected to node $i$ and $j$ and contains $z_1 \in Z$ on its edges $(b,i,z_1)$ and $(b,j,z_1)$.
Therefore, these edges are remapped to the newly introduced node $h$.
The remaining edge $(b,j,z_5)$ is invariant, since it stems from a monomial that does not contain the variable pair $x_ix_j$.
In summary, the graph structure can be evolved without specifically referring to monomials and propagating their changes. 
This is mainly due to the fact that (a) all changing monomials are identified by indices occurring on edges between nodes $i$ and $j$ (see \autoref{thm:c4alreadybetweenij}) and (b) a reduction acts on the local neighbourhood of nodes $i$ and $j$.

\section{Algorithm in Detail}
\label{sec:mechanism}
Any node-pair $\{i,j\}$ with $\beta_f(i,j) > 1$ is suited for a reduction step, since it is guaranteed to originate from a higher-degree monomial $m$ ($\operatorname{deg}(m) > 2$; see~\autoref{thm:MultiEdgeGuaranteedHigherOrderMonomial}).
We cannot state that for node-pairs connected by a single edge (\ie, $\beta_f(i,j) = 1$) in general, since they might stem from a degree-$2$ monomial.
Hence we introduce the \mygls{lsr} algorithm --- subdivided into two stages:
\begin{enumerate}
    \item[] \mygls{lsr} stage 1: Graph-based reduction
    \item[] \mygls{lsr} stage 2: Independent monomial-based reduction
\end{enumerate}
In stage 1, the graph structure is decisive in the performance gain and in stage 2, we no longer need the graph structure --- although it proves useful in the runtime analysis in the following section.
After first applying stage 1 and then stage 2 to a higher-degree \mygls{pbf} $f$, we arrive at a quadratic \mygls{pbf}, which adheres to the \emph{quadratisation} criteria in \autoref{eq:QuadratizationMinimumPreserving}.

Recall the basic structure for a \emph{quadratisation} algorithm (\autoref{alg:quark}), that is (a) choosing a variable pair according to certain criteria and (b) replacing it with a new variable in the higher-degree function and thereby adding a penalty term.
Motivated by the adaptability to degree-$2$ density in the resulting quadratic function, it is eminent to choose the next variable pair according to their occurrence in other monomials. 
Recall that this property is depicted by multiplicities in the graph.
Hence, we alter the existing algorithmic structure in part (a) while making sure to advance the graph structure as shown in \autoref{sec:GraphRepresentation}.
Part (b) still results in the variable pair being replaced in all occurring monomials and is therefore left unchanged in terms of results.

Our proposed algorithm chooses the next variable pair in stage 1 according to a sorted set of multiplicities in the graph via percentile $q$.
Recall that $B \coloneq \{\beta_1, \ldots, \beta_{|B|}\}$ is the set of multiplicities in $G$ and $\beta_1 < \ldots < \beta_{|B|}$, as in \autoref{sec:GraphRepresentation}.
We define $\tilde{\beta}_q \coloneq \beta_{\lceil q \cdot |B| \rceil}$ for a percentile $q$ (see \autoref{eq:BetaTildePercentile}).
A percentile $q=1$ results in a variable pair that occurs most often in all monomials, while a percentile $q=0.5$ results in choosing the median.
Recall that we exclude multiplicities $0$ and $1$ (see \autoref{sec:GraphRepresentation}) from function $R$ and multiplicity set $B$.
Hence, choosing $q=0$ will result in choosing a variable pair, that occurs in at least two monomials\footnote{It may be the case that $m_1 = x_ix_j$ and $\deg(m_2) > 2$.} $m$ with $\operatorname{deg}(m) \geq 2$.
Since function $R(\tilde{\beta}_q)$ returns a set of node-pairs with multiplicity $\tilde{\beta}_q$, we can choose a random element from it as the next variable pair.

After choosing a variable pair, we can alter the graph structure independently from the monomial reduction according to the introduced inductive step in \autoref{sec:GraphRepresentation}.
Function \textsc{update\_graph\_data(.)} firstly selects monomial indices from edges between nodes $i$ and $j$ in set $Z$ (\autoref{alg:lsr}: l. 18), since all changing monomials occur on these edges (see \autoref{thm:c4alreadybetweenij}). 
According to \autoref{thm:NodeOnlyConnectedToEitherIorJEdgesInvariant} we can restrict the search for changing edges to nodes connected to both node $i$ and $j$.
This property applies to nodes that we eventually save in set $N$ (\autoref{alg:lsr}: l. 19) --- the set of nodes with changing edges.
Without explicitly accessing the set of neighbours for nodes $i$ and $j$, we can pre-compute changing edges in $G$.
This method depicts a lower bound on finding changing edges in $G$ for an iteration step, as it is linear in the number of changing edges.
To clarify this point, let $m_z = x_1x_2...x_kx_ix_j$ ($k<i<j$) be a monomial, let $z$ be its index in $Z$, let $x_ix_j$ be the chosen reduction pair and let $x_1x_2\ldots x_ky_h$ be its reduced version (see \autoref{thm:ChangingEdgesMustConnectToiOrj} for a similar argument).
Then $E_\text{removed}$ denotes the set of removed edges from $G$ and $E_\text{added}$ denotes the set of new edges to $G$ or in other words the remapping of a subset of edges induced by monomial $m_z$ \footnote{Take into consideration that for simplicity in pseudocode, we do not enforce a total ordering of indices in edges, as introduced in \autoref{sec:GraphRepresentation}.}:
\begin{equation*}
    \begin{aligned}
        E_\text{removed} \coloneq \{&(1,i,z), (2,i,z), \ldots , (k,i,z)\\
        &(1,j,z), (2,j,z), \ldots , (k,j,z), (i,j,z)\}\\
        E_\text{added} \coloneq \{&(1,h,z), (2,h,z), \ldots , (k,h,z)\}.
    \end{aligned}
\end{equation*}
As a consequence of replacing edges $(k,i,z)$ and $(k,j,z)$ with $(k,h,z)$ (\autoref{alg:lsr}: ll. 24-25; 29), multiplicities change according to:
\begin{align}
\label{eq:changingbeta}
    &\beta(k,i) \gets \beta(k,i) - 1 \nonumber\\
    &\beta(k,j) \gets \beta(k,j) - 1 \\
    &\beta(k,h) \gets \beta(k,h) + 1. \nonumber
\end{align}
Also, edges between nodes $i$ and $j$ are removed\footnote{We exclude the penalty term (see \autoref{thm:PenaltyInvariant}).} (\autoref{alg:lsr}: l. 29).
Last but not least, function $R$ changes, when the graph is altered (see \autoref{eq:changingbeta}).
Since node-pairs are mapped from the set of former multiplicities in $G$, set $M$ saves their values and corresponding node-pairs.
Therefore, any node-pair in $M$ can firstly be deleted from $R$ and secondly be added with its new multiplicity again (\autoref{alg:lsr}: l. 28):
In code, $R$ can be implemented by a sorted dictionary of sets, which leads to logarithmic access times on multiplicities and to constant average access times on elements of the sets of node-pairs.
Although the sorted property is not needed when updating the graph, it is required to select the next variable pair via percentile $q$, which deliberately influences properties of the resulting \mygls{pbf}.

Since we independently evolve the graph structure, it is necessary to update the polynomial representation separately. 
As an improvement to the standard implementation that needs to go over all monomials of $f$, we can use the graph structure to directly address only changing monomials.
Function \textsc{replace\_var\_pair(.)} (\autoref{alg:lsr}: l. 6) takes advantage of \autoref{thm:c4alreadybetweenij} to identify all changing monomial indices by gathering all indices on edges between nodes $i$ and $j$.
Based on these indices, a dictionary implementation of \autoref{tab:QuarkPolyDict} leads to the monomial representation of changing monomials.
Since edges induced by the penalty term are invariant under reduction of further steps (see \autoref{thm:PenaltyInvariant}), it is excluded from the graph structure.
Furthermore, it is saved in a separate \mygls{pbf} (\autoref{alg:lsr}: l. 8) to be able to later scale it properly and therefore adhere to the \emph{quadratisation} criteria given in \autoref{eq:QuadratizationMinimumPreserving}.
For as long as there are remaining multi-edges in the graph, stage 1 continues this process.
We know that this stage terminates, as we have shown in the supplementary material in a prior publication, by showing that the total number of edges in multi-edges decreases monotonically~\cite{schmidbauer_2024_reductions}.

As a prerequisite of stage 2 (\autoref{alg:lsr}: l. 11), we know that there are no more multi-edges in the graph:
\begin{equation*}
    \forall i,j \in V: \beta(i,j) \leq 1.
\end{equation*}
This means that there is no monomial left that shares a variable pair with other monomials.
Since remaining reduction steps do not introduce multi-edges again~\cite{schmidbauer_2024_reductions}, any two monomials will not share variable pairs for the remaining reduction steps. 
Hence, it is possible to reduce monomials independently of each other.
Function \textsc{multi\_reduce(.)} (\autoref{alg:lsr}: l. 13) quadratises a monomial $m$ and adds the necessary penalty terms to $p$.
Hereafter (\autoref{alg:lsr}: l. 16), input function $f$ is now a quadratic \mygls{pbf}, constrained by the penalty term $p$.

\begin{algorithm}[htbp]
\caption{Local Structure Reduction (LSR).}
\label{alg:lsr}
\SetAlgoLined
\LinesNumbered
\DontPrintSemicolon

\KwInput{\mygls{pbf} $f$, percentile $q$ \tcp*{\normalfont $\operatorname{deg}(f) > 2$, $q \in [0,1]$}}
\KwOutput{quadratic \mygls{pbf} $f'$, penalty \mygls{pbf} $p$}

\tcp*{\normalfont Stage 1: Graph-based reduction}
$h \gets 1$\;
$p \gets 0$\;
$G = (V,E), R \gets G_f = (V_f, E_f), R_f$\;

\While{$G$ {\normalfont contains multi-edges ($\exists i,j \in V: \beta(i,j) > 1$)}}{
    $\{i, j\} \gets \textsc{choose\_random\_element}(R(\tilde{\beta}_q))$\;
    $f \gets \textsc{replace\_var\_pair}(G, f, x_i, x_j, y_h)$\;
    $G, R \gets \textsc{update\_graph\_data}(G, R, i, j, h)$\;
    $p \gets p + \operatorname{p}(x_i, x_j, y_h)$\;
    $h \gets h+1$\;
}

\tcp*{\normalfont Now we have $\forall i,j \in V: \beta(i,j) \leq 1$}
\tcp*{\normalfont Stage 2: Independent monomial-based reduction}
\For{$m \in f$ with $\deg(m) > 2$}{
    
    \While{$\operatorname{deg}(m) > 2$}{
        $f, p \gets \textsc{multi\_reduce}(f, p, m)$\;
    }
}

\Return{$f, p$}

\BlankLine

\SetKwProg{Fn}{Procedure}{}{\Return{$G$, $R$}}
\Fn{{\normalfont \textsc{update\_graph\_data}}($G$, $R$, $i$, $j$, $h$)}{
    $Z \gets \big\{z: (i, j, z) \in E \big\}$\;
    $N \gets \{h\},\; E_\text{removed} \gets \{\},\; E_\text{added} \gets \{\}$\;

    \For{$z \in Z$}{
    
        \For{$k \in m_z \land k \neq h$\tcp*{\normalfont $m_z$ retrieves monomial indexed by $z$}}{
            
            $N \gets N \cup \{k\}$\;
            $E_\text{removed} \gets E_\text{removed} \cup \{(k, i, z)\} \cup \{(k, j, z)\}$\;
            $E_\text{added} \gets E_\text{added} \cup \{(k, h, z)\}$\;
        }
    }

    $M \gets \big\{(\beta(n,k), n, k) : n \in N, k \in \{i,j\}\big\}$\;
    $E \gets \big(E \setminus (E_\text{removed} \cup E^{i,j})\big) \cup E_\text{added}$\;
    $R \gets \textsc{update\_R}(G, R, N, M, i, j, h)$\;    
}

\end{algorithm}
\clearpage

\section{Analysis}
\label{sec:Analysis}
\subsection{Correctness}
\label{ssec:Correctness}
Let $f: \{0,1\}^n \to \mathbb{R}$ be a \mygls{pbf} with $\operatorname{deg}(f) > 2$. 
We argue that the quadratised function $f'$ resulting from $f$ through our introduced \mygls{lsr} method adheres to the \emph{quadratisation} criteria, given in \autoref{eq:QuadratizationMinimumPreserving}.
The graph structure improves performance when finding the next variable pair, but does not affect $f$.
Since the replacement of variable pairs in $f$ and introduction of penalty term $\operatorname{p}$ adheres to the standard method (see \autoref{ssec:MBackQuadratisations}), the introduced \mygls{lsr} algorithm adheres to the \emph{quadratisation} criteria in \autoref{eq:QuadratizationMinimumPreserving} \cite{Boros_2002}.
Take into consideration that $p$ may need to be scaled by a constant $c$. 
The minimum value for $c$ is problem dependent and thus we leave this last step to the user by returning $p$ unscaled.

\subsection{Runtime Analysis}
\label{ssec:RuntimeAnalysis}
In this section, we characterise the runtime complexity of the existing-monomial-based reduction as implemented in \texttt{\href{https://gitlab.com/quantum-computing-software/quark/-/tree/v1.1}{quark}} (link in pdf) and the new \mygls{lsr} graph-based reduction.
The basis for the main comparison is the runtime per iteration, but we also include estimations for complete runtimes.
The main difficulty is that these algorithms can lead to different \emph{quadratised} functions $f$ depending on the lexicographical order in each step.
Let $f_0 \vcentcolon= f: \{0,1\}^n \to \mathbb{R}$ be a \mygls{pbf} such that $\operatorname{deg}(f) > 2$ and let $T_f$ denote the number of monomials in $f$.
Furthermore, let $f_t$ denote a \mygls{pbf} after reduction step $t \in \mathbb{N}$ \footnote{$f_0 \overset{1. \text{ reduc step}}{\longrightarrow} f_1 \overset{2. \text{ reduc step}}{\longrightarrow} f_2 \longrightarrow \; \ldots \; \longrightarrow f_{t-1} \overset{t \text{-th reduc step}.}{\longrightarrow} f_t$.}.
We define the size of the input function $f_0$ by the number and size of it's monomials, that is, $\sum_{m \in f_0} |m|$, where we write $|m|$ as a short version of $\operatorname{deg}(m)$.

\subsubsection{Monomial-based Reduction}
The monomial-based reduction has 3 variants: \textit{Sparse}, \textit{Medium} and \textit{Dense}.
Each iteration consists of a two-stage process, that is, (a) searching for the next variable pair and (b) replacing that pair in every occurring monomial (see \autoref{alg:quark}).
Part (b) is variant-independent and its runtime for a reduction step from $f_{t-1}$ to $f_{t}$ is given by 
\begin{equation}
    \begin{split}
        \sum_{m\in f_{t-1}}{2 |m|} 
        &\leq T_{f_{t-1}} \cdot 2 \max_{m \in f_{t-1}}{|m|} 
        = T_{f_{t-1}} \cdot 2 \operatorname{deg}(f_{t-1}) \\
        &\leq T_{f_{t-1}} \cdot 2 \operatorname{deg}(f_0) 
        \leq T_{f_{t-1}} \cdot 2n.
    \end{split}
\end{equation}
Here, it is required to check if the two candidate variables are present in each monomial of $f_{t-1}$. 
In the monomial-based array implementation of monomials~\cite{Lobe_2023, Windgtter_2025} the search for the next variable pair leads to a full traversal of each monomial.
We also use the fact that a reduction step does not increase the degree of $f$.

In the following $\mathrm{RT}_\textit{Sparse}$ and $\mathrm{RT}_\textit{Dense}$ refer to the search in a single iteration for the \textit{Sparse} and \textit{Dense} variants.
Part (a) is variant-dependent. 
In particular, the \textit{Sparse} method searches for the highest-degree monomial.
Without caching monomials, the search for the highest-degree monomial takes 
\begin{equation}
    \mathrm{RT}_\textit{Sparse}(f_{t-1}) = T_{f_{t-1}}
\end{equation} steps\footnote{As for an implementation in python, the \textsc{len(.)} function accesses a cached attribute and therefore has a time complexity of $\mathcal{O}(1)$.}.
The \textit{Dense} method searches for the variable pair that appears most often among all monomials.
It therefore computes all \myglspl{pc} of every monomial in $f_t$.
Let $\mathrm{\mygls{pc}}_m$ denote the set of pair combinations resulting from a monomial $m \in f_{t-1}$.
For any degree-$k$ monomial $m$ (\ie,~$|m| = k$) there are $\binom{|m|}{2}$ variable pairs and hence, the number of pairs to consider is $\sum_{m\in f_{t-1}}{|\mathrm{\mygls{pc}}_m|} = \sum_{m\in f_{t-1}} {\binom{|m|}{2}}$.
Let \myglspl{upc} denote the set of unique variable pairs: $\mathrm{\mygls{upc}}_{f_t} = \bigcup_{m \in f_t}{\mathrm{\mygls{pc}}_m}$.
By iterating over variable pairs $u \in \mathrm{\mygls{upc}}$ and calculating the pair combinations of monomials in $f_{t-1}$, we get the count for $u$.
Sorting the resulting list of counts by $u$, gives the total runtime for searching for the most occurring pair among all monomials $\mathrm{RT}_\textit{Dense} = \log\big(|\mathrm{\mygls{upc}}_{f_{t-1}}| \cdot \sum_{m\in f_{t-1}} {\binom{|m|}{2}}\big) \cdot |\mathrm{\mygls{upc}}_{f_{t-1}}| \cdot \sum_{m\in f_{t-1}} {\binom{|m|}{2}}$.
In an earlier work \cite{schmidbauer_2024_reductions}, we show that the total size of multi-edges decreases with every reduction step.
Although the monomial-based method does not use a graph structure, we can still apply our theorem: Since the \textit{Dense} variant searches for the most occurring pair, the preliminary condition of selecting multi-edges in the graph applies. 
Therefore, $|\mathrm{\mygls{upc}}_{f_{t-1}}| \geq |\mathrm{\mygls{upc}}_{f_{t}}| \, \forall t \in \mathbb{N}$ and the runtime for the \textit{Dense} search in reduction step $t$ is given by
\begin{equation}
    \begin{split}
        \mathrm{RT}_\textit{Dense}(f_{t-1})
        &\leq \log\big(|\mathrm{\mygls{upc}}_{f_0}| \cdot \sum_{m\in f_{t-1}} {\binom{|m|}{2}}\big) \cdot |\mathrm{\mygls{upc}}_{f_0}| \cdot \sum_{m\in f_{t-1}} {\binom{|m|}{2}} \\
        &\leq \log\big(n^2 \cdot \sum_{m\in f_{t-1}} {\binom{|m|}{2}}\big) \cdot n^2 \cdot \sum_{m\in f_{t-1}} {\binom{|m|}{2}}.
    \end{split}
\end{equation}

\subsubsection{Graph-based Reduction}

\paragraph{Stage 1: Graph-based Reduction}
Firstly, we show that the preparatory effort to compute the needed data structures does not exceed $\mathcal{O}(T_{f_0} \cdot \binom{\operatorname{deg}(f_0)}{2} + n^2 \cdot \log(T_{f_0}))$.
Secondly, we examine the effort to compute the inner part of the while loop (\ie, a reduction iteration; see \autoref{alg:lsr}: ll. 5-9). 

Creating the monomial index dictionary, as in \autoref{tab:QuarkPolyDict}, iterates over all monomials present in ${f_0}$, that is, the size of the input dictionary: $\mathcal{O}(\sum_{m\in f_0}{|m|}) \subseteq \mathcal{O}(T_{f_0} \cdot \binom{\operatorname{deg}(f_0)}{2})$.
Creating the graph structure requires iterating over all variable pair combinations per degree-$k$ monomial $m$, of which there are $\binom{|m|}{2} = \binom{k}{2}$ many. 
In total, the number of edges in the graph is upper bounded by $\mathcal{O}(\sum_{m \in f_0}{\binom{|m|}{2}}) \subseteq \mathcal{O}(T_{f_0} \cdot \binom{\operatorname{deg}(f_0)}{2})$.
Last but not least, creating function $R$, as in \autoref{ssec:Fundamentals}, originates from the graph's connected node-pairs, and therefore needs $\mathcal{O}(|V_{f_0}|^2) = \mathcal{O}(n^2)$ computational steps. 
Recall that function $R$ is implemented as a sorted dictionary of sets.
Each insertion has the potential to generate a new entry in the sorted dictionary, for which the access and insertion times are in $\mathcal{O}(\log(n))$ \cite{SortedDictionary_2024}.
Let the set of multiplicities be denoted by $B_{f_0} = \{\beta_{f_0}(i,j)\, |\, i, j \in V_{f_0}\}$ (see \autoref{sec:GraphRepresentation}).
To be more precise, the access time in the sorted dictionary (or domain of $R_{f_0}$) is given by $\mathcal{O}(\log(|B_{f_0}|)) \subseteq \mathcal{O}(\log(T_{f_0}))$\footnote{Let $\beta_\mathrm{max}$ (compare to \autoref{thm:MaxNumberOfEdgesBetweenNodes}) denote the maximum number of edges between two nodes in $G$. Take into consideration that not every value in $\{0,\ldots ,\beta_\mathrm{max}\}$ needs to be present in $B_{f_0}$ --- thus improving the logarithmic access time.}.
Therefore, creating function $R$ is upper bounded by $\mathcal{O}(n^2 \cdot \log(T_{f_0}))$.
Hence, the preparatory effort to compute the needed data structures is bounded by:
\begin{equation}
    \mathcal{O}\left( T_{f_0} \cdot \binom{\operatorname{deg}(f_0)}{2} + n^2 \cdot \log(T_{f_0}) \right)
\end{equation}

Computing the index to access the desired subset of nodes in $R$ via percentile $q$ is done in constant time.
Using an iterator to access a random element of that subset also leads to a constant access time.
Thus, \textsc{choose\_random\_element($R(\tilde{\beta}_q)$)} (\autoref{alg:lsr}: l. 5) is upper bounded by $ \mathcal{O}(\log(|B_{f_{t-1}}|)) \subseteq \mathcal{O}(\log(T_{f_0}))$ with $B_{f_{t-1}}$ analogous to above.
Exploiting the local influence of a reduction's iteration, that is, restricting the number of processed edges to the local neighbourhood of nodes $i$ and $j$ (assuming $x_ix_j$ is going to be reduced), is decisive for the algorithm's performance gain.
Note that this not only includes extracting relevant edges, but also includes the update procedure on edges and monomials in the polynomial representation.
Let the set of indices referring to changing monomials be denoted by $Z$ (\autoref{alg:lsr}: l. 18).
These indices occur on edges between nodes $i$ and $j$ and thus it takes $\beta_{f_{t-1}}(i,j)$ steps to compute $Z$ --- assuming an average case access time on the graphs edges of $\mathcal{O}(1)$. 
The worst case runtime is given by $\mathcal{O}(\beta_{f_{t-1}}(i,j) \cdot |V_{f_{t-1}}|^2)$.

Ultimately, the set of edges that change during this reduction iteration is relevant.
Recall that these edges are split into sets $E_\text{removed}$ and $E_\text{added}$.
\autoref{thm:ChangingEdgesMustConnectToiOrj} states that any changing edge needs to be connected to $i$ or $j$, which allows us to compute the set of removed and new edges $E_\text{removed}$ and $E_\text{added}$ respectively in $\mathcal{O}(\sum_{z \in Z}{|m_z|})$, where $\beta_{f_{t-1}}(i,j) = |Z|$ by iterating over each affected monomial once. 
$m_z$ refers to the variable representation of a monomial $m$ indexed by $z$.
Due to --- on average --- constant access times in sets, this is an upper bound on the for-loop's runtime (\autoref{alg:lsr}: ll. 20-27).
Updating the edges of $G$ (\autoref{alg:lsr}: l. 29) is also bounded by above runtime, since the average runtime to additionally remove edges between nodes $i$ and $j$ is given by $\beta_{f_{t-1}}(i,j)$.

It is necessary to recalculate the multiplicity between node-pairs corresponding to edges that have changed during this iteration. 
These node-pairs can be extracted from set $N$ in $\mathcal{O}(\sum_{z \in Z}{|m_z|})$ (\autoref{alg:lsr}: l. 28). 
Due to the dictionary-like structure for the graph, computing the number of edges between two nodes is equal to accessing a cached property, namely the number of keys in the dictionary (\ie~$\mathcal{O}(1)$ (average case); $\mathcal{O}(|V_{f_{t-1}}|^2)$ (worst case)).
Let $B_{f_{t-1}} = \{\beta_{f_{t-1}}(i,j)\, |\, i, j \in V_{f_{t-1}}\}$ denote the set of multiplicities in the graph resulting from $f_{t-1}$.
Recall that each element in the sorted dictionary implementation of function $R$ can be accessed in $\mathcal{O}(\log(|B_{f_{t-1}}|)) \subseteq \mathcal{O}(\log(T_{f_{t-1}}))$.
This results in a total runtime of $\mathcal{O}(\log(|B_{f_{t-1}}|) \cdot \sum_{z \in Z}{|m_z|})$ for \textsc{update\_R(.)} (\autoref{alg:lsr}: l. 30).
Take into consideration that $|B_{f_{t}}|\, /\, |B_{f_{t-1}}| < c_{f_0}$ where $c_{f_0}$ is a constant for input function $f_0$, since a reduction step acts locally (see \autoref{thm:ChangingEdgesMustConnectToiOrj} and \autoref{fig:InductiveGraphEvolutionCases}).
Let $\omega$ be the index of the last reduction iteration in \mygls{lsr} stage 1. 
Then, $|B_{f_{\omega}}| = 0$.

As a side note, suppose that the graph of a \mygls{pbf} $f$ contains most of its edges between nodes $i$ and $j$. 
One can argue that the effort to remove and add edges in this iteration scales with $\mathcal{O}(|E_{f}|)$. 
However, recall that the total number of iterations is bounded (see \autoref{ssec:PropertiesLemmata}). 
Since every edge $E_f^{i,j}$ (see \autoref{sec:GraphRepresentation}) is removed\footnote{We exclude the penalty term from the graph structure.}, this special case is in fact efficient in terms of introducing less variables.

Assuming an array implementation of monomials, it takes $\mathcal{O}(|m|)$ steps to replace a variable pair in a monomial.
Since we store a monomial an edge stems from indirectly (see \autoref{tab:QuarkPolyDict}), changing monomials in the reduction process do not alter the graph's structure --- thus saving significant access time\footnote{Note that this statement assumes the specific algorithm's reduction process.}.
It suffices to change monomials in the index dictionary of $f$ (see \autoref{tab:QuarkPolyDict}) with an average access time of $\mathcal{O}(1)$, whereas the worst case access time is $\mathcal{O}(T_{f_{t-1}}) \subseteq \mathcal{O}(T_{f_0})$.
Thus, the runtime of \textsc{replace\_var\_pair(.)} (see \autoref{alg:lsr}: l. 6) is upper bounded by $\mathcal{O}(T_{f_0} \cdot \sum_{z \in Z}{|m_z|})$ (worst case) and on average $\mathcal{O}(\sum_{z \in Z}{|m_z|})$.
On average, dictionary insertion is achieved in $\mathcal{O}(1)$ and in the worst case in $\mathcal{O}(n)$.
Thus, adding the penalty term (\autoref{alg:lsr}: l. 8) has an average runtime of $\mathcal{O}(1)$ and a worst case runtime of $\mathcal{O}(t)$, since the penalty term introduces $4$ monomials per iteration (see \autoref{eq:PenaltyTermIntro}).

In a complete reduction iteration (\autoref{alg:lsr}: ll. 5-9), function \textsc{update\_R(.)} (\autoref{alg:lsr}: l. 30) has the highest runtime --- leaving an average case runtime of
\begin{equation}
    \mathrm{RT}_\text{LSR1} \in \mathcal{O}\left( \log(|B_{f_{t-1}}|) \cdot \sum_{z \in Z}{|m_z|} \right)
\end{equation}
per full iteration.
Since the number of multi-edges in the graph strictly decreases with every reduction iteration \cite{schmidbauer_2024_reductions}, we can bound the number of iterations in \autoref{alg:lsr} to $|E_{f_0}|$.

\paragraph{Stage 2: Independent Monomial-based Reduction}
Let $f_t: \{0,1\}^n \rightarrow \mathbb{R}$ be a \mygls{pbf} resulting from stage 1, that is, a function sharing no variable pairs among its monomials.
Let $m = x_1x_2x_3x_4x_5 \in f_t$ be a monomial. 
In contrast to stage 1, we can no longer exploit replacing the same variable pair in different monomials.
Therefore, using Boros' \cite{Boros_2002} reduction method, we can apply multiple reductions at once.
For example, applying $x_1x_2x_3x_4x_5 \rightarrow y_1y_2x_5$ in a single step by replacing $x_1x_2$ with $y_1$ and $x_3x_4$ with $y_2$. 
\textsc{multi\_reduce(.)} (\autoref{alg:lsr}: l. 13) uses this fact and therefore has a time complexity of $\Theta(|m|)$.
After applying this step once, only degree-$3$ and degree-$4$ monomials (\ie, $|m| \in \{3,4\}$) become quadratic in general.
Further steps are needed for higher-degree monomials. 
To be more precise, we differentiate two cases: 
\begin{equation}
    |m_\text{new}| = 
    \begin{cases}
        0.5 \cdot |m|, & \text{if $|m|$ is even}\\
        \lceil 0.5 \cdot |m|\rceil, & \text{if $|m|$ is odd}
    \end{cases},
\end{equation}
where $m_\text{new}$ represents the monomial after \textsc{multi\_reduce(.)} is applied to $m$.
Instead of computing the ceiling function, we can depict the number of iterations necessary for $m$ to become quadratic as $\tau = \log_2(\lfloor|m|\rfloor_2) = \lfloor \log_2(|m|) \rfloor$, where $\lfloor\cdot\rfloor_2$ rounds down to the nearest power of $2$ and $|m| \in \mathbb{N}, |m| > 4$.
Hence, $\tau$ is logarithmic in the monomial's size (\ie, $\tau \in \Theta(\log_2(|m|))$) --- leading to a partial geometric series for the runtime of the inner while-loop (\autoref{alg:lsr}: l. 12-14) and, due to its convergence, a total runtime of 
\begin{equation}
    \mathrm{RT}_\text{LSR2} \in \Theta(|m|)
\end{equation}
per full monomial \emph{quadratisation}. 
The summation over all left-to reduce monomials (\ie, $|m| > 2, m \in f_t$) leads to the total runtime of stage 2.
As a side note, we are not required to quadratise monomials in \textsc{multi\_reduce(.)}, but can stop the reduction process at an arbitrary point --- leaving us with a degree-$k$, $k>2$ function.

Since there are no common variable pairs among monomials in $f_t$, there are at maximum $\mathcal{O}(|E_{f_t}|) \subseteq \mathcal{O}(|V_{f_t}|^2)$ many monomials left to reduce in stage 2 and therefore as many necessary iterations in the for loop (\autoref{alg:lsr}: l. 11-15)\footnote{Note that, in stage 2, the graph does not contain multi-edges.}.
Take into consideration that the previous stage 1 algorithm changes monomials through replacements. 
Thus, it reduces the size of monomials for stage 2, which lowers the runtime of the inner while loop (\autoref{alg:lsr}: l. 12-14).

Despite changing the number of variables, stage 1 does not increase the number of monomials in the polynomial dictionary of $f$, since the penalty term is saved separately. 
On the one hand, this upper bounds the iteration count of the for loop in stage 2 (\autoref{alg:lsr}: l. 11-15) to $T_{f_0}$.
On the other hand, the total runtime for \mygls{lsr} stage 2 is given by
\begin{equation}
    \Theta\Big(\sum_{\substack{m\in f_t,\\|m| >2}}{|m|}\Big).
\end{equation}
Let $m = x_1 \ldots x_k$ be a degree-$k$ monomial (\ie, $|m| = k$) that is left to reduce from stage 1. 
It introduces $\binom{k}{2}$ edges to the graph.
Since the graph for stage 2 has no multi-edges, the total runtime for stage 2 is upper bounded by
\begin{equation}
    \mathcal{O}(E_{f_t}) \subseteq \mathcal{O}(|V_{f_t}|^2).
\end{equation}

One can parallelise stage 2 for any monomial $m \in f_t$, thus reducing the runtime to $\mathcal{O}((T_{f_0}/W) \cdot \text{<inner loop>} + \text{<distribute/gather>)}$, where $W$ denotes the number of worker tasks and $T_{f_0}$ denotes the number of terms in $f_0$.
Parallelisation is also possible in stage 1 (see \autoref{thm:ChangingEdgesMustConnectToiOrj}): Node-pairs that are not connected to each other do not influence each other during a reduction step.

\subsection{Average Runtime Guarantees and Optimality}
\label{ssec:GuaranteeAvgCaseRuntime}
Recall that a reduction iteration is characterised by two steps, that are, (a) finding the next variable pair and (b) replacing that pair by a new variable in all monomials it occurs in (see \autoref{alg:quark}).
Let $f: \{0,1\}^n \mapsto \mathbb{R}$ be a \mygls{pbf} with $T_f$ many monomials, let $G_f(V_f,E_f)$ be the corresponding graph of $f$ and let $\beta_f(i,j)$ denote the number of occurrences of the variable pair $x_ix_j$ in monomials of $f$.
The trivial lower bound of performing an iteration (\ie~(a) and (b)) is $\Omega(\beta_f(i,j))$, since we need access to all monomials in which $x_ix_j$ occurs.
When accessing variables in a monomial $m$, where each variable is saved as an element of an array, the worst case access time is $\mathcal{O}(|m|)$. 
Using a sorted array, it reduces to $\mathcal{O}(\log(|m|))$, due to binary search. 
Another option is to use a lookup-table per monomial where variable index $x$ is saved at position $x$, that is, mapping each variable in a monomial to the space of variables in the function.
Therefore, we can guarantee an access time of $\mathcal{O}(1)$, but use extra space, that is, $\mathcal{O}(n)$ per monomial, where $n$ is the number of variables in $f$. 
Hence, this method needs $\mathcal{O}(n \cdot T_f)$ space in total to store a \mygls{pbf}. 
Recall that we defined the size of the input function $f$ as the total size of its monomials, that is, $\sum_{m \in f} |m|$. 
Therefore, the extra space ($\mathcal{O}(n \cdot T_{f})$) needed is at most quadratic in the input size --- assuming there are no unused variables in $f$.
For practically relevant input functions $f$ (\ie, $n \ll T_f$), the extra space is sub-quadratic in the input size.

So far, we focused on the lower bound of (b), which corresponds to replacing the variable pair. Now, we consider searching
for the next pair, which corresponds to finding a lower bound for (a), and requires a clear definition of required properties for the next pair.
The trivial lower bound for an ambiguous choice is $\Omega (\sum_{m \in f_0}{|m|})$ for the whole \emph{quadratisation} process, since every monomial has to be considered at least once.
Our proposed method needs to be able to differentiate between the number of occurrences of a variable pair in each iteration --- motivated by the influence to the quadratised function (\ie, its density, number of variables and its monomial distribution among variables).
Assuming a sorted array implementation of the number of occurrences, function $R$'s implementation already achieves optimal runtime (avg. case) in each iteration in the first stage, due to binary search.

Assuming an array implementation of each monomial, the lower bound for part (b) rises to $\Omega(\sum_{z \in Z} |m_z|)$, where $|Z| = \beta_{f_{t-1}}(i,j)$, since searching in an unsorted array with size $|m|$ takes $\mathcal{O}(|m|)$ steps. 
Take into consideration that, as mentioned above, an array implementation can be replaced by more efficient data-structures --- therefore lowering the lower bound to $\Omega(\beta_{f_{t-1}}(i,j))$.

Other custom hash functions can be used to guarantee average case runtime in the mentioned data structures in \autoref{ssec:RuntimeAnalysis}. 
On the one hand, indexing monomials in a dictionary with exactly $T_f$ (see \autoref{tab:QuarkPolyDict}) many entries (\ie, the number of monomials in the input function\footnote{The penalty dictionary is separable from it: see \autoref{thm:PenaltyInvariant}.}), enables us to use $f(x) = x$ as a hash function, without spacial overhead.
On the other hand, each monomial can at most introduce a single edge between a particular node-pair in the graph.
Hence, the maximum multiplicity is given by at most $T_f$ (see \autoref{thm:MaxNumberOfEdgesBetweenNodes}).
Therefore, $f(x) = x$ is a suitable hash function for the sorted dictionary of function $R$ --- taking no more space than the input function has monomials.
Take into consideration that the size of the input function additionally includes the size of its monomials.
Function $R$ maps to at most $|V_{f_{t-1}}|^2$ node-pairs\footnote{We exclude node-pairs, containing less than two edges.}.
Unfortunately, $|V_{f_{t-1}}|^2$ scales with the number of iterations. 
Recall that we can however bound the maximum number of iterations $I_f$ by $\sum_{\substack{m \in f_0,\\|m| \geq 2}}{(|m|-2)} = -2 T_{f_0} + \sum_{\substack{m \in f_0,\\|m| \geq 2}} {|m|}$.
Since the set of nodes $V_{f}$ is isomorphic to the set of variables in $f$, the space complexity is $ \mathcal{O}((n + I_f)^2) \subseteq \mathcal{O}((n -2 T_{f_0} + \sum_{m \in f_0} {|m|})^2)$.

Another type of perfect hash functions (\ie, functions used whenever $\mathcal{O}(1)$ worst case accesses are required) use two stage universal hashing \cite{CormenThird_2009}. 
This method guarantees a linear space complexity in the number of keys --- although requiring a static set of keys to carefully choose the primary and secondary stage hash functions.
The monomial index dictionary provides a static set of keys, since the penalty term is separately stored (see \autoref{thm:PenaltyInvariant}) and is therefore a candidate for this type of hashing.
Despite not having a static set of keys, the dictionary of the input polynomial provides a static number of keys.
Take into consideration that this type of hashing introduces randomised data structures.

\subsection{Experimental Results}
\label{ssec:ExperimentalResults}
Although the asymptotic performance is most relevant in a complexity theoretic analysis, we also want to characterise the practical performance by evaluating a concrete implementation and thereby incorporating constant factors, structural effects of input functions and taking into account that the theoretical analysis overestimates the runtime.
In the following, we test randomly chosen \myglspl{pbf} $f: \{0,1\}^n \mapsto \mathbb{R}$ such that $\deg(f) = 4$ and varying densities such that $d_1(f) = d_2(f) = d_3(f) = d_4(f)$ with the monomial-based implementation and the new \mygls{lsr} method (graph\_based).
Take into consideration that we do not use custom hash-functions and no parallelisation, as proposed in \autoref{ssec:GuaranteeAvgCaseRuntime} and \autoref{ssec:RuntimeAnalysis}, but a standard python dictionary implementation.
Therefore, we expect the following scaling behaviour in runtime to be an upper bound. 

\begin{figure}[htb]
    \centering
    \includegraphics[width=\linewidth]{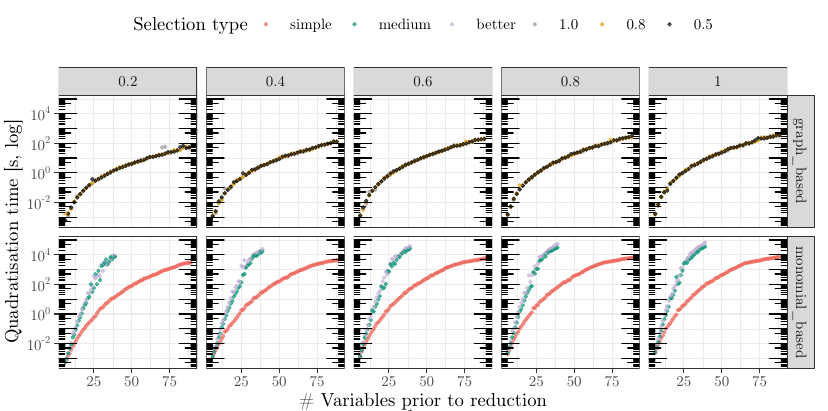}
    \caption{Time in seconds (y-axis) for the \emph{quadratisation} of a $\operatorname{deg}(f)= 4$ function $f$ vs problem size (x-axis: Number of variables). New \mygls{lsr} algorithm (top row) compared to existing monomial-based (bottom row).
    Vertical facets: Different base polynomial densities (\ie, $d_1(f) = d_2(f) = d_3(f) = d_4(f) \in \{0.2, 0.4, \ldots , 1.0\}$). 
    Selection type $1.0$ (\mygls{lsr} algorithm) comparable to \textit{Dense / better} (monomial-based). The \textit{Dense / better} and \textit{Medium} type are cut off at $39$ variables prior to reduction, which limits the runtime.
    }
    \label{fig:MBvsGBvariables}
\end{figure}
\begin{figure}[htb]
    \centering
    \includegraphics[width=\linewidth]{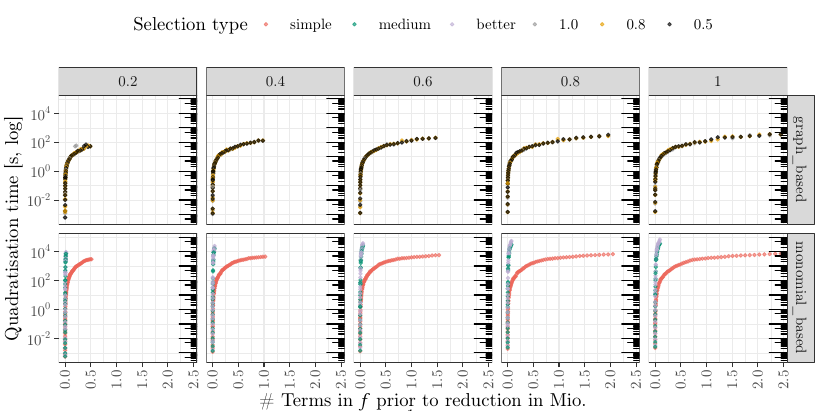}
    \caption{Time in seconds (y-axis) for the \emph{quadratisation} of a $\operatorname{deg}(f)= 4$ function $f$ vs problem size (x-axis: Number of Terms). New \mygls{lsr} algorithm (top row) compared to existing monomial-based (bottom row).
    Vertical facets: Different base polynomial densities (\ie, $d_1(f) = d_2(f) = d_3(f) = d_4(f) \in \{0.2, 0.4, \ldots , 1.0\}$). 
    Selection type $1.0$ (\mygls{lsr} algorithm) comparable to \textit{Dense / better} (monomial-based). The \textit{Dense / better} and \textit{Medium} type are cut off at $39$ variables prior to reduction, which limits the runtime.
    }
    \label{fig:MBvsGBterms}
\end{figure}

\autoref{fig:MBvsGBvariables} compares the runtime for the introduced \mygls{lsr} algorithm (graph\_based) and the monomial-based algorithm that uses a brute force search for the next variable pair.
Since the \textit{Dense / better} selection type of the monomial-based algorithm searches for a variable pair, which occurs most often among all monomials, it is comparable to the selection percentile $q = 1.0$ in the new algorithm.
The x-axis shows the number of variables prior to reduction (\ie, $n$) and the y-axis (log scale) shows the runtime in a logarithmic scale.
Different \mygls{pbf} densities are shown as vertical facets.
The \mygls{pbf}'s size increases when going from left to right. 
While the runtime of both implementations increases, we can see almost no differences in the percentiles $q \in \{0.5, 0.8, 1.0\}$, although the number of iterations increases with decreasing percentile $q$.
Furthermore, the \mygls{lsr} algorithm achieves a runtime even better than the simple method of the monomial-based implementation. 
Recall that the simple method uses $T_{f_{t-1}}$ steps in search for the next variable pair (\ie, step (a)) and $\sum_{m \in f_{t-1}} 2|m|$ steps for the replacement (\ie, part (b)). 
Our proposed algorithm outperforms this simple approach. 
Furthermore, $q = 1.0$ and the \textit{Dense / better} selection type are comparable in terms of selection strategy --- except for lexicographic sorting --- in each iteration. 
Thus, their comparison is the decisive one, when it comes to speedup.
At $39$ variables in the input function, the \textit{Dense / better} selection type already needed more than a day to compute the quadratised function ($d_4(f) = 1$). 
We therefore did not compute bigger functions for that type.
On the other hand, the new algorithm needs $\approx 10$ seconds for the same input polynomial.

When considering the number of terms $T_{f_0}$ in the input function $f_0$, \autoref{fig:MBvsGBterms} shows their influence on the runtime (y-axis; log scale) for the same input \myglspl{pbf} as in \autoref{fig:MBvsGBvariables}.
Take into consideration that the \textit{Dense / better} and \textit{Medium} variant for the monomial-based algorithm are also cut off at $39$ variables.
Thus, the number of terms is limited, although it increases with increasing density.
As before, we see the runtime benefit of the new algorithm.

\begin{figure}[htb]
    \centering
    \includegraphics[width=\linewidth]{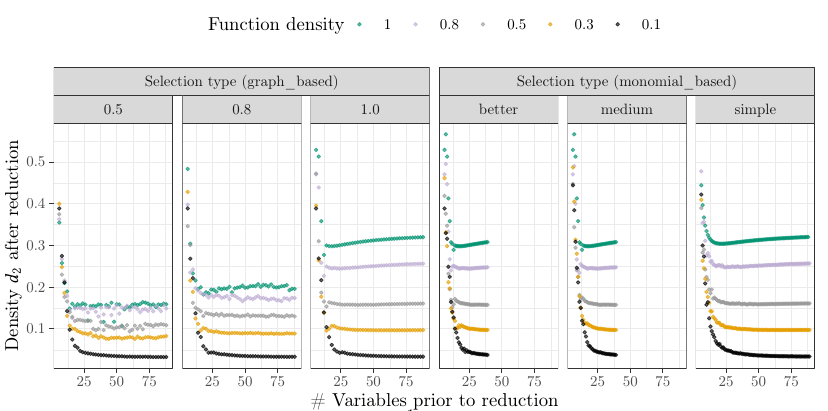}
    \caption{Degree-$2$ density (y-axis) for the \emph{quadratisation} of a $\operatorname{deg}(f)= 4$ function $f$ vs problem size (x-axis: Number of variables) for different base polynomial densities (\ie, $d_1(f) = d_2(f) = d_3(f) = d_4(f) \in \{0.1, 0.3, 0.5, 0.8, 1.0\}$). New algorithm on the left compared to monomial-based on the right. 
    Selection type $1.0$ (new algorithm) comparable to \textit{Dense / better} (monomial-based). The \textit{Dense / better} and \textit{Medium} type are cut off at $39$ variables prior to reduction, due to time limitations.
    }
    \label{fig:MBvsGBdensities}
\end{figure}

Apart from runtime, structural properties of quadratised \myglspl{pbf} are interesting in regard of their influence on further steps in a toolchain from a higher-level problem description to executing the problem on quantum hardware.
One aspect is the degree-$1$ and -$2$ density, which directly translate to \mygls{qubo} densities and therefore give insights on interactions in a quantum circuit, when for example using \mygls{qaoa}.
Assuming sufficiently many iterations in the algorithm, the degree-$1$ density tends towards its maximum value, that is, $d_1 \to 1$\cite{schmidbauer_2024_reductions}.
\autoref{fig:MBvsGBdensities} compares the degree-$2$ density (y-axis) for the introduced \mygls{lsr} algorithm (left) and the monomial-based algorithm (right) for different problem sizes (x-axis) and coloured by different \mygls{pbf}-densities (function density).
Interestingly the monomial-based search variants do not differ visually, although they implement different search variants.
As before, the \textit{Dense / better} variant compares to $q=1.0$, which also shows in the density plot for up to $39$ variables, where the \textit{Dense / better} variant is cut off due to runtime limitations.
Lowering the percentile $q$ lowers the degree-$2$ density in the quadratised function --- although, we see a greater difference when varying the function density of the input \mygls{pbf}.
Take into consideration that we generate the input function randomly, which encourages uniformly distributed variable pairs among monomials.
Thus, exploiting problem inherent structures in the input function is not possible, which is contrary to our previous work \cite{schmidbauer_2024_reductions}, where we analyse the effect of monomial-based reduction in the context of a Job-Shop Scheduling problem.

\begin{figure}[htb]
    \centering
    \includegraphics[width=\linewidth]{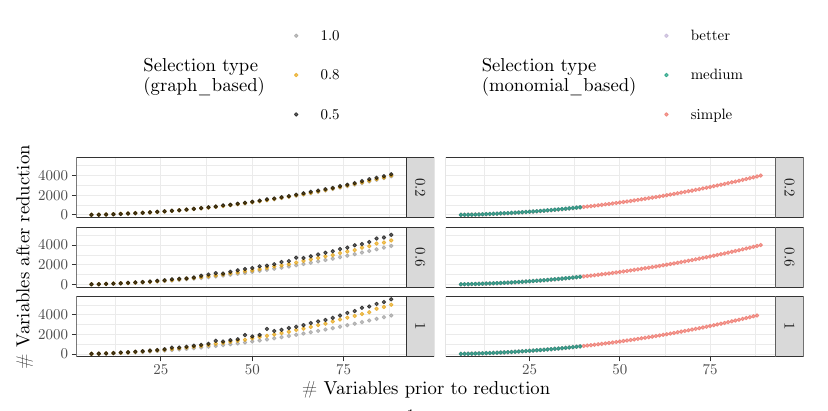}
    \caption{Number of variables after \emph{quadratisation} (y-axis) for the reduction of a $\operatorname{deg}(f)= 4$ function $f$ vs problem size (x-axis: Number of variables). New algorithm (left column) compared to monomial-based (right column). 
    Vertical facets: Different base polynomial densities (\ie, $d_1(f) = d_2(f) = d_3(f) = d_4(f) \in \{0.2, 0.6, 1.0\}$). 
    Selection type $1.0$ (new algorithm) comparable to \textit{Dense / better} (monomial-based). The \textit{Dense / better} and \textit{Medium} type are cut off at $39$ variables prior to reduction, due to time limitations. The number of variables after reduction (y-axis) minus the number of variables prior to reduction (x-axis) depict the number of iterations in the \mygls{lsr} algorithm and therefore the number of additional variables.
    }
    \label{fig:MBvsGBVarsBeforeAfter}
\end{figure}

The number of variables after \emph{quadratisation} of $f: \{0,1\}^n \mapsto \mathbb{R}$ translates to the number of qubits in a quantum circuit, when for example using \mygls{qaoa}.
\autoref{fig:MBvsGBVarsBeforeAfter} compares the number of variables prior to reduction (\ie, $n$: problem size) to the number of variables after reduction (\ie, $n + I_f$; y-axis), where $I_f$ corresponds to the number of reduction iterations.
\autoref{fig:MBvsGBVarsBeforeAfter} differentiates between the new graph\_based (left) and the monomial-based algorithm (right), as well as different horizontally faceted input function densities (\ie, $d_1(f) = d_2(f) = d_3(f) = d_4(f)$).
Recall that the degree-$2$ density depends on the selection type, which is connected to the number of variables each variant introduces.
While \autoref{fig:MBvsGBVarsBeforeAfter} shows no difference between different monomial\_based variants, the graph\_based variant manages to extend $I_f$ depending on selection type. 
Since the graph\_based variant implements a variety of selection types (via percentiles of multiplicities; see \autoref{tab:SortedDictofDicts1}), interpolation in an automated process is possible, which then influences properties of quantum circuits generated from the resulting function (\ie, \# qubits, gate distribution, 
circuit depth and runtime).
Consider that the baseline (\ie, $1.0$ and \textit{better}) overlaps predominately in both the graph\_based and monomial\_based reduction algorithms.

\begin{figure}[htb]
    \centering
    \includegraphics[width=\linewidth]{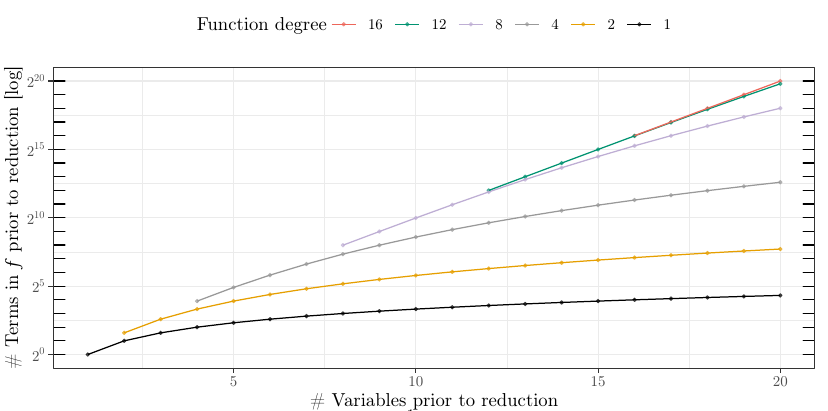}
    \caption{Number of variables (x-axis) vs number of terms (y-axis) prior to reduction for functions $f$ such that $deg(f) = \text{Function degree}$ and densities $d_1(f) = d_2(f) = \ldots  = d_{\operatorname{deg}(f)}(f) = 1$.
    }
    \label{fig:DegreeSweepVariables}
\end{figure}

We show how the number of terms in a function $f: \{0,1\}^n \mapsto \mathbb{R}$, where $f$ has all possible terms up to a specified degree, scales with the number of variables.
Recall that there are, $\binom{n}{k} \in \mathcal{O}(n^k)$ possible degree-$k$ monomials in $f$.
Hence, when we see $k$ as a constant and scale the number of variables $n$, the number of terms is bounded by a polynomial
for a sufficiently large value of \(n\).
However, when $k$ reaches $n$ or in other words, when $f$ has all possible monomials, their number scales exponentially with the number of variables in $f$, which is visualised in \autoref{fig:DegreeSweepVariables}.
Therefore, as the number of variables (typically the problem size) grows in practically relevant problems, problem formulations tend towards sparse \myglspl{pbf}, since exponential space requirements to represent that function are infeasible.
Thus, for larger problem sizes, this leads to less edges in the graph representation and potentially less connected nodes --- further reducing the runtime of a reduction step.

\section{Industrial Utility}
\label{sec:EffectsOnKSat}
\subsection{SAT and its Connection to Quantum Computing}
\label{ssec:Con2QC}

\mygls{sat} is the seminal NP-complete problem that influenced and shaped major parts of complexity theory~\cite{Cook_71,Karp_72}.
\mygls{sat} can naturally express logically constrained problems, and is relevant in many applications:
This includes interdisciplinary verification problems (\eg, checking against a formal specification, as in electronic circuits or software development) or scheduling problems (\eg, in manufacturing or timetable planning) that are highly relevant industrial problems~\cite{Bofill2020, Marques-Silva2008}. For the following analysis, we
consider SAT instances that arise from industrially inspired problems.

Recall that a \mygls{sat} formula in \mygls{cnf} consists of $m$ clauses $C_i$ that are conjoined by logical \textit{and} ($\land$) operations:
\begin{equation*}
    \psi(\vec{x}) = \bigwedge_{i=1}^m C_i,\; \vec{x} \in \{T,F\}^n,
\end{equation*}
where each clause consists of literals $l_i$ (\ie, possibly negated variables $x_i$) that are conjoined by logical \textit{or} ($\lor$) operations.
For instance, \(C_1 = (x_1 \lor x_2 \lor \overline{x_3})\)
is a clause that consist of variables $x_1$, $x_2$ and $x_3$ and literals $x_1$, $x_2$ and $\overline{x}_3$ ($\overline{x}_3$ denotes the negation of $x_3$). 
For any given input $\vec{x} \in \{T,F\}^n$ of a \mygls{sat} formula $\psi(\vec{x})$, $\psi(\vec{x})$ evaluates to either true (T) or false (F) (we substitute $1$ for T and $0$ for F in the following).
\mygls{sat} is the task of deciding whether or not there exists an input $\vec{x}$ such that $\psi(\vec{x})$ is true. $|C|$ denotes the number of literals or variables in a clause $C$ \footnote{Note that the number of literals in a clause is equal to the number of variables in the same clause. However, the number of literals for a given formula might be (and usually is) larger than the number of variables.}.
We call $\psi(\vec{x})$ a $k$-\mygls{sat} instance if $k$ is the maximum number of literals in a clause (\ie, $k = \max_{i \in \{1,\ldots,m\}} |C_i|$).
If each clause in a \mygls{sat} instance $\psi(\vec{x})$ has the same number of literals $k$ (\ie, $|C_i| = k, i \in \{1,\ldots,m\}$), we call $\psi(\vec{x})$ an exact $k$-\mygls{sat} instance.

\begin{figure}[tbh]
    \centering
    \includegraphics[page=1,width=\linewidth]{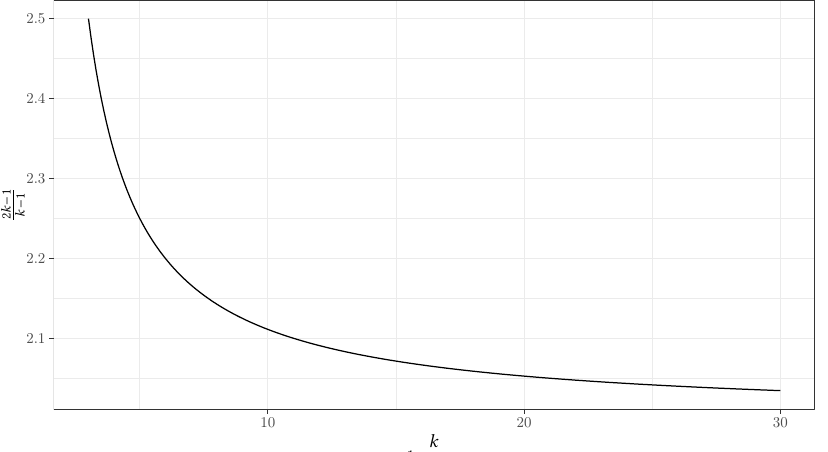}
    \caption{Location of satisfiable to non-satisfiable phase transition for power law distributed $k$-\mygls{sat} instances for varying $k$ and small enough ratios $\frac{m}{n}$.}
    \label{fig:PowerLawPhaseTransition}
\end{figure}

The structural properties of \mygls{sat} formulae are known to strongly influence the performance of \mygls{sat} solvers. 
For instance, there are uniformly random generated \mygls{sat} formulas, where exact $k$-\mygls{sat} clauses are randomly populated with literals --- each with identical probability\footnote{Clauses that contain a variable and its negation are trivially satisfiable and can be excluded.}. 
By fine tuning the number of clauses $m$ and variables $n$, the generation process delivers instances with a well-known phase transition for $k=3$ at $\alpha = \frac{m}{n} \approx \alpha_C = 4.267$ from satisfiable to non-satisfiable instances~\cite{Mzard2002}.
While such instances are of fundamental interest~\cite{Froleyks_2021},
practical (industry-relevant) problems usually exhibit substantially different structures.
This gives rise to targeted \mygls{sat} solvers~\cite{Ansotegui_09_Towards_Industrial, Sundermann2023, Morgado2013}.
Since our proposed \mygls{lsr} method is not independent of the input structure, it is interesting what characteristics arise from different input structures, particularly ones close to industry-relevant problems.
Ansótegui~\etal~\cite{Ansotegui_09_Towards_Industrial} propose several methods to generate industrial-like \mygls{sat} instances.
In particular, they uses a power law distribution for variables in clauses; the resulting instances come close to industrial \mygls{sat} instances, as Friedrich~\etal show~\cite{Friedrich_17_phase, Friedrich_17_power_law}.
Similar to uniform random \mygls{sat} instances, they also show a phase transition, whose location however not only depends on the ratio of clauses to variables, but also on the power law distribution that variables are sampled from --- defined by $\beta_S$.
In particular, a phase transition from satisfiable to non-satisfiable instances occurs at $\beta_S = \frac{2k-1}{k-1}$ for a small enough constant $\alpha = \frac{m}{n}$~\cite{Friedrich_17_power_law}. 
\autoref{fig:PowerLawPhaseTransition} shows that the location of the phase transition assumes smaller values for higher $k$.
In particular, it converges to $2$ for high values of $k$: $\lim_{k\to \infty} \frac{2k-1}{k-1} = 2$.
For example, fixing $\beta_S = 2.5$ and increasing $k > 3$ lets instances tend towards satisfiable instances, provided a small enough clause to variable ratio.
Take into consideration that this cannot be stated for \mygls{sat} instances within a certain region that is formed by higher clause to variable ratios and $\beta_S > \frac{2k-1}{k-1}$, since the location of a (possible) phase transition is not known precisely.

Transforming \mygls{sat} instances into representations 
compatible with quantum solvers
involves many choices and parameters (\eg, concrete transformation method, (intermediate) optimisation steps, choice of (quantum) algorithm and its parameters, mapping and routing problem, hardware platform) that influence properties of the resulting hardware executable representation. 
We zoomed into this pictured tree of transformation paths by setting the starting point at uniform random $k$-\mygls{sat} instances and their transformation to \mygls{qubo} in a previous work~\cite{schmidbauer_25_satstrikesback}.
In contrast, the approach introduced in this paper goes beyond uniform random $k$-\mygls{sat} instances as well as the \mygls{qubo} model by explicitly considering effects on the Ising model and \mygls{qaoa} circuits.
We showed that the number of monomials in a particular $k$-\mygls{sat} to \mygls{pubo} transformation scales exponentially in the number of positive variables in a clause, which quickly prevails positive effects of this transformation path when increasing $k$.
In this work, we introduce an additional prior optimisation step that prevents this unfavourable exponential scaling.

\begin{figure}[htb]
    \centering
    \includegraphics[page=1]{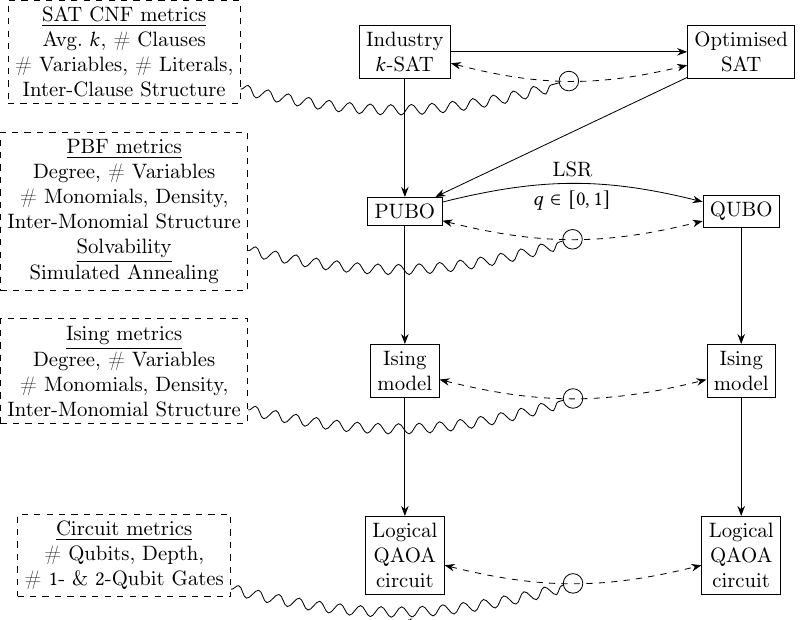}
    \caption{Experiment setup: Starting at industrial like $k$-\acrshort{sat} formulas, their transformation (over an optimised \acrshort{sat} formulation) to \acrshort{pubo}, \acrshort{qubo}, the Ising model and logical quantum circuits for \acrshort{qaoa} with special focus on properties at and between each respective horizontal representation (cf.~\autoref{tab:TransPathsNamingConvention}). Solid arrows represent transformations (\eg, from \mygls{pubo} to \mygls{qubo}); dashed arrows depict comparisons between two (similar) representations and are associated with specific properties (\eg, metrics or solvers) as shown on the left.}
    \label{fig:SAT_overview}
\end{figure}

\autoref{fig:SAT_overview} gives an overview of our experiments: Solid rectangles represent representations at different levels of  abstraction --- starting at $k$-\mygls{sat} (top) over \mygls{pubo} and \mygls{qubo} (\ie, \myglspl{pbf}), the Ising model and \mygls{qaoa} circuits (bottom).
Solid arrows represent transformations (in-)between levels of abstraction. 
An industrial-like $k$-\mygls{sat} instance,
is either directly mapped to \mygls{pubo} (referred to as \acrshort{dPubo}) or first subjected to an optimisation step and then cast as \mygls{pubo} (referred to as \acrshort{optPubo}) for comparison.
These two \mygls{pubo} formulations can then be used to either directly transform into Ising models or to first transform to \mygls{qubo} (referred to as \acrshort{dQubo} or \acrshort{optQubo}) and then a mapping to Ising models. 
We then create logical \mygls{qaoa} circuits from each of the four Ising models.
Logical quantum circuits are independent of the hardware platform and would need to undergo further transformations to be hardware executable (\eg, mapping problem, qubit assignment problem or transpilation to the hardware gate set).
In \autoref{fig:SAT_overview}, we list important properties that are associated to the representation in each abstraction layer on the left. 
The main objective of the following experiments is to characterise how and to what extent they change when applying transformations in the same abstraction layer (\ie, horizontal transformations) or applying transformations between abstraction layers (\ie, vertical transformations).
In the following sections, we first introduce each representation, necessary concepts and transformations and then characterise their respective influence --- using the aforementioned $k$-\mygls{sat} instances as input.

\subsection{(Optimised) $k$-SAT to PBF}
\label{ssec:OptSAT2PBF}
\subsubsection{Optimised $k$-SAT}
Recall that a \mygls{pbf} is a function  $f: \{0,1\}^n \rightarrow \mathbb{R}$ that assumes a multi linear representation. 
Since the domain of $f$ consists of all binary vectors with length $n$ and is isomorphic to the domain of a \mygls{sat} formula with $n$ variables, it is natural to directly map variables from \mygls{sat} to \mygls{pubo}.
By using De-Morgan's laws, it is possible to map clauses without introducing additional variables~\cite{Dobrynin_2024}. 
This is a local\footnote{In the sense of this isolated transformation --- without considering effects of subsequent transformations.} benefit compared to other methods (\eg, via the maximum independent set problem) that use additional variables~\cite{choi_2010, schmidbauer_25_satstrikesback}.

More technically, let $\psi(\vec{x}) = C_1 \land C_2 \land C_3$ be an exact $4$-\mygls{sat} instance, with 
\begin{align*}
        C_1 &= (\overline{x_1} \lor \overline{x_2} \lor \overline{x_3} \lor \overline{x_4}), &
        C_2 &= (x_2 \lor x_3 \lor \overline{x_4} \lor x_5),&
        C_3 &= (x_2 \lor \overline{x_3} \lor x_4 \lor x_5).
\end{align*}
While negative literals $\overline{x_i}$ can be directly mapped to variables $x_i$, positive literals $x_i$ are mapped to terms $(1-x_i)$, resulting in the following higher-order\footnote{Note that this mapping automatically results in a quadratic respectively linear \mygls{pbf} for $2$- or $1$-\mygls{sat}, respectively.} \myglspl{pbf} of degree $4$:
\begin{equation*}
\label{eq:PBFforClauseExample}
    \begin{split}
        f_{C_1}(x_1,x_2,x_3,x_4) &= 1 - x_1x_2x_3x_4,\\
        f_{C_2}(x_2,x_3,x_4,x_5) &= 1 - (1-x_2)(1-x_3)x_4(1-x_5),\\
        f_{C_3}(x_2,x_3,x_4,x_5) &= 1 - (1-x_2)x_3(1-x_4)(1-x_5).\\
    \end{split}
\end{equation*}
The objective function encoding $\psi(\vec{x})$ is obtained by:
\begin{equation}
\label{eq:SATSummationOptimisation}
    f(\vec{x}) = \sum_{m \in \{1,2,3\}} f_{C_m}.
\end{equation}
$\psi(\vec{x})$ represents a maximisation problem,
and the equivalent minimisation problem $-f(\vec{x})$, is easy to obtain.
Let $C = (x_1 \lor x_2 \lor \ldots \lor x_k)$ be a $k$-\mygls{sat} clause with $k \geq 2$ positive literals. 
Then its corresponding degree-$k$ \mygls{pbf} is given by $f_C(x_1,x_2,\ldots,x_k) = 1 - (1-x_1)(1-x_2)\ldots(1-x_k)$. 
However, this leads to an exponential amount of monomials in $f_C$ after term expansion. 
To be more precise, the total number of monomials for function $f_C$ is given by 
\begin{equation}
\label{eq:SATExponentialPositiveSimplemapping}
    T_{f_C} = 2^t -1,    
\end{equation}
with $t$ (for this example: $t=k$) being the number of positive literals in clause $C$.
As mentioned in \autoref{ssec:ExperimentalResults} (see also \autoref{fig:DegreeSweepVariables}), this is intractable for practical instances beyond a certain $k > k_\alpha$.
Such clauses are formed naturally when variables are subject to an \textit{at least one} relationship (\eg, \textit{at least} one machine in a factory has to fulfil conditions specified by the meaning of variables).
Consider that the difference between positive and negative literals in clauses is not simply a matter of inverting the corresponding function, but a matter of shifting the minima (for maximisation) or maxima (for minimisation), which we illustrate in \autoref{tab:SATMinMaxShiftingPositiveVsNegative}.

\begin{table}[htb]
    \centering
    \caption{Effect of literal negation in \mygls{sat} for $C_1=(x_1 \lor x_2 \lor x_3)$, $C_2 = (\overline{x_1} \lor \overline{x_2} \lor \overline{x_3})$ and corresponding functions $f_{C_1}(x_1,x_2,x_3) = 1-(1-x_1)(1-x_2)(1-x_3)$ and $f_{C_2}(x_1,x_2,x_3) = 1 - x_1x_2x_3$.}
    \label{tab:SATMinMaxShiftingPositiveVsNegative}
    \begin{tabular}{ccccc}\toprule
            $x_1x_2x_3$    & $C_1$   & $f_{C_1}$ &  $C_2$ & $f_{C_2}$ \\ \midrule
            $000$ & $0$ & $0$ & $1$ & $1$ \\
            $001$ & $1$ & $1$ & $1$ & $1$ \\
            ... & ... & ... & ... & ... \\
            $110$ & $1$ & $1$ & $1$ & $1$ \\
            $111$ & $1$ & $1$ & $0$ & $0$ \\
                    
         \bottomrule
    \end{tabular}
\end{table}

Hence, we desire an optimised input $k$-\mygls{sat} instance with less positive variables. 
Therefore, we iteratively replace a positive literal in all clauses that it occurs in by a new negative literal and constrain it by two new clauses.
More technically, let $\psi(\vec{x}) = C_1 \land C_2$ be an exact $5$-\mygls{sat} instance with 
\begin{equation*}
    \begin{split}
        C_1 &= (x_1 \lor x_2 \lor x_3 \lor x_4 \lor x_5),\\
        C_2 &= (\overline{x_1} \lor \overline{x_2} \lor \overline{x_3} \lor x_4 \lor x_5).
    \end{split}
\end{equation*}
Then, $x_5$ can be replace by a new negative literal \textcolor{lfdblue}{$\overline{x_6}$}:
\begin{equation}
\label{eq:SATExampleReplaceWithNegative}
    \begin{split}
        C_1 &= (x_1 \lor x_2 \lor x_3 \lor x_4 \lor \textcolor{lfdblue}{\overline{x_6}}),\\
        C_2 &= (\overline{x_1} \lor \overline{x_2} \lor \overline{x_3} \lor x_4 \lor \textcolor{lfdblue}{\overline{x_6}}),\\
        C_3 &= (x_5 \lor \textcolor{lfdblue}{x_6}),\\
        C_4 &= (\overline{x_5} \lor \textcolor{lfdblue}{\overline{x_6}}),
    \end{split}
\end{equation}
where $C_3$ and $C_4$ are constraint clauses that ensure $x_5 = \overline{x_6}$.
These steps could be repeated until no more positive literals occur in $C_1$ or $C_2$ \footnote{As a side note, whenever a variable either is only present as positive or negative literals in a \mygls{sat} formula, corresponding clauses can be trivially excluded (since they are trivially satisfiable), as it would be the case for $x_4$ and $x_5$ in clauses $C_1$ and $C_2$ (see \autoref{eq:SATExampleReplaceWithNegative}).}.
However, introducing (and constraining) new variables creates new clauses and therefore introduces new monomials. 
To be more precise, constraint clauses $C_3$ and $C_4$ from (\autoref{eq:SATExampleReplaceWithNegative}) are mapped to:
\begin{equation*}
    \begin{split}
        C_3 &= (x_5 \lor \textcolor{lfdblue}{x_6}) \rightarrow f_{C_3}(x_5, x_6) = 1 - (1-x_5)(1-x_6) = x_5 + x_6 - x_5x_6,\\
        C_4 &= (\overline{x_5} \lor \textcolor{lfdblue}{\overline{x_6}})\rightarrow f_{C_4}(x_5, x_6) = 1 - x_5x_6,
    \end{split}
\end{equation*}
which in total leads to $f_{C_3}(x_5,x_6) + f_{C_4}(x_5,x_6) = 1 + x_5 + x_6 - 2x_5x_6$ and therefore four monomials. 
Since the optimisation problem for $\psi(\vec{x})$ is obtained by summation (see \autoref{eq:SATSummationOptimisation}), monomials that are introduced by the constraint clauses might be already present in other functions $f_{C_i}$ and hence would not increase the total number of monomials. 
Since $x_6$ is a newly introduced variable that replaces all occurrences of positive literal $x_5$, monomials $x_6$ and $-2x_5x_6$ can only originate from the two constraint clauses.
Monomial $x_5$ can occur in other functions $f_{C_i}$(\eg, when $\overline{x_5}$ is the only negative literal in a clause $C_i$).
We can now combine these observations with the exponential scaling behaviour of the simple mapping (see \autoref{eq:SATExponentialPositiveSimplemapping}):
Let $C$ be a $k$-\mygls{sat} clause with $t$ positive literals:
\begin{equation}
\label{eq:SATClauseCparmterisedbyTandA}
    C = (\underbrace{\overbrace{\ldots \lor}^{a} \; |\; \overbrace{\ldots \lor}^{t-a}}_{t \text{ positive}} \;|\; \underbrace{\ldots \lor \ldots}_{k-t \text{ negative}}),
\end{equation}
where $t-a$ are the number of positive literals that are replaced according to aforementioned optimisation method --- leaving $a$ positive literals in the optimised clause.
The number of monomials introduced for clause $C$ (including constraint terms) is approximately\footnote{We deliberately leave out the constant monomial, as it is introduced by more than one function $f_{C_i}$ for practically relevant problem instances.} given by $2^a + 3(t-a)$.

\begin{figure}[htb]
    \centering
    \includegraphics[page=1, width=\linewidth]{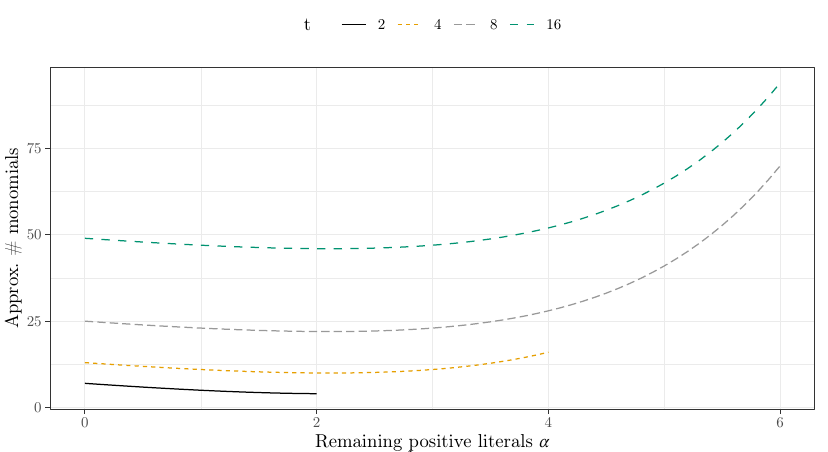}
    \caption{Approximate number of monomials introduced by mapping a clause $C$ (see \autoref{eq:SATClauseCparmterisedbyTandA}). The x-axis shows the number of remaining positive literals after optimisation, while $t$ is the number of positive literals prior to optimisation.}
    \label{fig:SATCompareExpLinearMapping}
\end{figure}

\autoref{fig:SATCompareExpLinearMapping} shows this function over $a$ (x-axis) and different values of $t \in \{2,4,8,16\}$. 
The minimum of each function ($t > 2$) is at $ a \approx 2.11$. 
Since $2^2 + 3(t - 2) < 2^3 + 3(t - 3)$, leaving two positive literals results in approximately the least amount of monomials for an isolated clause.
However, take into consideration that with every introduced additional variable, the (brute-force) search-space doubles in size. 
Hence, we choose to leave four positive literals, since the number of additional monomials is close to the optimum (see \autoref{fig:SATCompareExpLinearMapping}).
Also, the optimisation strategy replaces a positive literal in all clauses that it occurs in --- further reducing the number of introduced variables.

For the following experiments, we use power-law distributed ($\beta=2.5$) $k$-\mygls{sat} instances, with $k \in \{3,5,7,11\}$ and varying numbers of clauses (\ie, $|C| \in \{13, 53, 107, 163, 263\}$).
We also vary the number of possible variables $N \in \{7, 13, 23, 37, 47, 59, 71, 83, 101, 113\}$, while ensuring $N>k$.
However, the actual number of variables $n$ is lower than $N$, whenever variables are not chosen during the random clause sampling process. 
When applicable, we always show the actual number of variables $n$.

\begin{figure}[htb]
    \centering
    \includegraphics[page=1, width=\linewidth]{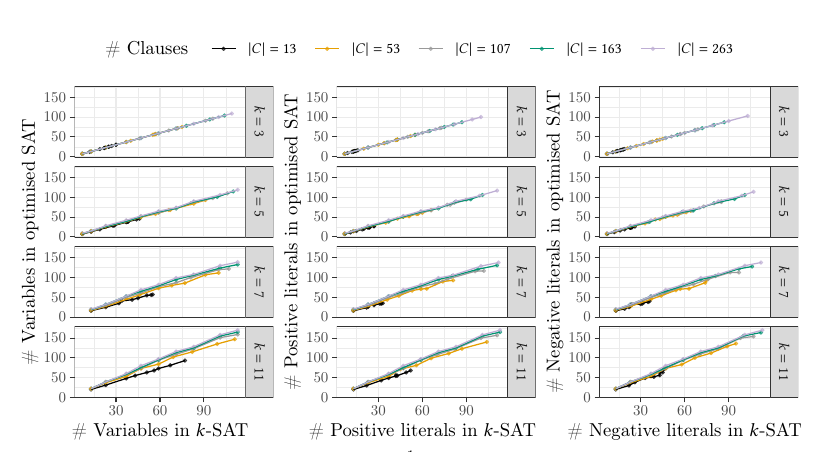}
    \caption{Influence of optimisation strategy (see \autoref{eq:SATExampleReplaceWithNegative}) on important \mygls{sat} metrics (y-axis) based on \mygls{sat} metrics prior to optimisation (x-axis). Number of Variables (left), number of positive literals (middle) and number of negative literals (right) --- coloured by the number of clauses and horizontally faceted by $k$ in the \mygls{sat} instance prior to optimisation.}
    \label{fig:SATVarsLits}
\end{figure}

\autoref{fig:SATVarsLits} shows the effect of the optimisation strategy on three \mygls{sat} metrics (\ie, left: the number of variables,  middle: the number of positive literals and right: the number of negative literals). 
Metrics for the optimised \mygls{sat} instance are shown on the y-axis, while metrics for the instance prior to optimisation are shown on the x-axis. 
Furthermore, we differentiate between the number of clauses by colour and $k$ by horizontal facets.
From a birds eye view, we can see that the scaling behaviour follows a linear relationship for $k=3$ with larger differences at higher $k$.
This is due to the fact that we leave four positive variables per clause (see \autoref{fig:SATCompareExpLinearMapping}) and thus $3$-\mygls{sat} instances are unaffected by this strategy. 
In contrast, for $k=5$, only clauses that strictly contain positive literals are eligible for optimisation.
Also, since we only replace positive literals, the number of additional variables is upper bounded by the number of positive literals, which is at maximum the number of variables.
\autoref{fig:SATVarsLits} confirms the limited effect on the number of variables (and literals) for $k=5$.
For $k \in \{7,11\}$, we see that instances with less clauses lead to less introduced variables (and literals).
Furthermore, smaller number of variables have a similar effect. 
The effect of an optimisation step (\ie, replacing a positive literal $x_i$ by a new negative literal $\overline{x_j}$) can be generalised: 
\begin{enumerate}
    \item The number of variables increases by one.
    \item Two additional $2$-\mygls{sat} clauses (\ie, the constraint clauses) are added.
    \item The number of positive literals increases by one, since literals $x_i$ and $x_j$ only occur in one constraint clause.
    \item The number of negative literals increases by at least one ($\overline{x_j}$ in former clauses and in one constraint clause).
If the negative literal $\overline{x_i}$ was not part of the \mygls{sat} instance prior to optimisation, the number of negative literals (due to one constraint clause) additionally increases by one.
\end{enumerate}
From (2), we can deduce that the average $k$ for the optimised \mygls{sat} instance depends on the number of optimisation steps (\ie, the number of additional variables).
Let $\psi(x_1, \ldots, x_n) = C_1 \land \ldots \land C_m$ be an exact $k$-\mygls{sat} instance prior to optimisation.
Then, after $\gamma$ optimisation steps,
\begin{equation*}
    \mathrm{Avg.-}k = \frac{1}{m + \gamma}(\sum_{m} |C_m| + 2\gamma).
\end{equation*}

\begin{figure}[htb]
    \centering
    \includegraphics[width=.4\linewidth]{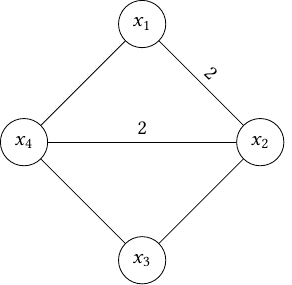}
    \caption{Graph representation of $\psi(\vec{x}) = (x_1 \lor \overline{x_2}) \land (x_2 \lor \overline{x_3} \lor \overline{x_4}) \land (\overline{x_1} \lor \overline{x_2} \lor x_4)$.}
    \label{fig:SATGraphExample}
\end{figure}

Although the mentioned and analysed metrics (see \autoref{fig:SATCompareExpLinearMapping}) are relevant to characterise the scaling behaviour for larger \mygls{sat} instances, they do not give a detailed insight into the inner structure of an (optimised) \mygls{sat} instance.
Following our graph definition for \myglspl{pbf} (see \autoref{sec:GraphRepresentation}), we now introduce a multi-graph representation for \mygls{sat} formulas.
Let $\psi(x_1,\ldots,x_n) = C_1 \land \ldots \land C_m$ be a $k$-\mygls{sat} instance with $m$ clauses and $n$ variables.
For the variable incidence graph $G_\psi(V_\psi,E_\psi)$, we define $V_\psi = \{x_1, \ldots, x_n\}$ as the set of variables of $\psi(\vec{x})$.
In contrast to the graph representation of \myglspl{pbf}, we no longer consider where edges stem from (\ie, which monomials in the case of \myglspl{pbf} or which clauses in the case of \mygls{sat}).
If two variables $x_i$ and $x_j$ (can also be negated literals) both occur in a clause $C_i, i \in \{1,\ldots,m\}$, then edge $e = (x_i, x_j)$ is added to $E_\psi$.
For example, let $\psi(x_1, x_2,x_3,x_4) = C_1 \land \textcolor{lfdblue}{C_2} \land \textcolor{lfdyellow}{C_3}$ with $C_1 = (x_1 \lor \overline{x_2})$, $\textcolor{lfdblue}{C_2 = (x_2 \lor \overline{x_3} \lor \overline{x_4})}$ and $\textcolor{lfdyellow}{C_3 = (\overline{x_1} \lor \overline{x_2} \lor x_4)}$.
Then for its corresponding graph $G_\psi(V_\psi,E_\psi)$, the set of nodes $V_\psi =\{x_1,x_2,x_3,x_4\}$ and the multi-set of edges 
\begin{equation*}
    E_\psi = \big\{(x_1,x_2),
    \textcolor{lfdblue}{(x_2,x_3), (x_2,x_4), (x_3,x_4)}, 
    \textcolor{lfdyellow}{(x_1,x_2), (x_1,x_4), (x_2,x_4)}\big\}.
\end{equation*}
Note that edges $(x_1,x_2)$ and $(x_2,x_4)$ occur two times in $E_\psi$ (\ie, have multiplicity $2$). 
For simplicity and better visual representation, we show the multiplicity on edges as an edge label, instead of drawing multiple edges, which leads to the graph shown in \autoref{fig:SATGraphExample}.

\begin{figure}[htb]
    \centering
     \begin{subfigure}[b]{0.48\textwidth}
         \centering
         \includegraphics[page=1, width=\textwidth]{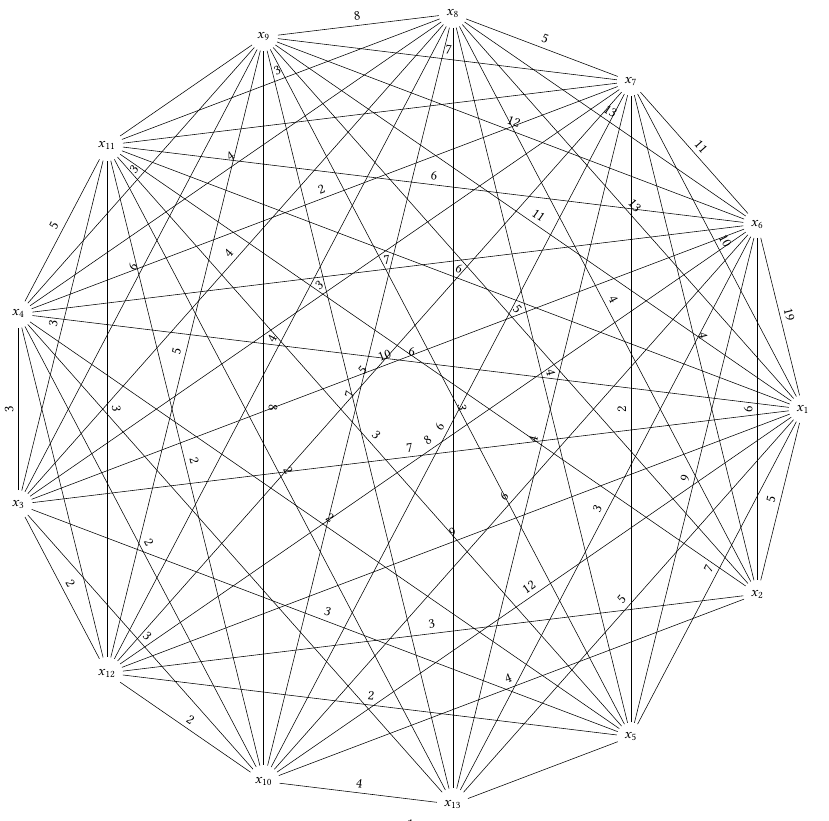}
         \caption{Graph $G_\psi$.}
         \label{fig:SATStructureVsOptSAT_direct}
     \end{subfigure}
     \hfill
     \begin{subfigure}[b]{0.48\textwidth}
         \centering
         \includegraphics[page=1, width=\textwidth]{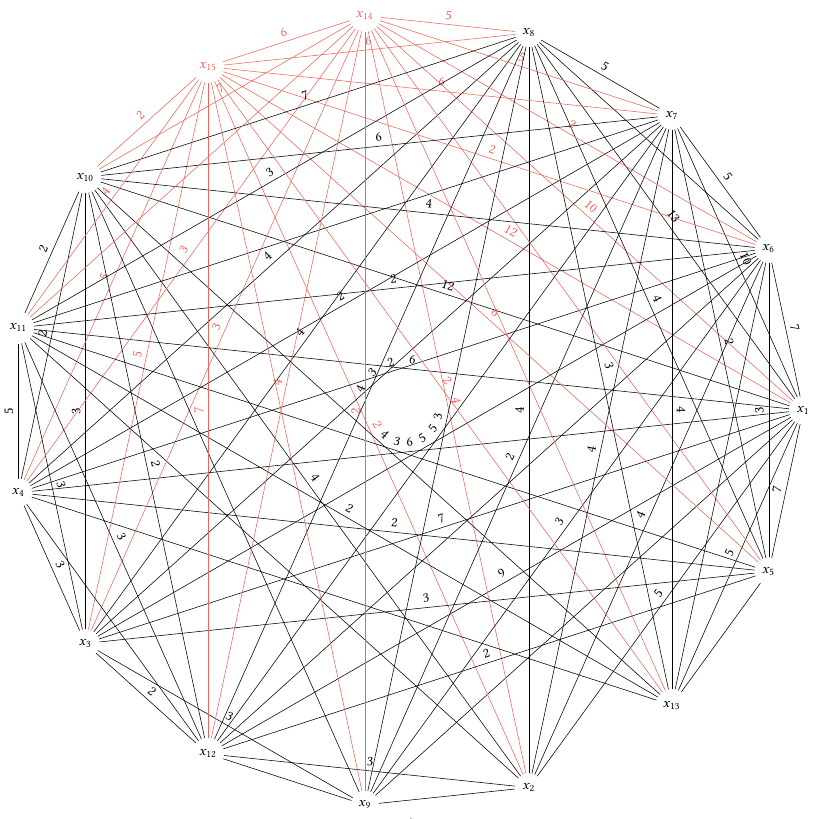}
         \caption{Graph $G_{\psi'}$ of optimised \mygls{sat}.}
         \label{fig:SATStructureVsOptSAT_opt}
     \end{subfigure}
        \caption{Inner structure of a power law distributed $5$-\mygls{sat} instance with $13$ variables and $37$ clauses (left) and inner structure of its optimised \mygls{sat} formula (right). New nodes and edges are coloured in \textcolor{lfdred}{red}.}
        \label{fig:SATStructureVsOptSAT}
\end{figure}

We now want to show the effect of the optimisation strategy on the inner structure of an exemplary power law distributed $5$-\mygls{sat} instance $\psi(\vec{x})$ with $13$ variables and $37$ clauses.
Firstly, \autoref{fig:SATStructureVsOptSAT_direct} shows the graph representation of this instance. 
Secondly, \autoref{fig:SATStructureVsOptSAT_opt} shows the graph representation of the optimised \mygls{sat} instance $\psi'(\vec{x})$, where new nodes and edges are coloured in \textcolor{lfdred}{red} (see also the transformation paths in \autoref{fig:SAT_overview}). 
Note that an increase or decrease in multiplicity does not lead to a different colour on that edge.
We can see that two new variables $x_{14}$ and $x_{15}$ are introduced that are both connected to all other variables via edges. 
Therefore, there are four new constraint clauses that introduce two edges with multiplicity $2$, that are $(x_{14}, x_j)$ and $(x_{15}, x_i)$, where $x_i$ and $x_j$ are the positive literals that were replaced\footnote{There might be additional edges $(x_{14}, x_j)$ or $(x_{15}, x_i)$ that increase the multiplicity between these nodes, since we do not replace negative literals.}.
There are $6$ connections between the two new variables and therefore there are exactly $6$ clauses that contain these two variables. 
Recall that we choose to leave $4$ positive variables per clause and hence there should be no connections between these two additional variables, since we started at a $5$-\mygls{sat} instance. 
However, positive literals are replaced in every clause they occur in (which further reduces the amount of monomials introduced in the \mygls{pbf} mapping).
Hence, there must have been $6$ clauses containing both positive literals $x_i$ and $x_j$.
We will use these two graph representations and their underlying \mygls{sat} instances as the basis for the remaining transformation steps (see \autoref{fig:SAT_overview}) and figures down to a logical \mygls{qaoa} circuit.

In summary, the effects of the optimisation strategy on important metrics for \mygls{sat} are predictable (analytically or quantitatively), which is an important factor when integrating this strategy into (automatic) toolchains.
However, the question remains to what extent these \mygls{sat} instances affect properties of following \myglspl{pbf}.
Since we also want to consider the effects of a \mygls{pubo} to \mygls{qubo} transformation (see \autoref{fig:SAT_overview}), we refer to each path of transformation as in \autoref{tab:TransPathsNamingConvention}.

\begin{table}[htb]
    \centering
    \caption{Naming convention for transformation paths (see \autoref{fig:SAT_overview}).}
    \label{tab:TransPathsNamingConvention}
    \begin{tabular}{cc}\toprule
        Transformation path & Description       \\ \midrule
            \glstarget{dPubo}{\glsentryshort{dPubo}} & \mygls{pubo} from a non-optimised \mygls{sat} instance.     \\
            \glstarget{optPubo}{\glsentryshort{optPubo}} & \mygls{pubo} from an optimised \mygls{sat} instance.    \\
            \glstarget{dQubo}{\glsentryshort{dQubo}}   & \mygls{qubo} from a \mygls{pubo} from a non-optimised \mygls{sat} instance.    \\
            \glstarget{optQubo}{\glsentryshort{optQubo}} & \mygls{qubo} from a \mygls{pubo} from an optimised \mygls{sat} instance.    \\
         \bottomrule
    \end{tabular}
\end{table}

\subsubsection{PUBO and QUBO}
\begin{figure}[htb]
    \centering
    \includegraphics[width=\linewidth]{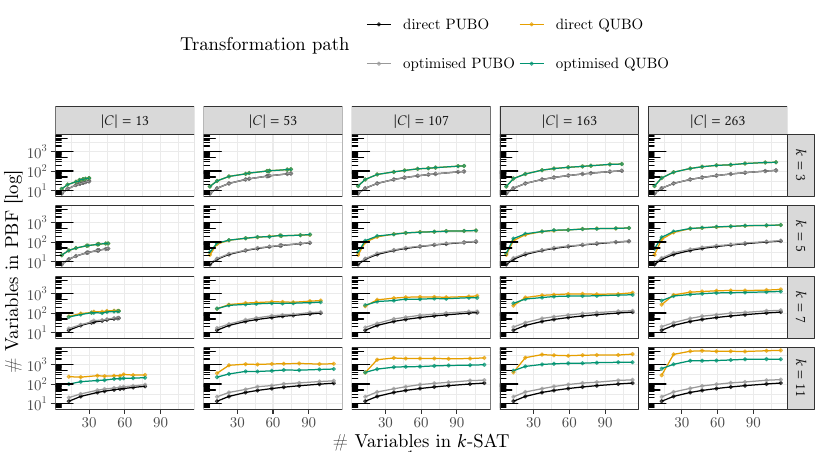}
    \caption{Number of variables in \mygls{pbf} (y-axis) vs number of variables in non-optimised $k$-\mygls{sat} (x-axis) horizontally faceted by $k$ and vertically faceted by the number of clauses. Colour represents the concrete transformation path (\ie, from non optimised $k$-\mygls{sat}: d. and from optimised $k$-\mygls{sat}: opt.; see \autoref{tab:TransPathsNamingConvention}). The difference between \acrshort{dPubo} and \acrshort{dQubo} depicts the number of iterations in the \mygls{lsr} algorithm and therefore the number of additional variables in \mygls{qubo}. The same applies when comparing \acrshort{optPubo} and \acrshort{optQubo}.}
    \label{fig:SATNumVarsPBF}
\end{figure}

\begin{figure}[htb]
    \centering
    \includegraphics[width=\linewidth]{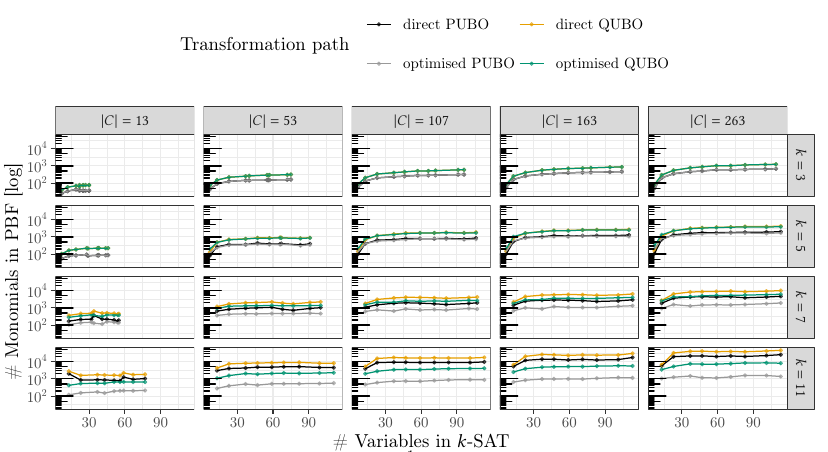}
    \caption{Number of monomials in \mygls{pbf} (y-axis) vs number of variables in non-optimised $k$-\mygls{sat} (x-axis) horizontally faceted by $k$ and vertically faceted by the number of clauses. Colour represents the concrete transformation path (\ie, from non optimised $k$-\mygls{sat}: d. and from optimised $k$-\mygls{sat}: opt.; see \autoref{tab:TransPathsNamingConvention}).}
    \label{fig:SATNumMonomialsPBF}
\end{figure}

For the \mygls{pubo} to \mygls{qubo} transformation via the \mygls{lsr} method, we use percentile $q=1$ (see \autoref{alg:lsr}) to introduce less monomials. 
The following figures show the number of actually used variables of the non-optimised \mygls{sat} instance on the x-axis.
\autoref{fig:SATNumVarsPBF} and \ref{fig:SATNumMonomialsPBF} show the number of variables and the number of monomials (y-axis) in \mygls{pbf} respectively. 
As in \autoref{fig:SATVarsLits}, they are horizontally faceted by $k$. 
Conversely, vertical facets show the the number of clauses in the non-optimised \mygls{sat} instance and colour encodes the specific path of transformation.
Recall that the transformation from \mygls{sat} to \mygls{pbf} does not introduce additional variables.
Therefore, the difference between \acrshort{dPubo} and \acrshort{optPubo} in \autoref{fig:SATNumVarsPBF} shows the number of introduced variables by the optimisation strategy.
Also, in \autoref{fig:SATNumVarsPBF}, \acrshort{dPubo} is a linear relation and can therefore be used as reference for the logarithmic y-axis. 
For \autoref{fig:SATNumVarsPBF}, we can clearly distinguish between \mygls{pubo} and \mygls{qubo}, with the latter using more variables. 
Also, as $k$ increases, the difference becomes larger, which is to be expected, since higher $k$ leads to higher-degree monomials in \mygls{pbf}. 
With higher-degree monomials, more reduction steps are necessary and therefore more variables are introduced. 
However, this trend can break if specific structures (\eg, if higher-degree monomials share many variable pairs) can be leveraged by the \mygls{lsr} method.
With higher number of clauses, the difference between the number of variables in \mygls{qubo} an \mygls{pubo} becomes larger --- meaning more variables are introduced.
Interestingly, both \mygls{qubo} formulations tend towards a relatively stable number of variables, when increasing the number of variables in \mygls{sat} (and \mygls{pubo}). 
We presume that for fixed number of clauses $|C|$ and fixed $k$, the number of variables in \mygls{qubo} (after transformation from \mygls{pubo}) is relatively stable for varying clause to variables ratios $\frac{m}{n} > 1$.
Another interesting aspect is the difference in both \mygls{qubo} models for $k=11$. 
Clearly, the \mygls{sat} optimisation strategy is beneficial for the number of variables in \mygls{qubo} --- even for smaller $k$, the optimisation strategy does not increase the number of variables in \mygls{qubo}, which is contrary to the \mygls{pubo} formulation.

Apart from the number of variables, the number of monomials in \myglspl{pbf} majorly impact the performance of (quantum) solvers. 
For example, in Simulated Annealing, the objective function needs to be evaluated for every variable flip, which mostly governs the computational effort per iteration.
Indubitably, finding a solution with Simulated Annealing depends on the structure and size of the search space, which (without further reduction) grows exponentially with increasing number of variables.
Analogously to \autoref{fig:SATNumVarsPBF}, \autoref{fig:SATNumMonomialsPBF} shows the number of monomials on its y-axis. 
Note that for $k=3$ the optimisation strategy has no effect and thus \acrshort{dPubo} and \acrshort{optPubo}, as well as \acrshort{dQubo} and \acrshort{optQubo} do not show differences.
In contrast to \autoref{fig:SATNumVarsPBF}, it is evident that with increasing $k$, \acrshort{optQubo} outperforms \acrshort{dPubo}, albeit \acrshort{optPubo} having the lowest number of monomials.
To put this into perspective, the \mygls{pubo} model can encode information into monomials ranging up to degree-$k$, while the \mygls{qubo} model is restricted to degree-$0$, -$1$ and -$2$ monomials. 
Since we count the number of monomials, this highlights the impact of the \mygls{sat} optimisation strategy and shows that upstream optimisation can have major impact on representations in lower layers.
Analogously to \autoref{fig:SATNumVarsPBF}, the number of monomials in \mygls{pubo} and \mygls{qubo} are relatively stable with increasing number of variables in \mygls{sat}, which means that (a) the optimisation strategy does not introduce specific structure that would significantly hinder or benefit the \mygls{lsr} method and (b) indicates why the number of variables for \mygls{qubo} in \autoref{fig:SATNumVarsPBF} is also relatively stable.

\begin{figure}
    \centering
    \includegraphics[width=.4\linewidth]{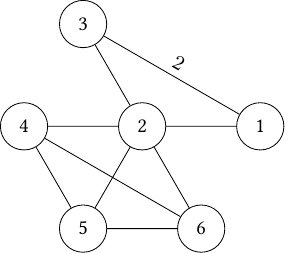}
    \caption{Multi-graph $G^f$ for $f(x_1, \ldots, x_6) = \pi x_1x_2x_3 -13 x_2x_4x_5x_6 + 7 x_1x_3$.}
    \label{fig:SATMultiGraphPBFExample}
\end{figure}

Apart from the degree, the number of variables and the number of monomials, we also want to characterise the inner structure of resulting \myglspl{pbf}. 
Hence, we define a similar multi-graph structure, as for \mygls{sat} (see \autoref{fig:SATGraphExample}). 
Let $f(x_1, \ldots, x_n)$ be a degree-$k$ \mygls{pbf} with $m$ monomials.
We define the set of vertices $V^f$ in its graph $G^f(V^f,E^f)$ as the set of variables from $f$ ($V^f = \{x_1, \ldots, x_n\}$).
Similarly, an edge $e=(x_i, x_j)$ is added to the multi-set of edges $E^f$, if both $x_i$ and $x_j$ occur in the same monomial. 
Note that this definition differs from the graph definition in \autoref{sec:GraphRepresentation} such that we no longer consider to which monomial an edge belongs to. 
As with the graph representation for \mygls{sat}, we do not show multiple edges between two nodes, but rather denote the number of edges between two nodes by an edge label.
For example, let $f(x_1, \ldots, x_6) = \pi x_1x_2x_3 -13 x_2x_4x_5x_6 + 7 x_1x_3$ be defined as for \autoref{fig:Graph_example} and let $G^f(V^f,E^f)$ be its corresponding graph, which we show in \autoref{fig:SATMultiGraphPBFExample}.

\begin{figure}[htb]
    \centering
     \begin{subfigure}[b]{0.48\textwidth}
         \centering
         \includegraphics[page=1, width=\textwidth]{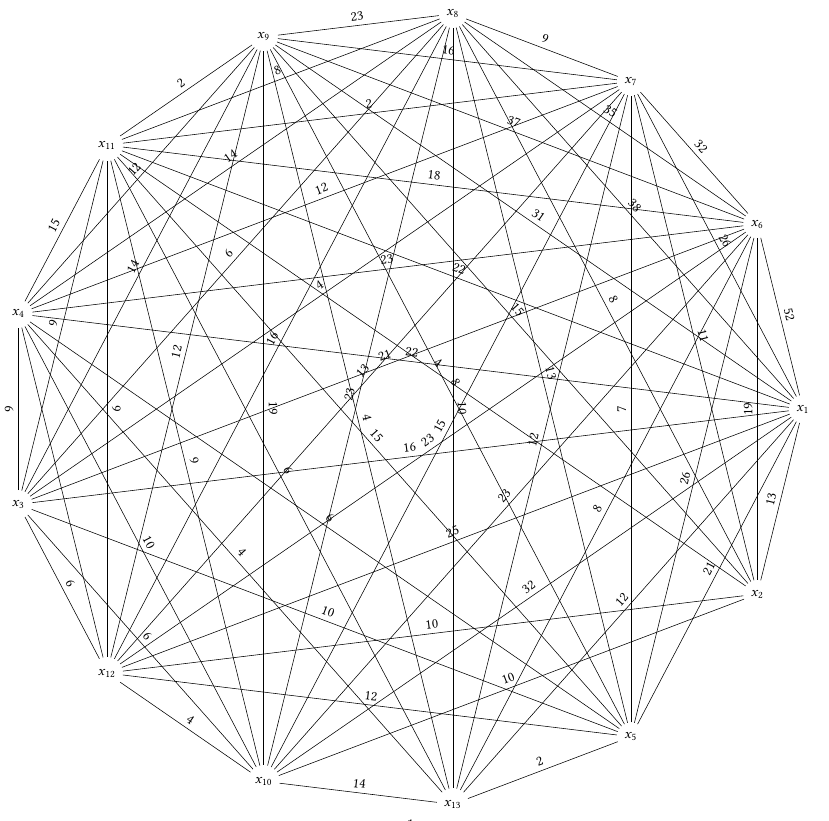}
         \caption{Graph for \acrshort{dPubo}.}
         \label{fig:SATStructurePBF_dPubo}
     \end{subfigure}
     \hfill
     \begin{subfigure}[b]{0.48\textwidth}
         \centering
         \includegraphics[page=1, width=\textwidth]{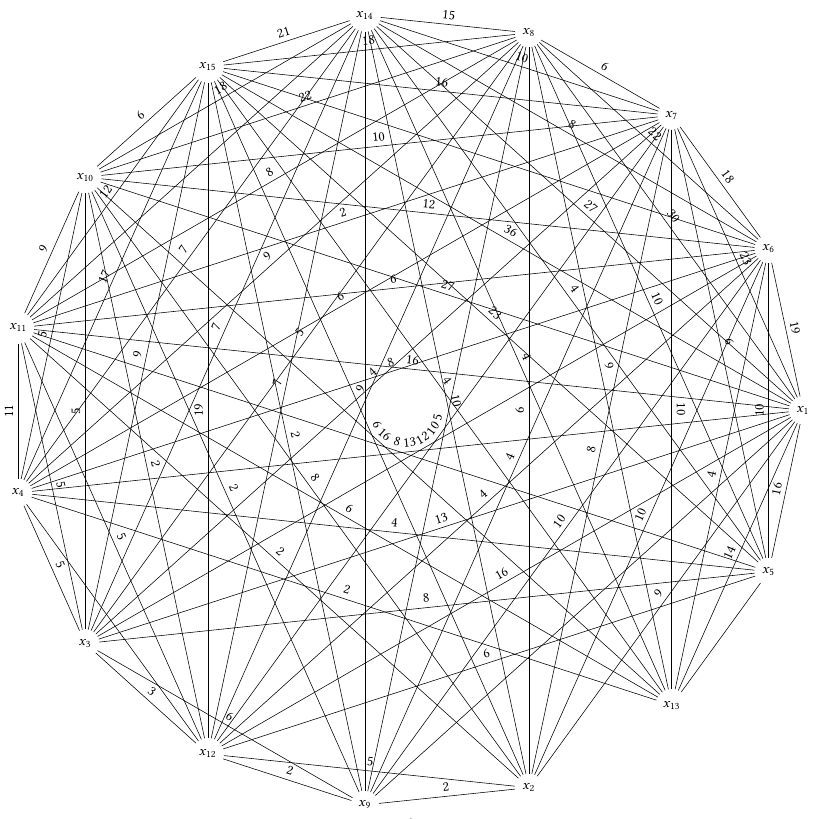}
         \caption{Graph for \acrshort{optPubo}.}
         \label{fig:SATStructurePBF_optPubo}
     \end{subfigure}
     \hfill
     \begin{subfigure}[b]{0.48\textwidth}
         \centering
         \includegraphics[page=1, width=\textwidth]{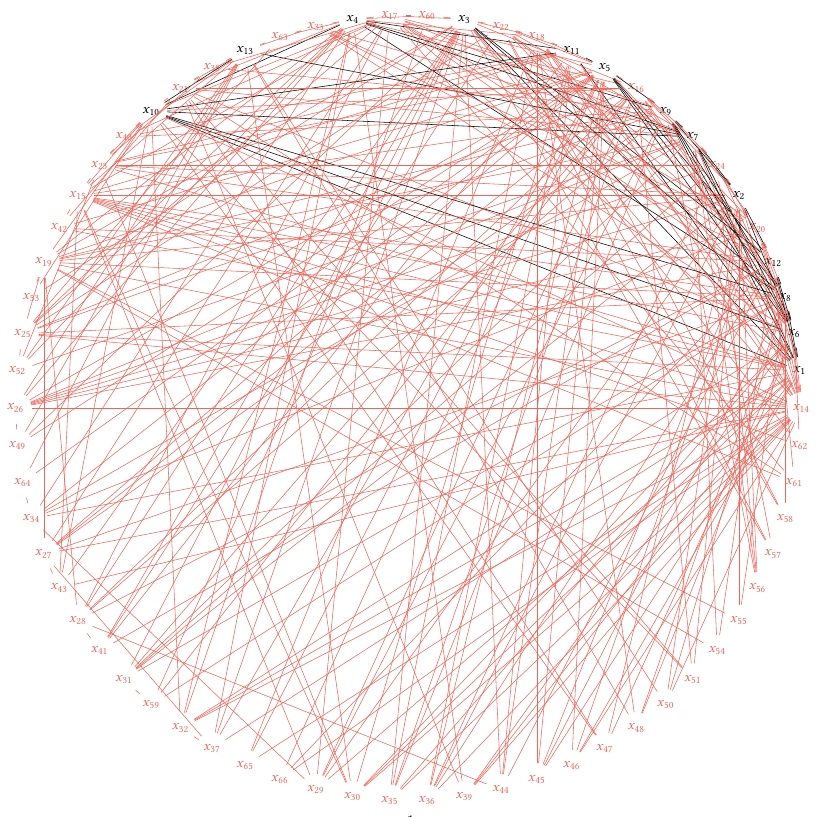}
         \caption{Graph for \acrshort{dQubo}.}
         \label{fig:SATStructurePBF_dQubo}
     \end{subfigure}
     \hfill
     \begin{subfigure}[b]{0.48\textwidth}
         \centering
         \includegraphics[page=1, width=\textwidth]{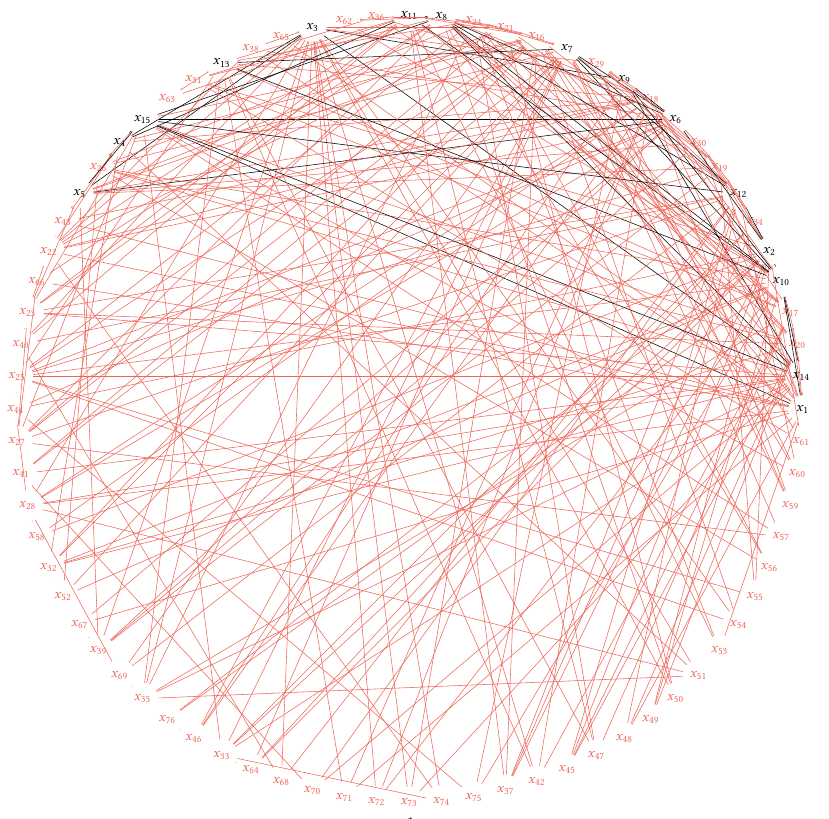}
         \caption{Graph for \acrshort{optQubo}.}
         \label{fig:SATStructurePBF_optQubo}
     \end{subfigure}
        \caption{Inner structure of \mygls{pubo} (top) and \mygls{qubo} (bottom) representations of a power law distributed $5$-\mygls{sat} instance with $13$ variables and $37$ clauses (left). Inner structure of its optimised \mygls{sat} formula and derived \mygls{pubo} (top) and \mygls{qubo} (bottom) representations (right). New nodes and edges compared to the previous representation (see \autoref{fig:SATStructureVsOptSAT} for \mygls{pubo} models) are coloured in \textcolor{lfdred}{red}. See \autoref{fig:SAT_overview} and \autoref{tab:TransPathsNamingConvention} for more information about the concrete transformation paths.}
        \label{fig:SATStructurePBF}
\end{figure}

Following this definition, \autoref{fig:SATStructurePBF} shows the graphs for \acrshort{dPubo}, \acrshort{optPubo}, \acrshort{dQubo} and \acrshort{optQubo} for the same $5$-\mygls{sat} instance as in \autoref{fig:SATStructureVsOptSAT} with the non-optimised \mygls{sat} instance as basis on the left and the optimised instance on the right.
As with the previous graphs, we colour new nodes and edges in \textcolor{lfdred}{red} (changing multiplicities alone do not lead to the colour \textcolor{lfdred}{red}).
We can see that both \mygls{pubo} representations do not introduce additional variables and do not connect previously unconnected node pairs compared to \autoref{fig:SATStructureVsOptSAT}. 
This is a result of the mapping procedure:
Let $\psi(\vec{x}) = (x_1 \lor x_2 \lor x_3)$ be a $3$-\mygls{sat} instance, $G_\psi$ its corresponding graph and let $f(x_1,x_2,x_3) = 1 - (1-x_1)(1-x_2)(1-x_3) = x_1 + x_2 + x_3 -x_1x_2 -x_1x_3 -x_2x_3 + x_1x_2x_3$ be its corresponding degree-$3$ \mygls{pbf} and $G^f$ its corresponding graph.
$G_\psi$ has exactly three edges, that connect all nodes --- forming a $3$-clique.
Since $f$ contains $x_1x_2x_3$ as a monomial, this monomial alone also leads to the same $3$-clique in $G^f$. 
However, since (due to term expansion) lower-order monomials of at least degree-$2$ are present in $f$, the multiplicities increase. 
This can also be seen in \autoref{fig:SATStructurePBF_dPubo} and \ref{fig:SATStructurePBF_optPubo}, where a larger increase in multiplicities is an indicator of lower-order terms, which originate from term expansion of positive literals in the \mygls{sat} instance. 
Therefore, we can see that in total, the increase in multiplicities is lower for \acrshort{optPubo} than for \acrshort{dPubo} compared to \autoref{fig:SATStructureVsOptSAT}.
Take into consideration that this effect is more pronounced for higher $k$.
\autoref{fig:SATStructurePBF_dQubo} and \ref{fig:SATStructurePBF_optQubo} show the respective graphs for \mygls{qubo} that originate from \mygls{pubo} (top) via our introduced \mygls{lsr} algorithm ($q=1$).
We can see that, compared to the graphs for \mygls{pubo}, many new edges and variables are introduced. 
For each new variable at least three new edges (stemming from the penalty term) are introduced.
However, take into consideration that graphs $G^f$ for quadratic \myglspl{pbf} cannot have multiple edges between nodes $x_i$ and $x_j$, since there is only a single degree-$k$, $k\leq2$ monomial containing both $x_i$ and $x_j$ (\ie, $\alpha x_ix_j$).
Hence, each edge directly visualises a degree-$2$ monomial.
We can see a slightly higher concentration of edges among the former variables in the graph for \acrshort{dQubo} than for \acrshort{optQubo}, although the differences are not clearly visible from this small $5$-\mygls{sat} instance. 
For higher $k$, larger number of variables and larger number of clauses, the graphs for \mygls{qubo} (due to the direct visualisation of degree-$2$ monomials) will follow the observations of \autoref{fig:SATNumVarsPBF} and \ref{fig:SATNumMonomialsPBF} --- meaning that there are less edges and variables for \acrshort{optQubo}\footnote{Note that we also count the number of degree-$0$ and -$1$ monomials in \autoref{fig:SATNumMonomialsPBF}. However, there is at maximum only one degree-$0$ monomial and the number of degree-$1$ monomials is at maximum the number of variables.}.

\subsubsection{Energy Landscape}
To this point, we thoroughly analysed metrics and how they change when transforming \mygls{sat} to \mygls{pbf}.
What remains in this abstraction layer, is a characterisation of the energy landscape of resulting \myglspl{pbf}.
For instance, Simulated Annealing is a classical heuristic to solve optimisation problems (encoded as \myglspl{pbf}) \cite{Nikolaev_2010, McGeoch_2014, Kirkpatrick_1983}. 
In essence, it is a probabilistic algorithm that iteratively proposes changes to the current solution and either trivially or probabilistically accepts the proposed change.
More technically, let $f(x_1, \ldots, x_n)$ be a \mygls{pubo} that encodes a minimisation problem. 
Then, \autoref{alg:SA} shows the inner workings of a variant of Simulated Annealing, where function \textsc{accept\_anyway($\Delta E, \kappa$)} performs a random experiment to accept proposed changes that increase the energy of $f$ (\ie, $f(\vec{x}') > f(\vec{x})$; see \cite{McGeoch_2014}). 
The idea is to potentially escape local minima at the beginning of the algorithm with high probability (high temperature $\kappa'$). 
Note that the probability also depends on the change in energy (\autoref{alg:SA}: l. 17) --- resulting in lower probability to accept the change if the energy change is larger.
Also, the probability to accept a proposed change (\autoref{alg:SA}: l.17) decreases as the temperature decreases.
Take into consideration that we deliberately trivially accept changes that do not change the energy (\autoref{alg:SA}: l. 6) to allow for exploration of flat energy landscapes.
This is due to the unique properties of \mygls{sat} to \mygls{pubo} formulations that can tend towards (partially) flat energy landscapes. 
For example, let $\psi(x_1,x_2,x_3)$ be an exact $3$-\mygls{sat} formula with $7$ out of $8$ possible unique clauses (meaning no two clauses contain the same literals).
For instance, let $C_\text{missing} = (\overline{x_1} \lor \overline{x_2} \lor x_3)$.
Then, $\psi(\vec{x})$ has exactly one satisfying solution $\vec{x} = 110$.
Its corresponding \mygls{pbf} $f$ also has exactly one optimum at $\vec{x} = 110$.
For any two bit vectors $\vec{x}, \vec{y} \in \{0,1\}^n \setminus {110}$, $f(\vec{x}) = f(\vec{y})$.
Hence, $f$ has a flat energy landscape, except for $\vec{x} = 110$, which encodes the optimum.

\begin{algorithm}[htb]
\caption{Basic steps for Simulated Annealing (minimisation).}
\label{alg:SA}
\SetAlgoLined
\LinesNumbered
\DontPrintSemicolon
    \KwInput{\mygls{pbf} $f(\vec{x})$, steps $S$, initial temperature $\kappa$}
    \KwOutput{Vector $\vec{x}^*$}
    
    $\vec{x} \gets \textsc{choose\_random\_element}(\{0,1\}^n)$, $s \gets 0$\;
    \While{$s < S$}{
        $\kappa' \gets -(s - S)\cdot \kappa$ \tcp*{\normalfont Current temperature}
        $i \gets \textsc{choose\_random\_element}(\{1, \ldots, n\})$\;
        $\vec{x}' \gets \textsc{flip\_bit}(i, \vec{x})$\;
    
        \eIf{$f(\vec{x}') \leq f(\vec{x})$}{
            $\vec{x} \gets \vec{x}'$ \tcp*{\normalfont Trivially accept flip}
        }{
            \If{${\normalfont \textsc{accept\_anyway}}(f(\vec{x}') - f(\vec{x}), \kappa')$}{
                $\vec{x} \gets \vec{x}'$ \tcp*{\normalfont Accept based on random experiment} 
            }
        }
    
        $s \gets s + 1$\;
    }
    \Return{$\vec{x}$}
    \BlankLine
    \SetKwProg{Fn}{Procedure}{}{\Return{False}}
    \Fn{{\normalfont \textsc{accept\_anyway}}($\Delta E$, $\kappa$)}{
        $p \gets \min \{1, e^{-\Delta E / \kappa}\}$\;
    
        \If{${\normalfont \textsc{choose\_random\_element}}([0,1]) < p$}{
            \Return{True}
        }
    }
\end{algorithm}

\begin{figure}[htb]
    \centering
    \includegraphics[width=\linewidth]{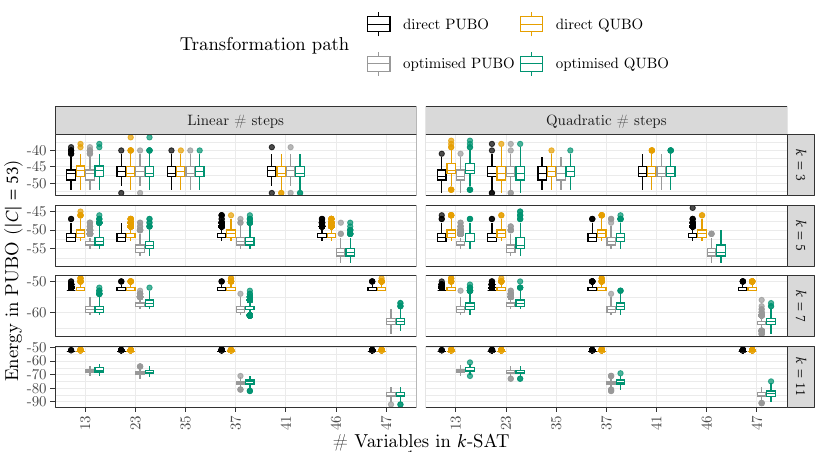}
    \caption{Energy in \mygls{pubo} (y-axis) vs number of variables in non-optimised $k$-\mygls{sat} (x-axis) horizontally faceted by $k$ and vertically faceted by the number of steps in Simulated Annealing (\ie, either linear or quadratic in the y-axis). Each box-plot shows $100$ runs of the best performing initial temperature for varying random initial assignments of variables. Colour represents the concrete transformation path (\ie, from non optimised $k$-\mygls{sat}: d. and from optimised $k$-\mygls{sat}: opt.; see \autoref{tab:TransPathsNamingConvention}). Since \acrshort{optPubo} results from a \mygls{sat} instance with a higher number of clauses, its energy minimum is lower (except for $k=3$) than for \acrshort{dPubo}. Since both stem from the same non-optimised \mygls{sat} instance, the difference in energy for \acrshort{optPubo} and \acrshort{dPubo} has no meaning.}
    \label{fig:SATSABox}
\end{figure}

Since the performance of Simulated Annealing depends on the structure of the energy landscape of a \mygls{pbf}, we use it to compare \mygls{pubo} and \mygls{qubo} formulations. 
This is also interesting in view of a recent study by Dobrynin \etal \cite{Dobrynin_2024} who compare the energy landscapes of \mygls{pubo} and \mygls{qubo} for combinatorial optimisation problems.
Recall that a \textit{quadratisation} has to adhere to \autoref{eq:QuadratizationMinimumPreserving}, which ensures that (under the minimisation of newly introduced variables) each value of the original function is preserved. 
This is usually achieved by multiplying penalty terms (introduced through \mygls{lsr}) by a large positive factor. 
For the Simulated Annealing experiment, we choose Boros penalty factor \cite{Boros_2002} for a \mygls{pbf} $f$ in its multi-linear representation (see \autoref{eq:multi_linear_polynomial}):
\begin{equation}
\label{eq:BorosPenaltyTermQubo}
    1+ 2 \cdot \sum_{S \subseteq \{1, \ldots, n\}} |\alpha_S|.
\end{equation}
Note that this penalty factor differs in value for the non-optimised and optimised \myglspl{pubo}.
Our primary goal is to compare \acrshort{dPubo} with \acrshort{dQubo} and \acrshort{optPubo} with \acrshort{optQubo}.
Since a \textit{quadratisation} does not alter the meaning of previous variables $\vec{x}$ in \mygls{pubo} $f(\vec{x})$, we can evaluate \mygls{pubo} $f(\vec{x})$ with solutions of \mygls{qubo} for a fair comparison.
More precisely, solutions from  \acrshort{dQubo} can be evaluated in \acrshort{dPubo} and solutions from \acrshort{optQubo} can be evaluated in \acrshort{optPubo}.
If, for example, a penalty term is not satisfied, this leads to a large increase in energy for \mygls{qubo} (see \autoref{eq:BorosPenaltyTermQubo}), but not necessarily for \mygls{pubo}.
For the Simulated Annealing experiments, we use a set $K = \{\kappa_i | \kappa_i \geq 1, i \in \mathbb{N}\}$ of initial temperatures to accommodate for different energy landscapes with a recursive sequence $(\kappa_i)_{i \in \mathbb{N}}$, such that $\kappa_1 = 1+ 2 \cdot \sum_{S \subseteq \{1, \ldots, n\}} |\alpha_S|$, as in \autoref{eq:BorosPenaltyTermQubo}, and $\kappa_{i+1} = \frac{\kappa_{i}}{2}$.
We then show $100$ runs for the initial temperature that achieves the lowest energy in \mygls{pubo} for either a linear or a quadratic number of steps in the number of actually used variables in $k$-\mygls{sat} in \autoref{fig:SATSABox}.
Similar to previous \autoref{fig:SATNumVarsPBF} and \ref{fig:SATNumMonomialsPBF}, in \autoref{fig:SATSABox}, we show the number of actually used variables in $k$-\mygls{sat} on the x-axis, use colour for the concrete transformation path (see \autoref{tab:TransPathsNamingConvention}) and show $k$ as horizontal facets.
In contrast, the y-axis shows the energy in \mygls{pubo} and we use vertical facets to differentiate between a linear and quadratic amount of steps.
Additionally, we only show instances with $|C|=53$ clauses (prior to \acrshort{sat} optimisation).
For $k=3$ the energy for the optimised and non-optimised \mygls{pubo} and \mygls{qubo} are similar --- except for random fluctuations ---, since our optimisation strategy does not affect $3$-\mygls{sat} instances (see \autoref{fig:SATCompareExpLinearMapping}).
For higher $k$, the minimum of \acrshort{optPubo} is generally lower than for \acrshort{dPubo}, whenever the optimisation strategy introduces additional clauses.
Hence, \acrshort{dPubo} and \acrshort{optPubo} are not comparable in energy.
When comparing \acrshort{dPubo} with \acrshort{dQubo} and \acrshort{optPubo} with \acrshort{optQubo}, we can see that overall \mygls{pubo} performs (slightly) better. 
Take into consideration that \myglspl{qubo} are at a disadvantage over \myglspl{pubo}, due to an increased number of variables (see \autoref{fig:SATNumVarsPBF}), since the number of steps is fixed and Simulated Annealing considers a random variable flip per iteration.
Since Simulated Annealing --- in essence --- heuristically traverses the node-weighted Hamming Graph of a \mygls{pbf} $f$, which not only depends on the structure of $f$, but also on the cooling schedule, we cannot conclude that \myglspl{qubo} are inferior to \myglspl{pubo} in Simulated Annealing from \autoref{fig:SATSABox}.

\subsection{PBF to Ising}
\label{ssec:PBF2Ising}
Ising models and \myglspl{pbf} are similar representations with the key difference being their domain.
Both models can be used to encode NP-hard optimisation problems \cite{Lucas_2014, cipra2000ising}.
While \myglspl{pbf} are functions $f: \{0,1\}^n \to \mathbb{R}$, Ising models are functions $f_I: \{-1,1\}^n \to \mathbb{R}$.
Hence, \myglspl{pbf} can be transformed to Ising models via the following relation between their variables \cite{Glover_2018}:
\begin{equation*}
    x_i = \frac{1 - s_i}{2},
\end{equation*}
where $x_i \in \{0,1\}$ represent variables in the domain of \myglspl{pbf} and $s_i \in \{-1,1\}$ represent variables in the domain of Ising models.
This mapping leads to equivalent models such that the energy landscape is persevered with $x_i = 0 \iff s_i = 1$ and $x_i = 1 \iff s_i = -1$.
For \myglspl{pbf}, $x^i = x$ ($x \in \{0,1\}$, $i \in \mathbb{N}$) and similar for Ising models ($s \in \{-1,1\}$, $i \in \mathbb{N}$) \footnote{$s_1s_2^4 = s_1\cdot 1 = s_1$. Therefore, monomial size can potentially decrease in Ising models. However, this is not the case for the transformation from \mygls{pbf}, since the transformation does not multiply \mygls{pbf} monomials.}:
\begin{equation*}
    s^i =
    \begin{cases}
        1, &\text{if $i$ even}\\
        s, &\text{if $i$ odd},
    \end{cases}
\end{equation*}
which allows for an analogous representation to multi-linear \myglspl{pbf} (see \autoref{eq:multi_linear_polynomial}).
However, the subtle difference in the domains of \myglspl{pbf} and Ising models has key implications for the metrics shown in \autoref{fig:SAT_overview}, when mapping from \mygls{pbf} to Ising.
For instance, let $f(x_1,x_2,x_3) = x_1x_2x_3$ be a \mygls{pbf}. 
Then its corresponding Ising 
\begin{equation}
\label{eq:IsingExpScalingExample1}
    \begin{split}
        f_I(s_1,s_2,s_3) &= \frac{1 - s_1}{2} \frac{1 - s_2}{2} \frac{1 - s_3}{2}\\
        &= \frac{1}{8}\Big((1-s_1)(1-s_2)(1-s_3)\Big) \\
        &=  \frac{1 -s_1 - s_2 - s_3 + s_1s_2 + s_1s_3 + s_2s_3 - s_1s_2s_3}{8}
    \end{split}
\end{equation}
contains all possible monomials up to degree-$3$. 
As illustrated in \autoref{fig:DegreeSweepVariables}, with degree-$k$ functions $f$, the number of possible monomials up to degree-$k$ grows exponentially in $k$, which points to a similar problem as in the non-optimised mapping from \mygls{sat} to \mygls{pbf} (see \autoref{ssec:OptSAT2PBF}).
Hence, using lower-degree monomials in \mygls{pbf} $f$ (\eg, via \textit{quadratisation}) can be beneficial in terms of the number of monomials in Ising $f_I$.
However, when lower-degree monomials are present in $f$, they might be subsumed by higher-degree monomials. 
For example, when \mygls{pbf} $f(x_1,x_2,x_3) = x_1x_2x_3 + x_1x_2$, then Ising 
\begin{equation}
\label{eq:PBF2IsingExampleNoAdditionalMonomials}
    \begin{split}
        f_I(s_1,s_2,s_3) &= \frac{1 - s_1}{2} \frac{1 - s_2}{2} \frac{1 - s_3}{2} + \frac{1 - s_1}{2} \frac{1 - s_2}{2}\\
        &= \frac{1}{8}\Big((1-s_1)(1-s_2)(1-s_3)\Big) + \frac{1}{4}\Big((1-s_1)(1-s_2) \Big) \\
        &=  \frac{1 -s_1 - s_2 - s_3 + s_1s_2 + s_1s_3 + s_2s_3 - s_1s_2s_3}{8} + \frac{1 - s_1 - s_2 + s_1s_2}{4} \\
        &= \frac{3 -3s_1 - 3s_2 - s_3 + 3s_1s_2 + s_1s_3 + s_2s_3 - s_1s_2s_3}{8}
    \end{split}
\end{equation}
has the same number of monomials as in \autoref{eq:IsingExpScalingExample1}.
Hence, if \mygls{pbf} $f$ has all possible monomials for a subset of variables, the mapping to Ising does not introduce additional monomials for this subset of variables. 
However, as we illustrated in \autoref{fig:DegreeSweepVariables}, an exponential amount of monomials in \mygls{pbf} is infeasible to represent when scaling to larger problems and therefore \myglspl{pbf} tend towards sparse formulations. 
Therefore, the question arises how this transformation to Ising affects the \mygls{qubo} and \mygls{pubo} models.

\begin{figure}[htb]
    \centering
    \includegraphics[width=\linewidth]{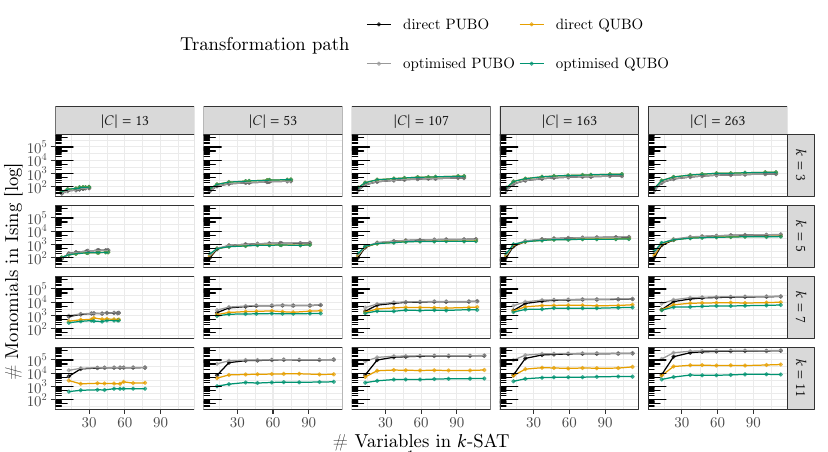}
    \caption{Number of monomials in Ising model (y-axis) vs number of variables in non-optimised $k$-\mygls{sat} (x-axis) horizontally faceted by $k$ and vertically faceted by the number of clauses. Colour represents the concrete transformation path (\ie, from non optimised $k$-\mygls{sat}: d. and from optimised $k$-\mygls{sat}: opt.; see \autoref{tab:TransPathsNamingConvention}).}
    \label{fig:SATNumMonomialsIsing}
\end{figure}

\autoref{fig:SATNumMonomialsIsing} shows the number of monomials in the Ising models (that  originate from the respective \mygls{pbf}; see \autoref{ssec:OptSAT2PBF}) on its y-axis for all four transformation paths (\ie, \acrshort{dPubo}, \acrshort{optPubo}, \acrshort{dQubo}, \acrshort{optQubo}).
For the following observations, take into consideration that both Ising models that originate from \mygls{pubo}, feature higher-order monomials (\ie, of degree-$k$), while Ising models originating from \mygls{qubo} are at most quadratic, although both encode the same \mygls{sat} problem. 
This aspect is not captured in the mere number of monomials which we show on the y-axis.
While for $k=3$ and $k=5$ each path seems to perform equally well, for $k>5$ both Ising models that originate from \mygls{qubo} have significantly less monomials than those that originate from \mygls{pubo}. 
However, recall that even for $k=3$ in \autoref{fig:SATNumMonomialsPBF}, both \mygls{qubo} models have significantly more monomials than their corresponding \mygls{pubo} models.
Upon closer inspection, for $k=5$ in \autoref{fig:SATNumMonomialsIsing}, Ising models stemming from \mygls{qubo} already outperform those stemming from \mygls{pubo}.
For $k=11$ the difference is eminent with $\approx 66$ times less monomials for \mygls{qubo} in the case of $|C| = 263$ clauses, more than $101$ variables and path \acrshort{optQubo} vs \acrshort{optPubo}.
We can also see that reducing the number of monomials in \mygls{pubo} (see \autoref{fig:SATNumMonomialsPBF}) via previous optimisation of the \mygls{sat} instances also lowers the number of monomials in Ising.
We do not explicitly show the number of variables in Ising, since they are exactly the same as in \autoref{fig:SATNumVarsPBF} (\ie, the \textit{quadratisation} introduces new variables). 
However, the number of variables for $k=11$, $|C| = 263$ and paths \acrshort{optQubo} vs \acrshort{optPubo} increases by a factor of $\approx 11$ for \mygls{qubo} (vs factor $\approx 66$ for monomials in \autoref{fig:SATNumMonomialsIsing}).
Take into consideration that these effects are more pronounced with higher $k$ --- except for (specifically generated) instances with inner structures that can be exploited by these transformations.

\begin{figure}[htb]
    \centering
     \begin{subfigure}[b]{0.48\textwidth}
         \centering
         \includegraphics[page=1, width=\textwidth]{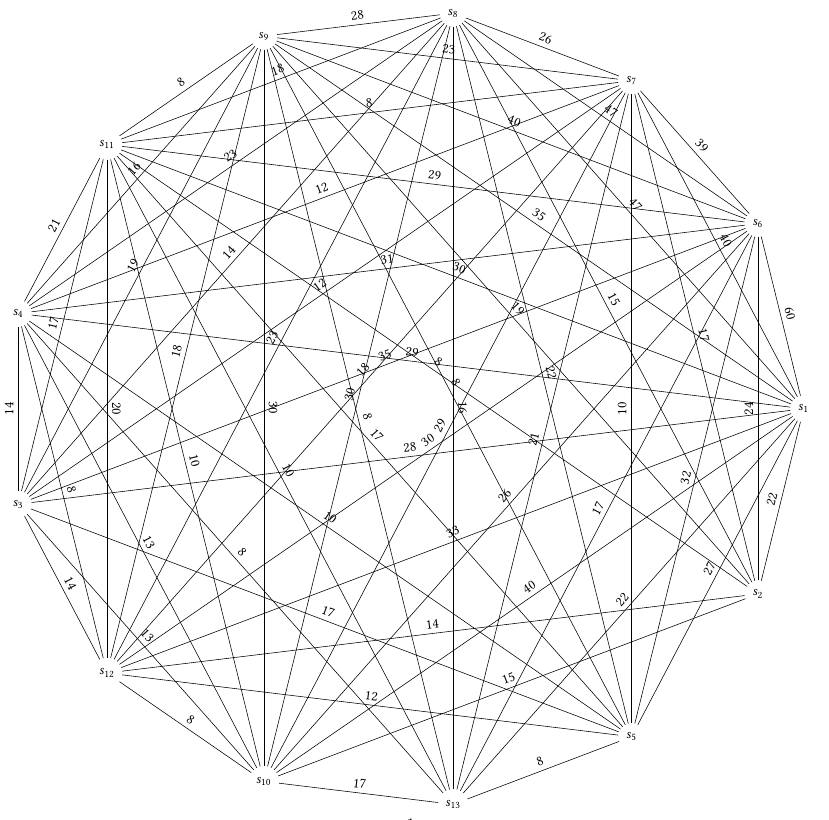}
         \caption{Graph for Ising via \acrshort{dPubo}.}
         \label{fig:SATStructureIsing_dPubo}
     \end{subfigure}
     \hfill
     \begin{subfigure}[b]{0.48\textwidth}
         \centering
         \includegraphics[page=1, width=\textwidth]{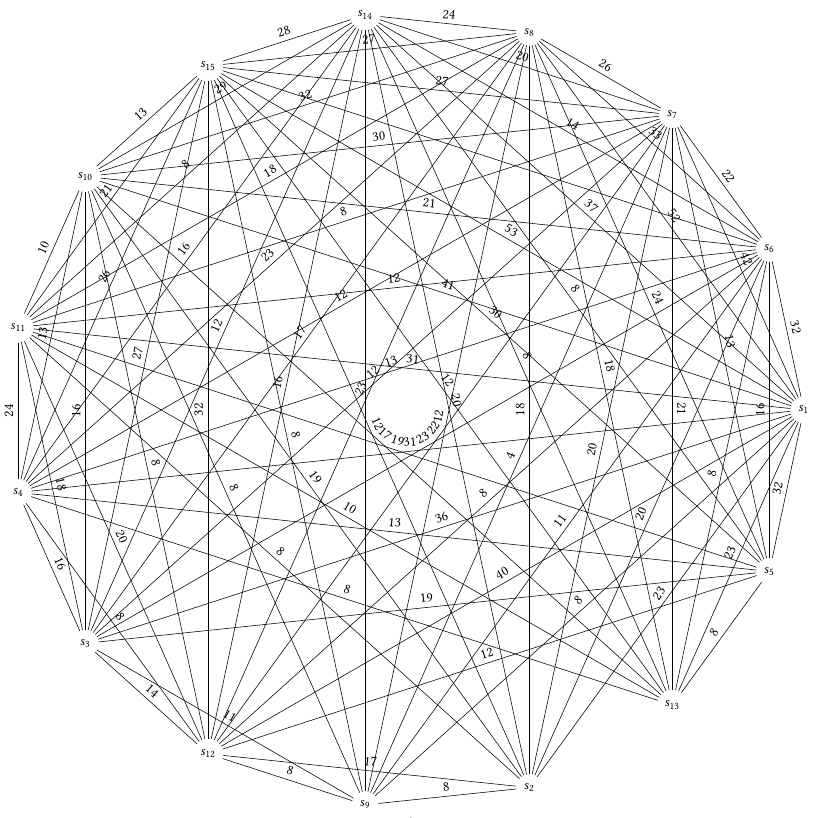}
         \caption{Graph for Ising via \acrshort{optPubo}.}
         \label{fig:SATStructureIsing_optPubo}
     \end{subfigure}
     \hfill
     \begin{subfigure}[b]{0.48\textwidth}
         \centering
         \includegraphics[page=1, width=\textwidth]{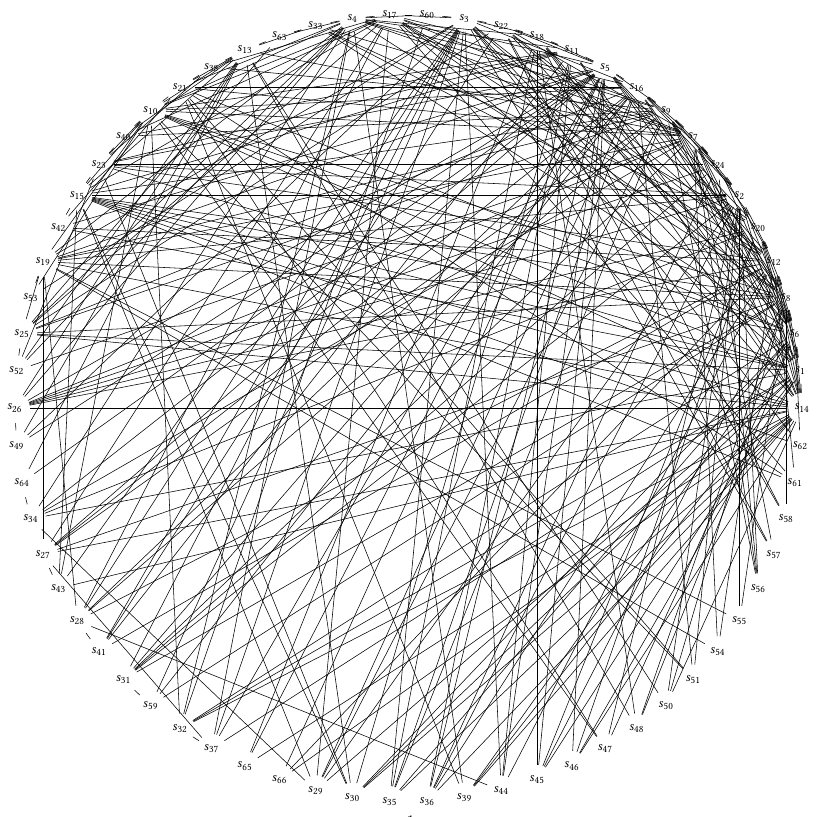}
         \caption{Graph for Ising via \acrshort{dQubo}.}
         \label{fig:SATStructureIsing_dQubo}
     \end{subfigure}
     \hfill
     \begin{subfigure}[b]{0.48\textwidth}
         \centering
         \includegraphics[page=1, width=\textwidth]{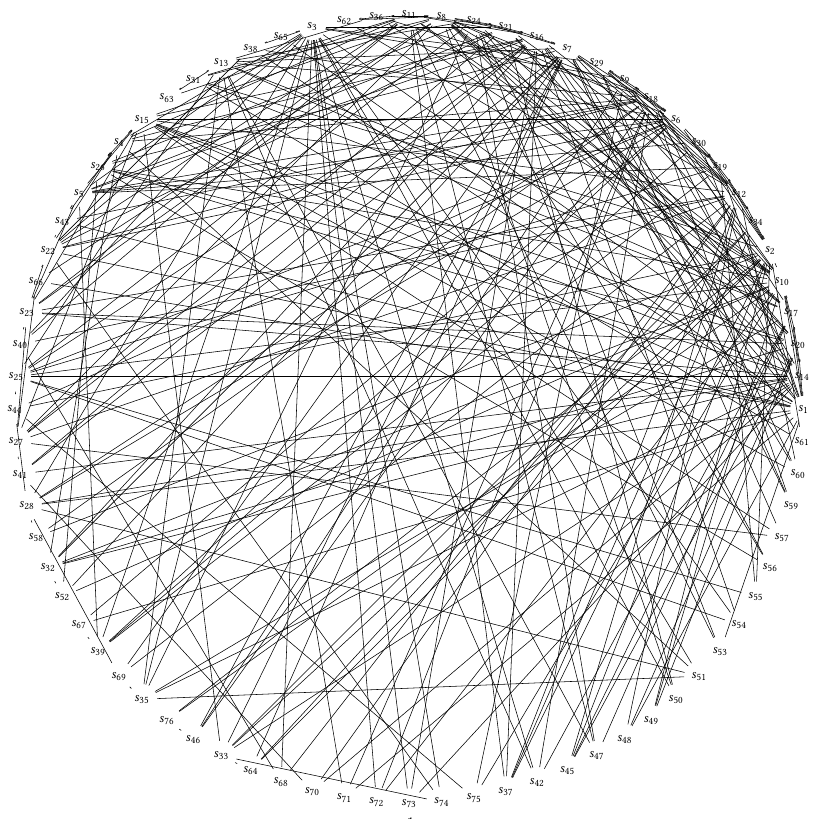}
         \caption{Graph for Ising via \acrshort{optQubo}.}
         \label{fig:SATStructureIsing_optQubo}
     \end{subfigure}
        \caption{Inner structure of Ising models stemming from \mygls{pubo} (top) and from \mygls{qubo} (bottom) that originate from a power law distributed $5$-\mygls{sat} instance with $13$ variables and $37$ clauses (left). Inner structure of its optimised \mygls{sat} formula and derived Ising models stemming from \mygls{pubo} (top) and from \mygls{qubo} (bottom) on the right. New nodes and edges compared to the previous representation (see \autoref{fig:SATStructurePBF} for \mygls{pubo} models) are coloured in \textcolor{lfdred}{red}. See \autoref{fig:SAT_overview} and \autoref{tab:TransPathsNamingConvention} for more information about the concrete transformation paths.}
        \label{fig:SATStructureIsing}
\end{figure}

Since Ising models can also be represented as multi-linear polynomials (over a different domain compared to \myglspl{pbf}), we reuse the graph definition for \myglspl{pbf} and show the inner structure of Ising models in \autoref{fig:SATStructureIsing}.
Since no edges or nodes in \autoref{fig:SATStructureIsing} are coloured red, there are no new connections between nodes and no new nodes in the Ising model compared to their previous \mygls{pubo} or \mygls{qubo} model. 
For Ising models that originate from \mygls{qubo}, we can generally state that no additional degree-$2$ monomials are introduced (see \autoref{eq:PBF2IsingExampleNoAdditionalMonomials} for an example), since there are no degree-$k$, $k>2$ monomials in \mygls{qubo}. 
Hence, their graphs are isomorphic. 
However, there are most certainly (new) degree-$1$ monomials in Ising that are not shown in these graphs. 
More generally, if a \mygls{qubo} graph has no unconnected nodes, then its corresponding Ising model $f_I$ (via the shown transformation) has all possible degree-$1$ monomials (or equivalently $d_1(f_I) = 1$).
This is a result of the term expansion, shown in \autoref{eq:IsingExpScalingExample1}: Every degree-$2$ monomial $\alpha x_ix_j$ in \mygls{pbf} generates two degree-$1$ monomials $s_i$ and $s_j$ in Ising (via the shown transformation).
Although, for the higher-order Ising models that originate from \mygls{pubo} (\autoref{fig:SATStructureIsing_dPubo} and \ref{fig:SATStructureIsing_optPubo}), no new connections are formed, the multiplicities increase, that stem from lower-order terms via term expansion. 
Take into consideration that, since we use a $5$-\mygls{sat} instance with $|C|= 37$ for these graphs (compare to \autoref{fig:SATNumMonomialsIsing}), the increase in multiplicity is rather limited. 
However, for higher $k$, the increase in multiplicities is going to follow the trend of \autoref{fig:SATNumMonomialsIsing} with a potentially different factor that depends on the metric under consideration.
For instance, the sum over all multiplicities will exceed the number of monomials for Ising models originating from \mygls{pubo}, since edges in the graph represent pairs of variables that are contained in monomials.

\subsection{Ising to QAOA} 
\begin{figure}[tbh]
    \centering
    \includegraphics[width=.7\linewidth]{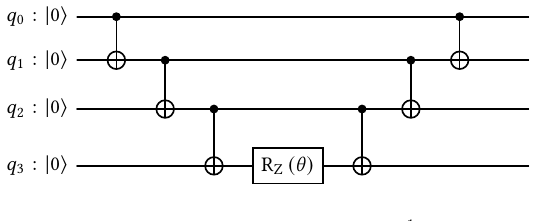}
    \caption{Decomposition of $R_{ZZZZ}(\theta) = e^{-i\frac{\theta}{2} Z \otimes Z \otimes Z \otimes Z}$ into two-qubit CX and single qubit $R_Z(\theta)$ \cite{Campbell_2021}.}
    \label{fig:QAOARZZZDecomposed}
\end{figure}

Ising models form the basis for quantum algorithms, such as \mygls{qaoa}, which can solve combinatorial optimisation problems \cite{Jung2025}.
Farhi \etal \cite{Farhi_2014} originally proposed \mygls{qaoa} as a hybrid quantum-classical algorithm.
Its quantum circuit has a layered structure, which consists of $p$ layers --- each composed of a problem specific unitary $U(H_C, \gamma_i) = e^{-i\gamma_i H_C}$ and a mixer $U(H_M, \beta_i) = e^{-i\beta_i H_M}$. 
Both $U(H_C, \gamma_i)$ and $U(H_M,\beta_i)$ are parametrised by $\gamma_i$ and $\beta_i$ that are subject to a classical optimiser (\eg, a gradient based approach).
The goal is to minimise the expectation value of the cost Hamiltonian $H_C$ that encodes the (optimisation) problem by starting in an equal superposition and then applying $U(H_C, \gamma_i)$ and $U(H_M, \beta_i)$ $p$-times in alternation:
\begin{equation*}
    \ket{\vec{\beta}, \vec{\gamma}} = U(H_M, \beta_p)U(H_C, \gamma_p)\ldots U(H_M, \beta_1)U(H_C, \gamma_1) H^{\otimes n} \ket{0}^{\otimes n},
\end{equation*}
where $H^{\otimes n} \ket{0}^{\otimes n}$ creates an equal superposition on all $n$ qubits.
$H_C$ can be derived from an Ising model by replacing variables $s_i$ with Pauli-Z operators acting on qubit $i$ \cite{pelofske2024short}.
Note that, due to exponentiation of $H_C$, Pauli-Z operators become rotation gates in $U(H_C, \gamma_i)$.
For instance, monomial $\alpha s_1$ is mapped to a rotation gate $R_{Z_1}(\alpha \gamma_i)$ that acts on qubit one. 
Higher-order monomials in the Ising model (\eg, $\alpha s_1s_2s_3$) lead to rotation gates acting on multiple qubits (\eg, $R_{Z_1Z_2Z_3}(\alpha \gamma_i)$ --- acting on qubit one, two and three). 
Since most currently available hardware natively supports up to two qubit interactions and usually does not directly support $R_{ZZ}$ gates, we decompose higher-order rotation gates as it is shown in \autoref{fig:QAOARZZZDecomposed} (although, in principle, it is possible to realise multi-qubit gates in trapped ion devices \cite{Shapira_2020}). 
Also, when generating quantum circuits from Ising models, this is in favour of a fair comparison between higher- and lower-order Ising models.
The mixer $U(H_M, \beta_i)$ can be represented as $U(H_M, \beta_i) = \sum_j R_{X_j}(\beta_i)$, where $R_{X_j}$ is a rotation around the $X$-axis acting on qubit $j$.

Recently, Montañez-Barrera and Michielsen \cite{MontaezBarrera_2025} analysed a fixed linear-ramp schedule for \mygls{qaoa} for combinatorial optimisation problems. 
In contrast to iterative \mygls{qaoa}, the linear-ramp variant does not require a classical optimiser to adjust parameters $\gamma_i$ and $\beta_i$ in each layer of the quantum circuit and thus the quantum circuit only needs to be executed once (with a specific number of shots).
The use of predetermined values for $\gamma_i$ and $\beta_i$ (\ie, linear-ramp) requires that the objective function is normalised, which would otherwise be delegated to the classical optimiser.

\begin{figure}[tbh]
    \centering
    \includegraphics[width=\linewidth]{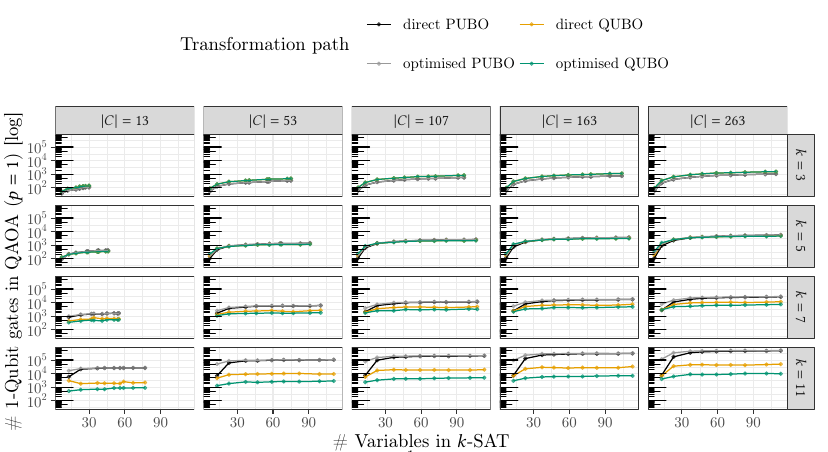}
    \caption{Number of single-qubit gates in quantum circuit for \mygls{lrqaoa} for $p=1$ (y-axis) vs number of variables in non-optimised $k$-\mygls{sat} (x-axis) horizontally faceted by $k$ and vertically faceted by the number of clauses. Colour represents the concrete transformation path (\ie, from non optimised $k$-\mygls{sat}: d. and from optimised $k$-\mygls{sat}: opt.; see \autoref{tab:TransPathsNamingConvention}).}
    \label{fig:QAOANumOneQubitGates}
\end{figure}

We will use \mygls{lrqaoa} for the following experiments. 
Since the structure of circuits in \mygls{lrqaoa} is equal to those in \mygls{qaoa}, the results --- featuring circuit metrics --- are also applicable to \mygls{qaoa}. 
\autoref{fig:QAOANumOneQubitGates} shows the number of single-qubit gates in a quantum circuit for \mygls{lrqaoa} with one layer ($p=1$). 
This includes all gates introduced by $U(H_M, \beta_i)$, all Hadamard gates $H^{\otimes n}$ ($n$ is the number of qubits) and all (decomposed) single-qubit gates from $U(H_C, \gamma_i)$.
Since we decompose higher-order gates --- leaving one single-qubit $R_Z$ gate per monomial\footnote{Except for the constant monomial.} ---, the number of monomials in Ising (see \autoref{fig:SATNumMonomialsIsing}) closely depicts the number of single-qubit gates introduced by $U(H_C, \gamma_i)$.
Moreover, since the number of single qubit gates introduced by $U(H_M, \beta_i)$ and $H^{\otimes n}$ is exactly two times the number of qubits, \autoref{fig:QAOANumOneQubitGates} closely depicts the sum of \autoref{fig:SATNumVarsPBF} and \ref{fig:SATNumMonomialsIsing} --- with analogous observations.

\begin{figure}[tbh]
    \centering
    \includegraphics[width=\linewidth]{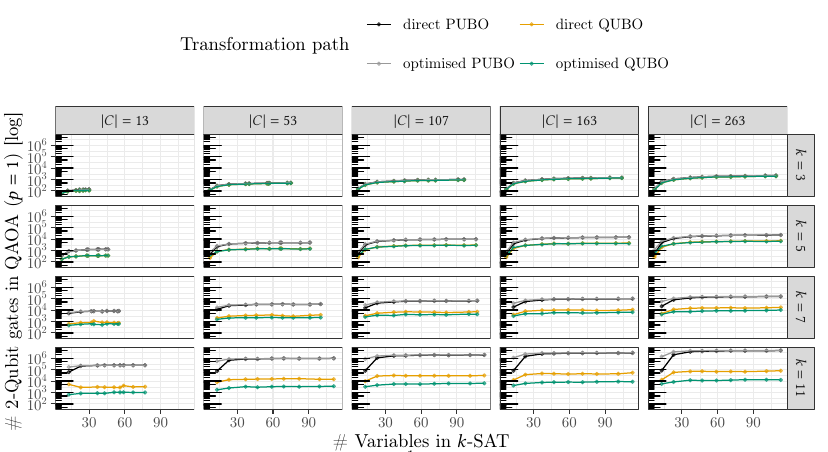}
    \caption{Number of two-qubit gates in quantum circuit for \mygls{lrqaoa} for $p=1$ (y-axis) vs number of variables in non-optimised $k$-\mygls{sat} (x-axis) horizontally faceted by $k$ and vertically faceted by the number of clauses. Colour represents the concrete transformation path (\ie, from non optimised $k$-\mygls{sat}: d. and from optimised $k$-\mygls{sat}: opt.; see \autoref{tab:TransPathsNamingConvention}).}
    \label{fig:QAOANumTwoQubitGates}
\end{figure}

Apart from single-qubit gates, \autoref{fig:QAOANumTwoQubitGates} shows the number of two-qubit gates in the aforementioned quantum circuit for \mygls{lrqaoa} (see \autoref{fig:QAOANumOneQubitGates}).
In contrast to \autoref{fig:QAOANumOneQubitGates}, (decomposed) gates only stem from (higher-order) monomials in the corresponding Ising model.
Recall that the mere number of monomials in Ising (\autoref{fig:SATNumMonomialsIsing}) does not depict the size of monomials.
Conversely, \autoref{fig:QAOANumTwoQubitGates} takes this into account due to the applied decomposition: Any degree-$k$ monomial in Ising results in $2(k-1)$ two-qubit gates in the quantum circuit. 
Hence, the beneficial effects of \mygls{qubo} models in terms of the number of gates are even more pronounced than in Ising (Note that \autoref{fig:SATNumMonomialsIsing} and \autoref{fig:QAOANumTwoQubitGates} have different y-axis scales).
Comparing the paths for optimised \mygls{qubo} and \mygls{pubo} for $k=11$, $|C|=263$ and the maximum tested amount of variables in $k$-\mygls{sat}, the path for \mygls{qubo} copes with $\approx 383$ times less gates in the quantum circuit, while requiring $\approx 11$ times more variables.
While for Ising models, these effects are more pronounced for higher $k$, they are especially pronounced for quantum circuits, due to the decomposition strategy (a more fair comparison).
Moreover, quantum circuits for \mygls{lrqaoa} that originate from quadratic Ising models can be depth optimised almost optimally (up to one additional layer) in polynomial time, due to Vizing's theorem \cite{Berge_1991}. 
More technically, a (at most) quadratic Ising model is mapped to a graph, for which a proper edge-colouring then determines the circuits depth.
In higher-order Ising models, this mapping results in a hypergraph (\ie, an edge can connect multiple nodes) for which (in general) it is difficult to find an optimal edge-colouring (see \cite{Herrman_2021} for more information).
Also, take into account that all tested models (via paths \acrshort{dPubo}, \acrshort{optPubo}, \acrshort{dQubo}, \acrshort{optQubo},) encode the same initial $k$-\mygls{sat} instances --- highlighting the relevance of transformation paths and their study, due to vastly different properties of resulting quantum specific representations (especially when considering scaling behaviour).

\section{Conclusion and Outlook}
\label{sec:concl}
We introduce an optimised algorithm, subdivided into two stages, to compute a \emph{quadratisation} for pseudo boolean functions (\myglspl{pbf}) that play a major role in the formulation of combinatorial optimisation problems (\myglspl{cop}) --- not limited to quantum computing.
We give a thorough mathematical analysis of its inner workings and prove properties related to the underlying graph structure, which ultimately leads to the performance gain. 
Furthermore, we give complexity theoretic bounds on its performance and show empirically that the proposed algorithm outperforms the existing monomial-based algorithm.
On top of that, the introduced algorithm is more versatile in terms of selecting specific characteristics of the quadratised function (\ie, the degree-$2$ density and the number of introduced variables) --- enabling it to be in turn used in an automatic transformation process that optimises quantum circuit metrics.
We prove that a reduction iteration acts locally on the graph representation, which paves the way for future parallel execution. 
Moreover, it is easy to extend our proposed algorithm to not only allow for \emph{quadratisations} (\ie, the reduction to a degree-$2$ function), but also for higher-degree reductions (\ie, degree-$k$, $k>2$).

Moreover, we show in detail how industrial-like \mygls{sat} instances are connected to quantum computing by a multitude of explicit transformation processes --- also putting into perspective where \emph{quadratisations} are located. 
We identify major limitations of scaling to larger problem instances and show explicitly how to circumvent them.
As a result of that, we identify that lower-order models (such as \mygls{qubo}) have advantages over higher-order models (\mygls{pubo}), when it comes to scaling behaviour in quantum circuits (\eg, for \mygls{qaoa}) and intermediate representations (\eg, Ising models). 
However, \mygls{pubo} models have better expressivity in higher abstraction levels and typical formulations of \mygls{sat} lead to \mygls{pubo} models.
Our introduced \mygls{lsr} algorithm allows for fast transformation from higher- to lower-order models (\eg, \mygls{qubo}), which on the one hand enables practitioners to use higher-order models in higher abstraction, while on the other hand taking advantage over the positive scaling behaviour of lower-order models in lower abstraction layers.

Ultimately, we desire an (automated) quantum toolchain that optimises metrics of hardware specific representations. 
Such a toolchain benefits from predictable transformation processes.
With this work, we show analytically and quantitatively to which degree and how transformations change properties of (intermediate) (quantum specific) representations.

\begin{small}
  \noindent\textbf{Acknowledgements}
  This work was supported by the German Research Foundation, grant MA \text{9739/1-1} and SCHA \text{1635/20-1}, as well as by the German Federal Ministry of Education and Research (BMBF), funding program ‘quantum technologies—from basic research to market’, grant number 13N16092 and by the European Union (Project Reference 101083427) and the European Funds for Regional Development (EFRE) (Project Reference 20-3092.10-THD-105) and the project ‘Algorithms for quantum computer development in hardware-software codesign’ (ALQU), \href{https://qci.dlr.de/en/alqu}{qci.dlr.de/en/alqu}, which was made possible by the DLR Quantum Computing Initiative (QCI) and the German Federal Ministry for Economic Affairs and Climate Action (BMWK). WM acknowledges support by the High-Tech Agenda of the Free State of Bavaria.
\end{small}

\bibliographystyle{ACM-Reference-Format}
\bibliography{references.bib}

\end{document}